\documentclass{article}
\usepackage[utf8]{inputenc}

\usepackage{amssymb}
\usepackage{listings}
\usepackage{amsmath,bm}
\usepackage{txfonts}
\usepackage{mathtools}
\usepackage{statmath}
\usepackage[dvipsnames]{xcolor}
\usepackage{xcolor}
\usepackage{caption}
\usepackage{nicematrix}
\usepackage{tikz}
    \usetikzlibrary{matrix, positioning}
\usepackage{multirow}
\usepackage[style=chem-angew,sorting=none]{biblatex}

\definecolor{codegreen}{rgb}{0,0.6,0}
\definecolor{codegray}{rgb}{0.5,0.5,0.5}
\definecolor{codepurple}{rgb}{0.58,0,0.82}
\definecolor{backcolour}{rgb}{0.95,0.95,0.92}

\lstdefinestyle{mystyle}{
    backgroundcolor=\color{backcolour},   
    commentstyle=\color{codegreen},
    keywordstyle=\color{magenta},
    numberstyle=\tiny\color{codegray},
    stringstyle=\color{codepurple},
    basicstyle=\ttfamily\scriptsize,
    breakatwhitespace=false,         
    breaklines=true,                 
    captionpos=b,                    
    keepspaces=true,              
    showspaces=false,                
    showstringspaces=false,
    showtabs=false,                  
    tabsize=2,
    escapeinside={\%*}{*)}
}
\makeatletter
\newsavebox{\@brx}
\newcommand{\llangle}[1][]{\savebox{\@brx}{\(\m@th{#1\langle}\)}%
  \mathopen{\copy\@brx\mkern2mu\kern-0.9\wd\@brx\usebox{\@brx}}}
\newcommand{\rrangle}[1][]{\savebox{\@brx}{\(\m@th{#1\rangle}\)}%
  \mathclose{\copy\@brx\mkern2mu\kern-0.9\wd\@brx\usebox{\@brx}}}
\makeatother

\makeatletter
\newcommand{\doublehat}[1]{% 
\begingroup%
  \let\macc@kerna\z@%
  \let\macc@kernb\z@%
  \let\macc@nucleus\@empty%
  \hat{\mathchoice%
    {\raisebox{.3ex}{\vphantom{\ensuremath{\displaystyle #1}}}}%
    {\raisebox{.3ex}{\vphantom{\ensuremath{\textstyle #1}}}}%
    {\raisebox{.16ex}{\vphantom{\ensuremath{\scriptstyle #1}}}}%
    {\raisebox{.14ex}{\vphantom{\ensuremath{\scriptscriptstyle #1}}}}%
    \smash{\hat{#1}}}%
\endgroup%
}
\makeatother

\def\measureaccent#1#2{%
   \setbox0=\vbox{$#1{#2}\hfil\break$\null\par
      \setbox0=\lastbox\unskip\unpenalty\global\setbox1=\lastbox}%
   \setbox0=\hbox{\unhbox1 \unskip\unpenalty\unskip \global\setbox3=\lastbox}%
   \setbox0=\vbox{\unvbox3 \setbox0=\lastbox}%
}
\def\doubleaccent#1#2{%
   \measureaccent{#1}{#2}\dimen0=\wd0 \measureaccent{#1}{\kern0pt#2}%
   \raise.3ex\rlap{\kern\dimexpr\dimen0-\wd0$#1{\phantom{#2}}$}{#1#2}%
}

\makeatother

\makeatletter
\newcommand{\heading}[1]% #1 = text
{\par\vskip 1.5ex \@plus .2ex
 \hangindent=1em
 \noindent\makebox[1em][l]{$\,\bullet$}\textbf{\normalsize #1}%
\par\vskip 1.5ex \@plus .2ex
\@afterheading}
\makeatother

\DeclareRobustCommand{\svdots}{% s for `scaling'
  \vbox{%
    \baselineskip=0.33333\normalbaselineskip
    \lineskiplimit=0pt
    \hbox{.}\hbox{.}\hbox{.}%
    \kern-0.2\baselineskip
  }%
}
\DeclareRobustCommand{\sddots}{% s for `scaling'
  \vbox{%
    \baselineskip=0.33333\normalbaselineskip
    \lineskiplimit=0pt
    \hbox{.\,\,\,\,}\hbox{\,\,.\,\,}\hbox{\,\,\,\,.}%
    \kern-0.2\baselineskip
  }%
}

\title{Essential Tools of Linear Algebra for Calculating Nuclear Spin Dynamics of Chemically Exchanging Systems}
\author{Jingyan Xu and Danila A. Barskiy}

\begin{document}

\maketitle

\begin{abstract}
In this work, we describe essential tools of linear algebra necessary for calculating the effect of chemical exchange on spin dynamics and polarization transfer in various nuclear magnetic resonance (NMR) experiments. We show how to construct Hamiltonian, relaxation, and chemical exchange superoperators in the Liouville space, as well as demonstrate corresponding code in Python. Examples of applying the code are given for problems involving chemical exchange between NH$_3$ and NH$_4^{+}$ at zero and high magnetic field and polarization transfer from parahydrogen relevant in SABRE (signal amplification by reversible exchange) at low magnetic field (0-20\,mT). The presented methodology finds utility for describing the effect of chemical exchange on NMR spectra and can be extended further by taking into account non-linearities in the master equation.
\end{abstract}

\section{Introduction}

%Understanding chemical exchange phenomena is extremely important for correct interpretation of data in nuclear magnetic resonance (NMR) and electron paramagnetic resonance (ESR) spectroscopy. Chemical exchange occurs when atoms or groups of atoms within a molecule (intramolecular exchange) or between different molecules (intermolecular exchange) switch their positions. This can lead to broadening (and even coalescence) of NMR peaks, making it difficult to interpret spectra and extract useful information about molecular structure and dynamics \cite{levitt2013spin,ernst1987principles,Bain2003}. 

Understanding chemical exchange phenomena is extremely important for the correct interpretation of data in nuclear magnetic resonance (NMR) and electron paramagnetic resonance (ESR) spectroscopy. Chemical exchange is a term used to describe a type of molecular motion that involves changes in molecular structure due to molecular conformations or the formation/breaking of chemical bonds. This can lead to broadening (and even coalescence) of NMR peaks, making it difficult to interpret spectra and extract useful information about molecular structure and dynamics \cite{levitt2013spin,ernst1987principles,Bain2003}. In addition, understanding NMR spectra in the presence of chemical exchange is crucial for developing new  magnetic resonance imaging (MRI) techniques that can provide insights into complex biological systems. For example, chemical exchange saturation transfer (CEST) \cite{Zhou2006,Zijl2011,liu2013nuts,ferrauto2017chemical} and magnetization transfer methods \cite{Bain2003,Zijl2018,Ward2000}, are actively being developed for investigating biological processes in living species.

Kaplan and McConnell were the first to demonstrate basic approaches for describing NMR spectra of chemically exchanging systems \cite{kaplan1958exchange,kaplan1958generalized,WOS:A1946UB26500003,WOS:A1958WA15400013}. Alexander and Binsch considered cases where the density matrix formalism is necessary to accurately describe the behavior of spin systems (for example, when spin-spin interactions cannot be disregarded) \cite{WOS:A19624143B00001,WOS:A19624143B00002,WOS:A1969C932900007}. As of today, many excellent papers, reviews, and books touch upon this topic \cite{kuhne1979study,Bain2003,kaplan2012nmr,bengs2021markovian,lindale2020infinite,jeener1982superoperators,abergel2003use}. However, the approach is typically described in a manner that is extremely difficult to follow for a chemist who is looking for clear visual representations of the underlying equations. In this paper, we describe linear algebra tools necessary for calculating nuclear spin dynamics using the density operator formalism focusing on constructing matrix representations of all necessary operators in both Hilbert and Liouville spaces. We substantiate the discussion with references to code in Python. The authors do not claim to invent new physics, however, we believe such a comprehensive account of linear algebra tools necessary for calculating chemical exchange in NMR will find their use by an interested reader who is capable of writing their own code for a particular system undergoing exchange.

\begin{figure}[ht]
    \centering
    \includegraphics[scale=1]{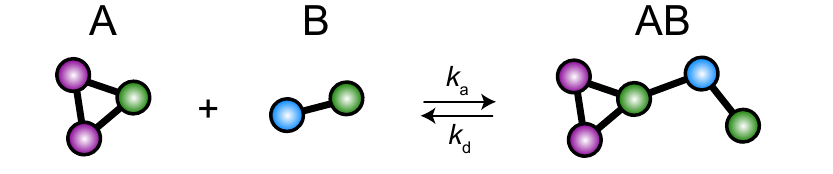}
    \caption{Example of a chemically exchanging system where a molecule A associates with a molecule B to generate a molecule AB, which can then dissociate back into A and B. The spin state of the molecule B is assumed to be constant.}
    \label{fig:fig1_reaction}
\end{figure}

In our work, we want to describe full spin dynamical evolution of the system subject to the chemical exchange reaction shown in Fig.~\ref{fig:fig1_reaction}. This definition allows us to focus on intermolecular exchange in which after dissociation of the molecule B from molecule AB, another molecule B can attach to A.

The differential equations describing the chemical kinetics of this system can be written in the matrix form as:
\begin{equation}
    \frac{d}{dt}\begin{pmatrix}
    [\mathrm{A}] \\
    [\mathrm{AB}]
    \end{pmatrix}
    =\begin{pmatrix}
    -k_a[\mathrm{B}] & k_d\\
     k_a[\mathrm{B}] & -k_d
    \end{pmatrix}
    \begin{pmatrix}
    [\mathrm{A}] \\
    [\mathrm{AB}]
    \end{pmatrix}
    =\begin{pmatrix}
    -\tilde{k}_a & +k_d\\
    +\tilde{k}_a & -k_d
    \end{pmatrix}
    \begin{pmatrix}
    [\mathrm{A}] \\
    [\mathrm{AB}]
    \end{pmatrix},
\label{Eq: Chemical kinetics}
\end{equation}
where [A], [B], and [AB] are concentrations of molecules A, B, and AB, respectively.

As we will see later, putting [B] in the kinetic matrix is reasonable if the spin state of B is not changing with time: for example, if the spin state of B is actively being refreshed, such as parahydrogen in signal amplification by reversible exchange (SABRE) experiments, or if B can be considered unpolarized at all times, as the case for H$^+$ exchanging with ammonia). It is also convenient to introduce the effective association rate $\tilde{k}_a=k_a[\rm B]$.

In the following, we will describe spin dynamics in the Liouville space (space of the quantum-mechanical operators) instead of the commonly used Hilbert space (space of the quantum-mechanical states)  \cite{WOS:000579258400001}. Working in the Liouville space is advantageous when considering both relaxation and chemical exchange as the related superoperators (denoted by double-hats) only need to be calculated once, as opposed to the Hilbert space treatment. In the absence of chemical exchange, the spin dynamics of molecule i (i=A,AB) is described by the well known Liouville-Von Neumann (LvN) equation:
\begin{equation}
    \frac{d}{dt}\hat{\rho}_{\mathrm{i}}=\doublehat{L}_{\rm i}
    \hat{\rho}_{\mathrm{i}}.
\label{Eq: LvN without exchange}
\end{equation}
Here $\hat{\rho}_{\mathrm{i}}$ represents the density operator of the corresponding to molecule i. The Liouvillian superoperator, $\doublehat{L}_{\rm i}=-i\doublehat{H}_{\rm i}+\doublehat{R}_{\rm i}$, acting on $\hat{\rho}_{\mathrm{i}}$, governs both the coherent evolution, dictated by the Hamiltonian superoperators $\doublehat{H}_{\rm i}= [\hat{H}_{\rm i},\cdot]$, and the incoherent relaxation, represented by the relaxation superoperator, $\doublehat{R}_{\rm i}$, of the nuclear spins within the molecule. Section~\ref{Section: rho in L sapce} shows how the superoperators act on density operators in Hilbert space. Moreover, a detailed explanation of the constructions of Hamiltonian superoperators and relaxation superoperators will be provided in Section~\ref{Section: Hamiltonian Superoperator} and Section~\ref{Section: Relaxation Superoperator}, respectively.

Regarding the effects of chemical exchange, the discussion below is based on the following assumptions: 1) individual molecular events of association and dissociation occur much faster than nuclear spin dynamics; 2) the state of molecule B remains unchanged; and 3) chemical kinetic rates are independent of the nuclear spin degrees of freedom. Under these assumptions, the master equation describing the effects of the chemical exchange can be derived by combining Eq.~\eqref{Eq: Chemical kinetics} and certain chemical exchange superoperatros.

\begin{figure}[!ht]
    \centering
    \includegraphics{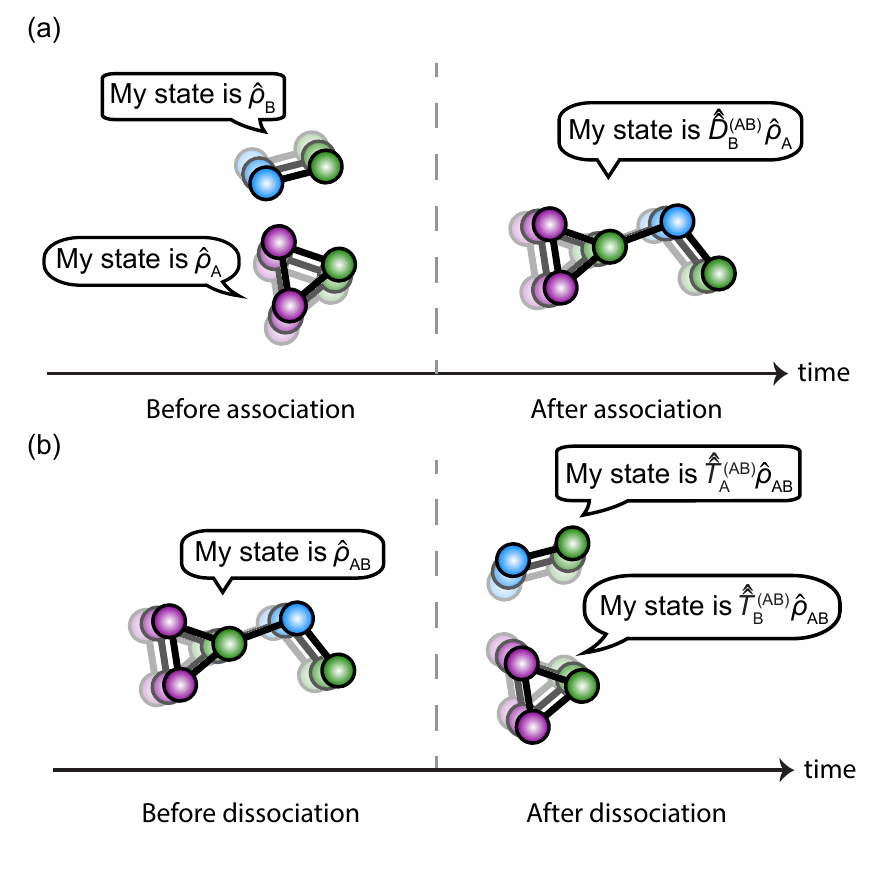}
    \caption{Spin states of individual molecules A, B, and AB before and after (a) the association event and (b) the dissociation event.}
    \label{fig:Chemical exchange superoperators visual}
\end{figure}

Let us begin by analyzing the association and dissociation processes at the single-molecule level using the density operator approach. Figure~\ref{fig:Chemical exchange superoperators visual}(a) illustrates an association event where molecules A and B associates to form molecule AB. The figure also presents the corresponding density operators of the involved molecules. Let $\hat{\rho}_{\mathrm{A}}$ and $\hat{\rho}_{\mathrm{B}}$ denote the density operators of molecules A and B, respectively, at the moment immediately prior to their association. The density operator of associated molecule AB is obtained by taking the Kronecker product between $\hat{\rho}_{\mathrm{A}}$ and $\hat{\rho}_{\mathrm{B}}$, resulting in $\hat{\rho}_{\mathrm{A}} \otimes \hat{\rho}_{\mathrm{B}}$. However, since our primary focus is on the evolution of $\hat{\rho}_{\mathrm{A}}$, we can express this Kronecker product in a form linearly dependent on $\hat{\rho}_{\mathrm{A}}$ as
\begin{equation*}
        \hat{\rho}_{\mathrm{A}} \otimes \hat{\rho}_{\mathrm{B}}=\doublehat{D}_{\mathrm{B}}^{\mathrm{(AB)}}\hat{\rho}_{\mathrm{A}},
\end{equation*}
where the association superoperator, $\doublehat{D}_{\mathrm{B}}^{\mathrm{(AB)}}$, acts on $\hat{\rho}_{\mathrm{A}}$ and transforms it into $\hat{\rho}_{\mathrm{A}} \otimes \hat{\rho}_{\mathrm{B}}$. Here, the subscript B emphasizes the dependence of the superoperator on  $\hat{\rho}_{\mathrm{B}}$, while the superscript (AB) indicates the order in which the Kronecker product between $\hat{\rho}_{\mathrm{A}}$ and $\hat{\rho}_{\mathrm{B}}$ takes place.

A similar analysis applies to the dissociation process. As shown in Fig.~\ref{fig:Chemical exchange superoperators visual}(b) where molecule AB dissociates into molecules A and B. Let $\hat{\rho}_{\mathrm{AB}}$ denote the density operator of molecule AB, at the moment immediately prior to its dissociation. The density operator of dissociated molecule A is obtained by taking the partial trace on $\hat{\rho}_{\mathrm{AB}}$ over the degrees of freedom associated with molecule B, resulting in $\mathrm{Tr}_{\mathrm{B}}(\hat{\rho}_{\mathrm{AB}})$. This partial trace can also be expressed in a form linearly dependent on $\hat{\rho}_{\mathrm{AB}}$ as
\begin{equation*}
    \mathrm{Tr}_{\mathrm{B}}(\hat{\rho}_{\mathrm{AB}}) = \doublehat{T}_{\mathrm{B}}^{\mathrm{(AB)}}\hat{\rho}_{\mathrm{AB}},
\end{equation*}
where the partial trace superoperator, $\doublehat{T}_{\mathrm{B}}^{\mathrm{(AB)}}$, acts on $\hat{\rho}_{\mathrm{AB}}$ by taking a partial trace over molecule B. Similarly, the density operator of dissociated molecule B is given by $\doublehat{T}_{\mathrm{A}}^{\mathrm{(AB)}}\hat{\rho}_{\mathrm{AB}}$, where the partial trace is taken over molecule A, as indicated by the subscript. We note that the order indicated by the superscript is still important to distinguish, for example, $\doublehat{T}_{\mathrm{A}}^{\mathrm{(AB)}}$ and $\doublehat{T}_{\mathrm{A}}^{\mathrm{(BA)}}$.

Now, we can examine the effect of chemical exchange on the density operator at the ensemble level, temporarily disregarding the effects of coherent evolution and incoherent relaxation. Let $\hat{\rho}_{\mathrm{A}}$ and $\hat{\rho}_{\mathrm{AB}}$ represent the ensemble-averaged density operators for molecules A and AB, respectively, at a certain time. After a short time interval $\Delta t$, these density operators evolve due to chemical exchange. The evolved density operators can be obtained by taking a concentration-weighted average between the ensembles of non-reacted molecules and those generated during $\Delta t$.

Assuming that $\Delta t$ is small enough such that each individual molecule reacts at most once, we can determine the concentrations and density operators of the non-reacted and reaction-produced molecules for A and AB after $\Delta t$. Let $n_a$ and $n_d$ denote the concentrations of molecules A and AB consumed during $\Delta t$, respectively. For molecule A, the concentrations of non-reacted and reaction-produced A molecules are $[\mathrm{A}]-n_a$ and $n_d$, respectively, while their density operators are $\hat{\rho}_{\mathrm{A}}$ and $\doublehat{T}_{\mathrm{B}}^{\mathrm{(AB)}}\hat{\rho}_{\mathrm{AB}}$. Similarly, for molecule AB, the concentrations of non-reacted and reaction-produced C molecules are $[\mathrm{AB}]-n_d$ and $n_a$, respectively, and their density operators are $\hat{\rho}_{\mathrm{AB}}$ and $\doublehat{D}_{\mathrm{B}}^{\mathrm{(AB)}}\hat{\rho}_{\mathrm{A}}$.

Consequently, the evolved density operators for molecules A and AB are determined by taking the concentration-weighted averages of the density operators of the non-reacted and the reaction-produced molecules, as discussed earlier. Let $\Delta (\hat{\rho}_{\mathrm{A}})$ and $\Delta (\hat{\rho}_{\mathrm{AB}})$ represent the changes of the density operators for molecules A and AB, respectively, during the time interval $\Delta t$. The evolved density operators are expressed as:
\begin{equation*}
    \begin{aligned}
        \hat{\rho}_{\mathrm{A}}+\Delta (\hat{\rho}_{\rm A}) & = \frac{([\mathrm{A}]-n_a)\hat{\rho}_{\mathrm{A}}+n_d\doublehat{T}_{\mathrm{B}}^{\mathrm{(AB)}}\hat{\rho}_{\mathrm{AB}}}{[\mathrm{A}]-n_a+n_d};\\
        \hat{\rho}_{\mathrm{AB}}+\Delta(\hat{\rho}_{\mathrm{AB}}) & = \frac{([\mathrm{AB}]-n_d)\hat{\rho}_{\mathrm{AB}}+n_a \doublehat{D}_{\mathrm{B}}^{\mathrm{(AB)}}\hat{\rho}_{\mathrm{A}}}{[\mathrm{AB}]-n_d+n_a}.
    \end{aligned}
\end{equation*}
Let $\Delta ([\mathrm{A}])$ and $\Delta ([\mathrm{AB}])$ denote the concentration changes for molecules A and AB, respectively, during the time interval $\Delta t$. Since $\Delta( [\mathrm{A}]) = -n_a + n_d$ and $\Delta ([\mathrm{AB}]) = n_a - n_d$, we can substitute $-n_a + n_d$ with $\Delta ([\mathrm{A}])$ and $n_a - n_d$ with $\Delta ([\mathrm{AB}])$ in the denominators. By multiplying both sides of the equations by the corresponding denominators, we can rearrange the equations as follows:
\begin{equation*}
    \begin{aligned}
        \Big([\mathrm{A}]+\Delta([\mathrm{A}])\Big)\Big(\hat{\rho}_{\mathrm{A}}+\Delta(\hat{\rho}_{\rm A})\Big) - [\mathrm{A}]\hat{\rho}_{\mathrm{A}} &=  -n_a \hat{\rho}_{\mathrm{A}}+n_d\doublehat{T}_{\mathrm{B}}^{\mathrm{(AB)}}\hat{\rho}_{\mathrm{AB}};  \\
        \Big([\mathrm{AB}]+\Delta ([\mathrm{AB}])\Big)\Big(\hat{\rho}_{\mathrm{AB}}+\Delta(\hat{\rho}_{\mathrm{AB}})\Big) - [\mathrm{AB}]\hat{\rho}_{\mathrm{AB}} &=  -n_d \hat{\rho}_{\mathrm{AB}} + n_a \doublehat{D}_{\mathrm{B}}^{\mathrm{(AB)}}\hat{\rho}_{\mathrm{A}}. \\
    \end{aligned}
\end{equation*}
Since the left sides of the equations above can be further simplified as
\begin{equation*}
    \begin{aligned}
         \Big([\mathrm{A}]+\Delta([\mathrm{A}])\Big)\Big(\hat{\rho}_{\mathrm{A}}+\Delta(\hat{\rho}_{\rm A})\Big) - [\mathrm{A}]\hat{\rho}_{\mathrm{A}} & = \Delta([\mathrm{A}]\hat{\rho}_{\mathrm{A}});\\
         \Big([\mathrm{AB}]+\Delta ([\mathrm{AB}])\Big)\Big(\hat{\rho}_{\mathrm{AB}}+\Delta(\hat{\rho}_{\mathrm{AB}})\Big) - [\mathrm{AB}]\hat{\rho}_{\mathrm{AB}} & = \Delta([\mathrm{AB}]\hat{\rho}_{\mathrm{AB}}).
    \end{aligned}
\end{equation*}
We introduce the concentration-normalized density operators $\hat{\sigma}_{\mathrm{A}}=[\mathrm{A}]\hat{\rho}_{\mathrm{A}}$ and $\hat{\sigma}_{\mathrm{AB}}=[\mathrm{AB}]\hat{\rho}_{\mathrm{AB}}$ for molecules A and AB, respectively. Indeed, the use of concentration-normalized density operators is particularly beneficial when discussing the equations of motion for systems undergoing non-equilibrium chemical reactions \cite{kuhne1979study}. In these systems, the concentrations of participating molecules are time-dependent. The changes of the concentration-normalized density operators during the time interval $\Delta t$ are readily obtained as:
\begin{equation*}
    \begin{aligned}
       \Delta(\hat{\sigma}_{\mathrm{A}}) &=  -n_a \hat{\rho}_{\mathrm{A}}+n_d\doublehat{T}_{\mathrm{B}}^{\mathrm{(AB)}}\hat{\rho}_{\mathrm{AB}};  \\
       \Delta(\hat{\sigma}_{\mathrm{AB}}) &=  -n_d \hat{\rho}_{\mathrm{AB}} + n_a \doublehat{D}_{\mathrm{B}}^{\mathrm{(AB)}}\hat{\rho}_{\mathrm{A}}. \\
    \end{aligned}
\end{equation*}

Based on Eq.~\eqref{Eq: Chemical kinetics}, we insert $n_a=\tilde{k}_a[\mathrm{A}]\Delta t$ and $n_d=k_d[\mathrm{AB}]\Delta t$, and divide both sides of the equations by $\Delta t$:
\begin{equation*}
    \begin{aligned}
        \frac{\Delta \hat{\sigma}_{\mathrm{A}}}{\Delta t} & = -\tilde{k}_a \hat{\sigma}_{\mathrm{A}} + k_d \doublehat{T}_{\mathrm{B}}^{\mathrm{(AB)}} \hat{\sigma}_{\mathrm{AB}};   \\
        \frac{\Delta \hat{\sigma}_{\mathrm{AB}}}{\Delta t} & = +\tilde{k}_a \doublehat{D}_{\mathrm{B}}^{\mathrm{(AB)}} \hat{\sigma}_{\mathrm{A}}-k_d \hat{\sigma}_{\mathrm{AB}} .
    \end{aligned}
\end{equation*}
In the limit $\Delta t \to 0$, we derive the evolution of the concentration-normalized density operators solely due to chemical exchange:
\begin{equation*}
    \begin{aligned}
        \frac{d}{dt} \hat{\sigma}_{\mathrm{A}} & = -\tilde{k}_a \hat{\sigma}_{\mathrm{A}} + k_d \doublehat{T}_{\mathrm{B}}^{\mathrm{(AB)}} \hat{\sigma}_{\mathrm{AB}};   \\
        \frac{d}{dt} \hat{\sigma}_{\mathrm{AB}} & = +\tilde{k}_a \doublehat{D}_{\mathrm{B}}^{\mathrm{(AB)}} \hat{\sigma}_{\mathrm{A}}-k_d \hat{\sigma}_{\mathrm{AB}} .
    \end{aligned}
\end{equation*}
Similarly to Eq.~\eqref{Eq: Chemical kinetics}, we can write the equations in matrix form as
\begin{equation}
    \frac{d}{dt}
    \begin{pmatrix}
    \hat{\sigma}_{\mathrm{A}} \\
    \hat{\sigma}_{\mathrm{AB}} 
    \end{pmatrix}
    =
    \begin{pmatrix}
    \!\!\!\!\!\!\!-\tilde{k}_a \doublehat{\textbf{\textit{1}}\,}_{\!\mathrm A} & +k_d \doublehat{T}_{\mathrm B}^{\mathrm {(AB)}}\\
    +\tilde{k}_a\doublehat{D}_{\mathrm B}^{\mathrm {(AB)}} & \!\!\!\!-k_d \doublehat{\textbf{\textit{1}}\,}_{\!\mathrm{AB}}
    \end{pmatrix}
    \begin{pmatrix}
    \hat{\sigma}_{\mathrm{A}} \\
    \hat{\sigma}_{\mathrm{AB}}
    \end{pmatrix}.
\label{Eq: LvN without L for AC}
\end{equation}
Here $\doublehat{\textbf{\textit{1}}\,}_{\!\mathrm{A}}$ and $\doublehat{\textbf{\textit{1}}\,}_{\!\mathrm{AB}}$ are identity superoperators that act on $\hat{\sigma}_{\mathrm{A}}$ and $\hat{\sigma}_{\mathrm{AB}}$, respectively. In practice, the coefficient matrix in Eq.~\eqref{Eq: LvN without L for AC} can be constructed based on the coefficient matrix in Eq.~\eqref{Eq: Chemical kinetics} by making the following substitution:
\begin{equation*}
    \begin{pmatrix}
    -\tilde{k}_a & +k_d\\
    +\tilde{k}_a & -k_d
    \end{pmatrix}
    \Longrightarrow
     \begin{pmatrix}
    \!\!\!\!\!\!\!-\tilde{k}_a \doublehat{\textbf{\textit{1}}\,}_{\!\mathrm A} & +k_d \doublehat{T}_{\mathrm B}^{\mathrm {(AB)}}\\
    +\tilde{k}_a\doublehat{D}_{\mathrm B}^{\mathrm {(AB)}} & \!\!\!\!-k_d \doublehat{\textbf{\textit{1}}\,}_{\!\mathrm{AB}}
    \end{pmatrix}\,,
\end{equation*}
where the identity superoperators are added to the diagonal positions and the chemical exchange superoperators accounting for the inter-conversion between different molecules are added to the off-diagonal positions.

In terms of the effects of coherent evolution and incoherent relaxation, the evolution of the concentration-normalized density operator in the absence of chemical exchange follows the same form as described in Eq.\eqref{Eq: LvN without exchange}. This is because Eq.\eqref{Eq: LvN without exchange} is unaffected by the specific normalization used. Building on Eq.~\eqref{Eq: LvN without exchange}, we can express the evolution of the concentration-normalized density operators for molecules A and C in matrix form as follows:
\begin{equation}
    \begin{aligned}
    \frac{d}{dt}
    \begin{pmatrix}
    \hat{\sigma}_{\mathrm{A}} \\
    \hat{\sigma}_{\mathrm{AB}} 
    \end{pmatrix}
    =
    \begin{pmatrix}
        \doublehat{L}_{\rm A} & \doublehat{\textbf{\textit{0}}}\\
        \doublehat{\textbf{\textit{0}}} &  \doublehat{L}_{\rm C}\\
    \end{pmatrix}
    \begin{pmatrix}
    \hat{\sigma}_{\mathrm{A}} \\
    \hat{\sigma}_{\mathrm{AB}}
    \end{pmatrix}.
    \end{aligned}
\label{Eq: LvN without chemical exchange for AC}
\end{equation}

We can now combine the effects of chemical exchange, coherent evolution, and incoherent relaxation by summing their respective coefficient matrices (as shown in Eq.\eqref{Eq: LvN without chemical exchange for AC} and Eq.\eqref{Eq: LvN without L for AC}). This enables us to express the evolution of the concentration-normalized density operators in a matrix form as follows:
\begin{equation}
    \frac{d}{dt}
    \begin{pmatrix}
    \hat{\sigma}_{\mathrm{A}} \\
    \hat{\sigma}_{\mathrm{AB}} 
    \end{pmatrix}
    =
    \left[
    \begin{pmatrix}
        \doublehat{L}_{\rm A} & \doublehat{\textbf{\textit{0}}} \\
        \doublehat{\textbf{\textit{0}}} &  \doublehat{L}_{\rm C}\\
    \end{pmatrix}
    +
    \begin{pmatrix}
    \!\!\!\!\!\!\!-\tilde{k}_a \doublehat{\textbf{\textit{1}}\,}_{\!\mathrm A} & +k_d \doublehat{T}_{\mathrm B}^{\mathrm {(AB)}}\\
    +\tilde{k}_a\doublehat{D}_{\mathrm B}^{\mathrm {(AB)}} & \!\!\!\!-k_d \doublehat{\textbf{\textit{1}}\,}_{\!\mathrm{AB}}
    \end{pmatrix}
    \right]
    \begin{pmatrix}
    \hat{\sigma}_{\mathrm{A}} \\
    \hat{\sigma}_{\mathrm{AB}} 
    \end{pmatrix}
    = \doublehat{M}
    \begin{pmatrix}
    \hat{\sigma}_{\mathrm{A}} \\
    \hat{\sigma}_{\mathrm{AB}} 
    \end{pmatrix}.
\label{Eq: LvN with chemical exchange}
\end{equation}

\begin{figure}[!ht]
    \centering
    \includegraphics{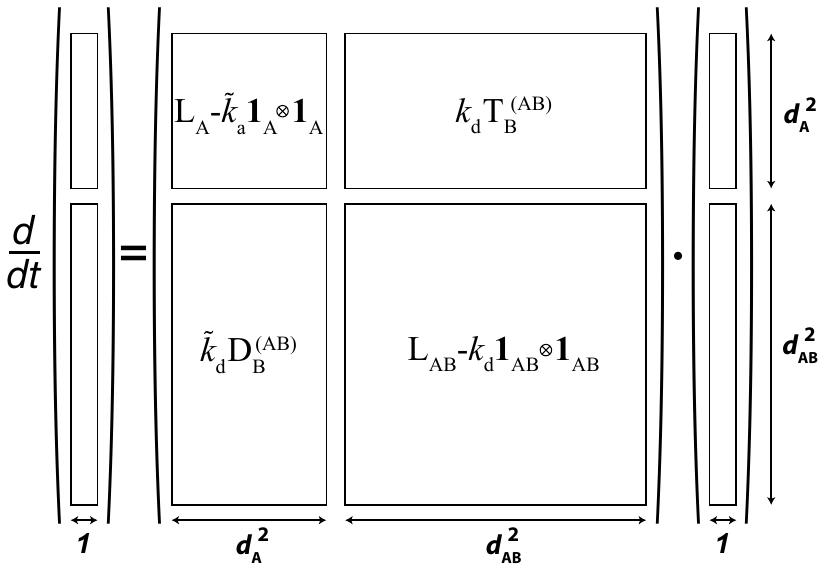}
    \caption{The matrix form of the LvN equation governing the evolution of the chemical exchanging system. The dimensions of Hilbert spaces for molecules A and AB are $d_{\rm A}$ and $d_{\rm AB}$, respectively.}
    \label{fig:LvN Eq}
\end{figure}

\section{Notations}

In this study, we use non-bold italic letters with a hat/double-hat (e.g., $\hat{I}_{x}$, $\hat{\rho}_{\mathrm{A}}$, $\doublehat{H}$, etc.) for operators/superoperators. One exception is angular momentum operator for which we use bold uppercase letter ``I'' with a hat (e.g., $\hat{\mathbf{I}}=(\hat{I}_x,\hat{I}_y,\hat{I}_z)$). In addition, the identity operator/superoperator and null-operator/superoperator will be represented as bold italic numbers with hats or double-hats, respectively (e.g., $\hat{\textbf{\textit{1}}\,}$ for identity operator and $\doublehat{\textbf{\textit{0}}\,}$ for null superoperator).

To represent the matrix representations of the aforementioned operators/superoperators, we use non-bold regular uppercase letters without hats or double-hats (e.g., A, B, I, H, etc.). 

However, identity matrices will be denoted using bold regular number one symbol for clarity. Specifically, $\textbf{1}_{k}$ represents the identity matrix with the dimensions $2^{k}\times 2^{k}$. In the context where the molecule notation (e.g., X) is used as the subscript, the matrix $\textbf{1}_{\mathrm{X}}$ %and $\textbf{1}_{\mathrm{2X}}$ 
denote the matrix representation of $\hat{\textbf{\textit{1}}\,}_{\!\mathrm{X}}$.%, and $\doublehat{\textbf{\textit{1}}\,}_{\mathrm{X}}$, respectively.

Scalars (numbers) are represented by italic lowercase letters.

By adhering to this notation, we aim to provide a coherent and unambiguous framework for discussing operators and matrices throughout the paper.

\section{Materials and Methods}
\subsection{Installation of JupyterLab}
JupyterLab \cite{jupyterlab} is an open-source web-based interactive development environment for working with notebooks, code, and data. It can be installed using the following command in the command prompt:
\begin{lstlisting}{language=bash}
    pip install jupyterlab
\end{lstlisting}

\subsection{Installation of packages}
NumPy \cite{numpy}, SciPy \cite{WOS:000510821500001}, and Matplotlib \cite{WOS:000245668100019} are the most commonly used scientific computing packages in Python. NumPy is a library for working with arrays and matrices, SciPy is a library for scientific and technical computing, and Matplotlib is a library for creating visualizations in Python. These packages can be installed using the following commands in the command prompt:
\begin{lstlisting}
    pip install numpy
    pip install scipy
    pip install matplotlib
\end{lstlisting}

\subsection{Running JupyterLab}
After installing JupyterLab and the necessary packages, users can initiate the application on their local machines. To launch JupyterLab, one must open the command prompt and enter the following command:
\begin{lstlisting}{language=bash}
    jupyter lab
\end{lstlisting}
This command will launch JupyterLab in the default web browser, providing access to a comprehensive user interface.

Upon opening JupyterLab, users can create a new notebook or open an existing one. To create a new notebook, simply click on the "File" menu located in the upper left corner of the screen. From the dropdown menu, select "New Notebook" and choose the preferred kernel for the simulations. Alternatively, users can create a new notebook by clicking on the "Python 3" button on the JupyterLab home page.

Once a new notebook is created, users can begin writing and executing their code. To execute a cell, simply press Shift + Enter or click on the ``Run'' button available in the toolbar. New cells can be added using the ``+'' button in the toolbar, while existing cells can be deleted by selecting them and clicking on the "scissors" button.

To halt JupyterLab, users can either close the web browser window or press Ctrl + C in the command prompt.

\subsection{Importation of packages}\label{Section: Packages}
For simulating nuclear spin dynamics under chemical exchange, certain packages must be imported at the beginning of the JupyterLab notebook. The importation can be performed as follows:
\begin{lstlisting}{language=bach}
    import numpy as np
    import scipy
    import matplotlib.pyplot as plt
    from math import sqrt
    from scipy.linalg import expm
    from scipy.fft import fft
\end{lstlisting}
The ``import'' statement is used to bring in the required packages or modules for use in the notebook. To simplify their usage in subsequent code, shorter aliases are assigned using the ``as'' keyword. For example, ``np'' and ``plt'' are used as aliases for NumPy and Matplotlib, respectively.

\section{Calculation}

\subsection{Matrix representations of bra- and ket-states}
In linear algebra and quantum mechanics, the Dirac's bra-ket notation is often used to represent states and operators in the Hilbert space. A ket-state $|a\rangle$ is represented by a column-vector in Hilbert space, while a bra-state $\langle b|$ is represented by a row-vector in the Hilbert space.

Matrix representation of the ket-vector $|a\rangle$ with respect to a given orthonormal basis $\{|i\rangle\}$ is a column-vector whose entries are the coefficients of $|a\rangle$ with respect to the basis. That is,
\begin{equation*}
    |a\rangle = \sum_i^{N} \alpha_i |i\rangle \implies
    \begin{pmatrix}
        \alpha_1 \\
        \alpha_2 \\
        \vdots \\
        \alpha_{N}
    \end{pmatrix},
\end{equation*}
where $\alpha_i = \langle i | a \rangle $.
% $$|a\rangle = \sum_i \alpha_i |i\rangle \implies [\psi]_{i} = \psi_i,$$
% where $[\psi]_{i}$ denotes the $i$-th entry of the column vector $|\psi\rangle$. 

Similarly, matrix representation of the bra-vector $\langle b|$ is a row-vector whose entries are complex conjugates of the coefficients of $|b\rangle$ with respect to the basis. That is,
\begin{equation*}
    \langle b| = \sum_i^{N} \beta_i^{*} \langle i | \implies
    \begin{pmatrix}
        \beta_1^{*} & \beta_2^{*} & \cdots & \beta_{N}^{*}
    \end{pmatrix},
\end{equation*}
where $\beta_i^{*} = \langle b | i \rangle $.
% $$\langle\phi| = \sum_i \phi_i^* \langle i | \implies [\phi]_{i} = \phi_i^*,$$
% where $[\phi]_{i}$ denotes the $i$-th entry of the row vector $\langle\phi|$. 

As an example, the matrix representation of $|i\rangle$ and $\langle j|$ are given below:\newline
\hspace*{60pt}
\begin{tikzpicture}[>=stealth,thick,baseline]
\matrix [matrix of math nodes, left delimiter={(},right delimiter={)}, outer sep=-6pt, row sep = -2pt](A){ 
0 \\ 
\svdots  \\
1 \\
\svdots  \\
0 \\
};
\node[right =25pt of A-3-1.east](L)  {$i^{\rm th}$ row};
\node[left = 8pt of A-3-1.west] {$|i\rangle = $};
\draw[->, shorten > =12pt](L.west)-- (A-3-1.east);
\end{tikzpicture}
\,\,\,\,\,\,\,\,
\begin{tikzpicture}[>=stealth,thick,baseline]
\matrix [matrix of math nodes,left delimiter=(,right delimiter=), outer sep=-6pt, column sep = -3pt](A){ 
0 & \cdots & 1  & \cdots & 0\\
};

\node[below=15pt of A-1-3.south](C) {$j^{\rm th}$ column};
\node[left = 8pt of A-1-1.west] {$\langle j| = $};
\draw[->](C.north)-- ([yshift=-4pt]A-1-3.south);
\end{tikzpicture}

Matrix representation of a linear operator $\hat{A}$ in Hilbert space with respect to the basis $\{|i\rangle\}$ is a matrix whose entries are given by ${\rm A}_{ij} = \langle i | \hat{A} | j\rangle$. Note that this representation is unique for a given basis. 

\subsection{Matrix multiplication}

The matrix multiplication is used to compose various quantum mechanical operators. The matrix representation of the product of two operators is defined as the matrix multiplication of their corresponding matrix representations. The  matrix product (C) of two matrices A and B is defined as:
\begin{equation*}
    {\mathrm C_{ij}} = \sum_{k} {\mathrm A}_{ik} {\mathrm B}_{kj},
\end{equation*}
where the inner dimension of matrix A and B should be the same. \newline

For example, matrix representation of the operator $|i\rangle\langle j |$ can be derived through the matrix multiplication $ |i\rangle \cdot \langle j |$:

\hspace*{10pt}
\begin{tikzpicture}[>=stealth,thick,baseline]
\matrix [matrix of math nodes, left delimiter=(,right delimiter=), outer sep=-6pt, row sep = -2pt](A){ 
0 \\ 
\svdots  \\
1 \\
\svdots  \\
0 \\
};

\matrix [matrix of math nodes,left delimiter=(,right delimiter=), outer sep=-6pt, column sep = -3pt, right = of A](B){ 
0 & \cdots & 1  & \cdots & 0\\
};

\matrix [matrix of math nodes,left delimiter=(,right delimiter=), outer sep=-6pt, column sep = -3pt, row sep = -2pt, right = of B](C){ 
0 & \cdots & 0  & \cdots & 0\\
\svdots & \sddots & \svdots  & \sddots & \svdots\\
0 & \cdots & 1  & \cdots & 0\\
\svdots & \sddots & \svdots  & \sddots & \svdots\\
0 & \cdots & 0  & \cdots & 0\\
};

\node[below=15pt of B-1-3.south](D) {$j^{\rm th}$ column};
\node[below=15pt of C-5-3.south](E) {$j^{\rm th}$ column};
\node[right =25pt of C-3-5.east](F)  {$i^{\rm th}$ row};

\draw[->](D.north)-- ([yshift=-4pt]B-1-3.south);
\node[left =25pt of A-3-1.west](L)  {$i^{\rm th}$ row};
% \node[left = 8pt of B-3-1.west] {$|i\rangle = $};
\draw[->, shorten > =12pt](L.east)--([xshift=3pt]A-3-1.west);
\draw[->](E.north)-- ([yshift=-4pt]C-5-3.south);
\draw[->](F.west)-- ([xshift=9pt]C-3-5.west);
\path (A) -- node {$\cdot$} (B)
              (B) -- node {$=$} (C);
\end{tikzpicture}

Another example of matrix multiplication is $\langle p| q\rangle$:

\hspace*{60pt}
\begin{tikzpicture}[>=stealth,thick,baseline]
\matrix [matrix of math nodes,left delimiter=(,right delimiter=), outer sep=-6pt, column sep = -3pt](A){ 
0 & \cdots & 1  & \cdots & 0\\
};
\matrix [matrix of math nodes, left delimiter=(,right delimiter=), outer sep=-6pt, row sep = -2pt,  right = of A](B){ 
0 \\ 
\vdots  \\
1 \\
\vdots  \\
0 \\
};
\node[below right = -15 pt and 15pt of B-5-1.east](L)  {$q^{\rm th}$ row};
\draw[->, shorten > =10pt](L.north)-- ([xshift=33.5pt]B-3-1.east) -- ([xshift=3pt]B-3-1.east);

\node[below=15pt of A-1-3.south](D) {$p^{\rm th}$ column };
\draw[->](D.north)-- ([yshift=-5pt]A-1-3.south);

% \node[left = 8pt of B-3-1.west] {$|i\rangle = $};
\path (A) -- node {$\cdot$} (B);

\node[right = 45 pt of B-3-1.east](L)  {= $\delta_{pq}$,};

\end{tikzpicture}

\noindent where $\delta_{pq}$ represents the Kronecker delta-function.

\heading{Implementation}
The Python code below returns the matrix multiplication of two input matrices A and B.
\begin{lstlisting}
    C = np.matmul(A,B)
\end{lstlisting}
Another code in Python that can produce the same result is ``einsum'' which is defined as the sum over the disappearing index \cite{einsum1}:
\begin{lstlisting}
    C = np.einsum('ik,kj->ij',A,B)
\end{lstlisting}
The ``einsum'' syntax is used heavily in this paper; An introduction to ``einsum" in NumPy will be given in Section~\ref{Section: einsum}. 

\subsection{Kronecker product}
The Kronecker product, also known as the matrix outer product, is a fundamental operation in linear algebra and plays a crucial role in quantum mechanics. We form the density operators of composite systems by taking the Kronecker product of the density operators associated with each individual subsystem.

In addition to its use in describing composite systems, the Kronecker product also finds extensive application in mapping linear operators from Hilbert space to corresponding superoperators in Liouville space.

% The matrix representation of the superoperator can be obtained similarly by taking the Kronecker product of the matrix representation of the operator with the identity matrix.

%In literature, the direct product is also termed as Kronecker-product or outer product. 
Given a $ m \times n$ matrix A and a $p \times q$ matrix B---where we, by convention, imply dimensions as (number of rows)$\times$(number of columns)---their Kronecker-product C$ = \mathrm A\otimes \rm B$ is a $pm \times qn$ matrix:
\begin{equation*}
    \mathrm{C}=
    \begin{pmatrix}
        a_{11}\mathrm{B} & a_{12}\mathrm{B} & \cdots & a_{1n}\mathrm{B}\\
        a_{21}\mathrm{B} & a_{22}\mathrm{B} & \cdots & a_{2n}\mathrm{B}\\
        \vdots  & \vdots & \ddots & \vdots \\
        a_{m1}\mathrm{B} & a_{m2}\mathrm{B} & \cdots & a_{mn}\mathrm{B}\\
    \end{pmatrix}.
\end{equation*}

Given this relationship, the elements of matrix C are given by
\begin{equation}
    \mathrm{C}_{IJ} =  \mathrm{A}_{ij}\mathrm{B}_{kl},
\label{Eq: Kronecker Product elements 1}
\end{equation}
where the indices $I,J$ are related to $i,j,k,l$ with $I=p(i-1)+k$ and $J=q(j-1)+l$.

It is worth noting that there are no restrictions on the dimensions of the two matrices when performing the Kronecker product.

The Kronecker product is bilinear and associative:
\begin{equation*}
\begin{split}
    \mathrm{A}\otimes(\mathrm{B}+\mathrm{C}) &=  \mathrm{A}\otimes\mathrm{B} + \mathrm{A}\otimes\mathrm{C};\\
    (\mathrm{B}+\mathrm{C})\otimes\mathrm{A} &= \mathrm{B}\otimes\mathrm{A} + \mathrm{C}\otimes\mathrm{A};\\
    (k\mathrm{A})\otimes\mathrm{B} &=\mathrm{A}\otimes( k \mathrm{B})=k(\mathrm{A}\otimes\mathrm{B});\\
        (\mathrm{A}\otimes\mathrm{B})\otimes\mathrm{C} &=\mathrm{A}\otimes(\mathrm{B}\otimes\mathrm{C}).\\
\end{split}
\end{equation*}

Additionally, if A, B, C, and D are matrices and their inner products are defined (AC and BD), i.e., the inner indices of two matrices are the same, then
\begin{equation}
    (\mathrm{A}\otimes\mathrm{B})\cdot(\mathrm{C}\otimes\mathrm{D}) = (\mathrm{A}\cdot\mathrm{C}) \otimes (\mathrm{B}\cdot\mathrm{D}),
\label{Eq:Mixed product property}
\end{equation}
which is called the mixed-product property.

The Kronecker product is used widely in Dirac's bra-ket annotations. For example,
\begin{equation*}
        |ij\rangle=|i\rangle\otimes |j\rangle .
\end{equation*}

\heading{Calculation}
The Kronecker product is computed by first creating a corresponding multidimensional matrix (denoted as T) using Numpy's einsum function, followed by a reshape operation. Three examples are given below. The codes in question are explained in detail in Section~\ref{Section: einsum}.  

\noindent\textbf{Example 1.} 
The following code calculates the Kronecker product C of two input matrices A and B (A has the dimensions $m \times n$, B has the dimensions $p \times q$, and the resulting matrix C has the dimensions $mp \times nq$). 
\begin{lstlisting}
    C = np.einsum("ij,kl->ikjl",%*\textcolor{cyan}{A}*),%*\textcolor{cyan}{B}*)).reshape(%*\textcolor{cyan}{m*p}*),%*\textcolor{cyan}{n*q}*))
\end{lstlisting}
where the subscript string follows from $\mathrm{C}_{IJ} = \mathrm{T}_{ikjl} =\mathrm{A}_{ij}\mathrm{B}_{kl}$. The \lstinline{reshape()} function is used to set the dimensions of matrix C, which is $mp \times nq$ as per the Kronecker product definition.

\noindent\textbf{Example 2.} 
The following code calculates the Kronecker product C of two input matrices A and B. A is a row matrix with $n$ columns and B has dimensions $p \times q$. The dimension of the resulting matrix C is $p \times nq$. 
\begin{lstlisting}
    C = np.einsum("i,jk->jik",%*\textcolor{cyan}{A}*),%*\textcolor{cyan}{B}*)).reshape(%*\textcolor{cyan}{p}*),%*\textcolor{cyan}{n*q}*))
\end{lstlisting}
where the subscript string are based on $\mathrm{C}_{IJ} =\mathrm{T}_{jik}= \mathrm{A}_{i}\mathrm{B}_{jk}$.

\noindent\textbf{Example 3.} 
The following code calculates the Kronecker product, D, of three input matrices A, B, and C. A has dimension $m_A \times n_A$, B is a column matrix with $m_B$ rows, and C has dimension $m_C \times n_C$. The dimension of the resulting matrix D is $m_A m_B m_C \times n_A n_C$. 
\begin{lstlisting}
    D = np.einsum("ab,c,de->acdbe",%*\textcolor{cyan}{A}*),%*\textcolor{cyan}{B}*),%*\textcolor{cyan}{C}*)).reshape(%*\textcolor{cyan}{mA*mB*mC}*),%*\textcolor{cyan}{nA*nC}*))
\end{lstlisting}
where the subscript string are based on $\mathrm{D}_{IJ} = \mathrm{T}_{acd,be} = \mathrm{A}_{ab}\mathrm{B}_{c}\mathrm{C}_{de}$.

\noindent\textbf{Additional notes.} The function \lstinline{np.kroncker(A,B)} could be used to calculate the Kronecker product between two matrices A and B. However, using the codes with \lstinline{np.einsum} can sometimes be more convenient when dealing with a Kronecker product between more than two matrices or making a summation over Kronecker products.

\subsection{Introduction to ``einsum'' in NumPy}\label{Section: einsum}

The ``einsum'' function in Numpy takes two main arguments: the subscripts string and the input arrays. The subscripts string defines the operation you want to perform, and the input arrays are the operands on which the operation is applied. The general syntax of einsum is as follows:

\begin{lstlisting}
    np.einsum(subscripts, operands)
\end{lstlisting}

The subscripts string is a comma-separated list of subscript labels and specifies how the arrays should be combined. Each label corresponds to a dimension of an input array. The subscripts string has the following format: '\lstinline{input_labels->output_labels}'. Here, \lstinline{input_labels} represents the labels for each input array, and \lstinline{output_labels} represents the labels for the output array. The input and output labels can be obtained by taking the indices of the corresponding mathematical expression of the operation.

In ``einsum", the summation is done through the summation labels which are determined by the indices that disappear in the output label. When constructing the subscripts string, the output labels represent the dimensions of the resulting array. If a dimension is missing in the output label, it signifies that the corresponding dimension in the input arrays should be summed over. To summarize, if a dimension is absent in the output label, it implies that the corresponding dimension in the input arrays will be summed over.
\newline
A few common examples was given below:
\heading{Matrix transpose}
\begin{equation*}
    \mathrm{B}_{ij}=\mathrm{A}_{ji}
\end{equation*}
The subscript string directly comes from the expression and they match the indices of the mathematical expression shown above. The only operand is matrix A. 
\begin{lstlisting}
    B=np.einsum('ji->ij',A)
\end{lstlisting}
The output array, denoted as B, has dimensions 'i' and 'j', indicating a 2-dimensional array. This is determined by the output label specified in the subscripts string.

\heading{Sum of matrix elements}
\begin{equation*}
    \mathrm{B} = \sum_{ij} \mathrm{A}_{ij}
\end{equation*}
\begin{lstlisting}
    B=np.einsum('ij->',A)
\end{lstlisting}
The absence of the dimensions 'i' and 'j' in the output string means the summation labels are 'i' and 'j' here. As a result, the dimensions 'i' and 'j' in matrix A  will be summed over. The output array, denoted as B, has no dimension, and thus a scalar.

\heading{Matrix column sum}
\begin{equation*}
    \mathrm{B}_{j} =\sum_{i} \mathrm{A}_{ij}
\end{equation*}
\begin{lstlisting}
    B=np.einsum('ij->j',A)
\end{lstlisting}
The summation label is 'i' here. The output array B has the dimension 'j' and thus a 1-dimensional array. 

\heading{Matrix multiplication}
\begin{equation*}
    \mathrm{C}_{ij} =\sum_{k} \mathrm{A}_{ik} \mathrm{B}_{kj}
\end{equation*}
\begin{lstlisting}
    C=np.einsum('ik,kj->ij',A,B)
\end{lstlisting}
The input and output labels in the subscript string match the mathematical expression of a matrix multiplication given above. The label 'ik' represents the dimensions of the first matrix A, 'kj' represents the dimensions of the second matrix B, and 'ij' represents the dimensions of the resulting matrix C. Both the input operands and the output array are 2-dimensional arrays. 

\heading{Kronecker product}
Assume matrix A has the dimensions $\rm m \times n$, and matrix B has the dimension $\rm p \times q$. The goal is to construct the two-dimensional array,
\begin{equation*}
    \rm \begin{pmatrix}
        \mathrm{C}_{11} &\cdots & \mathrm{C}_{1(n\times q)} \\
        \svdots &\sddots & \svdots \\
        \mathrm{C}_{(m\times p)1} & \cdots & \mathrm{C}_{(m\times p)(n\times q)}
    \end{pmatrix},
\end{equation*}
whose elements satisfies $\mathrm{C}_{IJ} = \mathrm{A}_{ij} \mathrm{B}_{kl}$ with $ I=p(i-1)+k$ and $ J=q(j-1)+l$. 

One way of calculating the matrix ${\rm C}_{\rm I,J}$ is to back convert the indices $i,j,k,l$ into $I,J$. This approach has been used in various contexts, for example, in the paper by Ivanov et al. \cite{knecht2016}. However, the backward procedure, i.e., determining the indices $\rm i,j,k,l$ from $\rm I,J$ can be a challenging task, especially when working with the Kronecker product between more than two matrices, which is frequently encountered for computing the chemical exchange superoperator. 

Alternatively, we can begin by constructing a multi-dimensional array T. To sort out the indices of T, we follow a general rule. Firstly, we arrange all the row indices in the order that corresponds to how the Kronecker product is formed, starting from the matrices on the left and progressing to those on the right. Next, we append the column indices in a similar order as the row indices. 

In this specific example, the array T is defined as $\mathrm{T}_{\rm ikjl} = \mathrm{A}_{\rm ij}\mathrm{B}_{\rm kl}$. The four indices of T, from the first to the last, represent the row index of A, the row index of B, the column index of A, and the column index of B, respectively. 
\begin{lstlisting}
    T = np.einsum("ij,kl->ikjl",A,B)
\end{lstlisting}
The subscript string is obtained based on the expression of T where 'ij' denotes the dimensions of the first matrix A, 'kl' represents the dimensions of the second matrix B, and 'ikjl' indicates the dimensions of the resulting array T. 

Second, calculate $\mathrm{C}_{IJ} = \mathrm{T}_{ikjl}$. The multi-dimensional array is internally stored as a 1-dimensional block in NumPy, with elements arranged in memory based on column-major order. This means that the elements along the last axis are stored adjacent to each other, followed by the elements along the second-last axis, and so on, with the elements along the first axis changing most slowly. The 4-D matrix $\mathrm{T}_{ikjl}$ is stored in memory following the order:
\begin{equation*}
    \mathrm{T}_{1111},\mathrm{T}_{1112},\cdots, \mathrm{T}_{111q}, \mathrm{T}_{1121},\cdots,\mathrm{T}_{11nq},\mathrm{T}_{1211},\cdots,\mathrm{T}_{mp11},\cdots,\mathrm{T}_{mpnq}
\end{equation*}
The elements $\rm T_{ikjl}$ can be equivalently expressed as $\rm C_{IJ}$:
\begin{equation*}
    \mathrm{C}_{11},\mathrm{C}_{12},\cdots, \mathrm{C}_{1q}, \mathrm{C}_{1(q+1)},\cdots, \mathrm{C}_{1(n\times q)},\mathrm{C}_{21},\cdots,\mathrm{C}_{(m\times p)1},\cdots,\mathrm{C}_{(m\times p)(n\times q)}
\end{equation*}
Upon the following reshape operation with the inputs as the dimension of the resulting matrix :
\begin{lstlisting}
    C = T.reshape(m*p,n*q)
\end{lstlisting}
The first $nq$ elements are taken as the first row, and the next $nq$ elements as the second row, and so on. The resulting 2-D matrix formed is exactly the Kronecker product of the two matrices.

Combing the two operations reaches the following code snippet for calculating the Kronecter product between two matrices:
\begin{lstlisting}
    C = np.einsum("ij,kl->ikjl",A,B).reshape(m*p,n*q)
\end{lstlisting}

As a detailed example, the calculation between a $1\times 2$ matrix and a $2 \times 2$ matrix,
\begin{equation*}
    \begin{pmatrix}
       \rm A_{11} & \rm A_{12}
    \end{pmatrix}
    \otimes
    \begin{pmatrix}
       \rm B_{11} & \rm B_{12}  \\
       \rm B_{21} & \rm B_{22}
    \end{pmatrix},
\end{equation*}
proceeds as follows:
First, calculate the 3-dimensional array $\rm T_{jik}=A_{i}B_{jk}$. Notice here that the matrix A only has a column index. The three indices of T, from left to right, represent the row index of B, the column index of A, and the column index of B. 
\begin{lstlisting}
    T=np.einsum('i,jk->jik',A,B)
\end{lstlisting}
The resulting array T is stored in memory as (The indices i, j, and k range from 1 to 2):
\begin{equation*}
    \rm T_{111}, T_{112}, T_{121}, T_{122}, T_{211}, T_{212}, T_{221}, T_{222}
\end{equation*}
The array T can be expressed using the elements of matrix A and matrix B with $\rm T_{jik}=A_{i}B_{jk}$ :
\begin{equation*}
   \rm  A_{11}B_{11}, A_{11}B_{12}, A_{12}B_{11}, A_{12}B_{12}, A_{11}B_{21}, A_{11}B_{22}, A_{12}B_{21}, A_{12}B_{22}
\end{equation*}
Then we apply the reshape operation according to the dimensions of the resulting matrix which is $2\times 4$ in this example:
\begin{lstlisting}
    C=T.resahpe(2,4)
\end{lstlisting}
which returns the following 2-D array,
\begin{equation*}
    \begin{pmatrix}
        \rm  A_{11}B_{11} & \rm A_{11}B_{12} & \rm A_{12}B_{11} & \rm A_{12}B_{12} \\
        \rm  A_{11}B_{21} & \rm A_{11}B_{22} & \rm A_{12}B_{21} & \rm A_{12}B_{22}
    \end{pmatrix},
\end{equation*}
which is the Kronecker product of the two matrices, as defined by the mathematical operation.

\subsection{The Hilbert spaces associated with nuclear spins of molecules}
To describe nuclear spin degrees of freedom in molecules, the very first thing to do is to construct the Hilbert space. In this work, we consider spin-1/2 nuclei such as protons or $^{15}$N. Spin-1/2 particles have two possible spin states denoted as $\alpha$ and $\beta$. Therefore, we assign a two-dimensional Hilbert space spanned by these states.

For a molecule with several nuclei, the total Hilbert space associated with the nuclear spins is the tensor product of the individual Hilbert spaces for each nucleus. Consider a simple example of a diatomic molecule consisting of two spin-1/2 nuclei, labeled as X and Y.  We denote the Hilbert space associated with nucleus X, and nucleus Y as $\mathbb{H}_{\mathrm{X}}$, and $\mathbb{H}_{\mathrm{Y}}$, respectively. The Hilbert space associated with the nuclear spins of this molecule could be given by
\begin{equation*}
     \mathbb{H}_{\mathrm{X}} \otimes \mathbb{H}_{\mathrm{Y}},
\end{equation*}
or 
\begin{equation*}
     \mathbb{H}_{\mathrm{Y}} \otimes \mathbb{H}_{\mathrm{X}}.
\end{equation*}

The relative order of the individual spin Hilbert spaces is important. To distinguish the two cases, we denote the molecule as XY if $\mathbb{H}_{\mathrm{X}}$ comes before $\mathbb{H}_{\mathrm{Y}}$, and YX if it is the other way around. Furthermore, the labels X and Y can also represent groups of nuclei, such that the Hilbert space of the molecule is the tensor product of the Hilbert spaces associated with its molecular fragments. This construction of the total Hilbert space is frequently encountered below in discussions related to chemical exchange superoperators.

The basis states of the composite Hilbert space are constructed by taking tensor products of the basis states of the individual Hilbert spaces. For example, consider a molecule denoted as XY, a basis state in its composite Hilbert space
could be written as:
\begin{equation*}
    |ij\rangle_{\mathrm{XY}}=|i\rangle_{\mathrm{X}}\otimes |j\rangle_{\mathrm{Y}}
\end{equation*}
Here, $|i\rangle_{\mathrm{X}}$ represents a Zeeman basis-state of the nucleus X, and $|j\rangle_{\mathrm{Y}}$ represents a Zeeman basis-state of the nucleus Y. The complete set of basis states in the total Hilbert space spans all possible combinations of the Zeeman basis states associated with each individual nuclear spin. Since the ordering of the individual spin Hilbert spaces have been fixed, the subscripts specifying the nuclei are typically omitted for the basis states. 

The Zeeman product basis enables selective addressing of individual spins, making it particularly advantageous for constructing chemical exchange superoperators. Consequently, in this paper, we consistently adopt this basis for the representations of a diverse range of operators required for the simulation.

\subsection{Matrix representations of angular momentum operators}\label{Section: Step1}
Angular momentum operators are important operators that they serve as building blocks for constructing other important operators Therefore, let us begin by discussing the matrix representations of angular momentum operators of a single spin-1/2 nucleus. Let $\hat{I}_x$, $\hat{I}_y$, and $\hat{I}_z$ denote the three components of the angular momentum operator.

The matrix forms of these operators represented in Zeeman basis ($\{|\alpha\rangle,\,\,|\beta\rangle\}$) is provided as follows:
\begin{equation*}
    {\rm I}_{x}= \frac{1}{2}
    \begin{pNiceMatrix}
    0 & 1\\
    1 & 0\\
    \end{pNiceMatrix}, \,\,\,\,
     {\rm I}_{y}= \frac{1}{2i}
    \begin{pNiceMatrix}
    0 & 1\\
    -1 & 0\\
    \end{pNiceMatrix}, \,\,\,\,
     {\rm I}_{z}= \frac{1}{2}
    \begin{pNiceMatrix}
    1 & 0\\
    0 & -1\\
    \end{pNiceMatrix}.
\end{equation*}

% These matrix representations will prove to be useful in describing the behavior of spin-1/2 systems in various physical applications.
In the context of multi-spin-1/2 systems, the spin angular momentum operators can be constructed by taking a tensor product of the spin-1/2 angular momentum operators and identity operators of the appropriate size. Let $\hat{I}_{i\alpha}$ denote the $\alpha$ component of the angular momentum operator associated with the $i$-th spin. In Zeeman product basis, $\hat{I}_{i\alpha}$ is represented with the following matrix:
\begin{equation}
 {\rm I}_{i\alpha}=\textbf{1}_{i-1} \otimes  {\rm I}_{\alpha} \otimes \textbf{1}_{n-i},
\label{Eq:AMO}
\end{equation}
where the matrix $\textbf{1}_{k}$ represents the identity matrix with the dimensions $2^{k} \times 2^{k}$.
% \begin{equation*}
% 1_{k}=\underbrace{\mathbf{1}_{1}\otimes\mathbf{1}_{1}\otimes \cdots \otimes \mathbf{1}_{1}}_{k\,\,\text{terms}}
% \end{equation*}
% with the convention $\mathbf{1}_{1}=\begin{pmatrix}  1 & 0
% \\ 0 & 1 \end{pmatrix}$ and $\mathbf{1}_{0}=1$.

For example, the matrix representation of $\hat{I}_{2x}$ acting on the second spin of a three spin-1/2 system is given by:
\begin{equation*}
    {\rm I}_{2x}=
    \begin{pmatrix}
    1 & 0\\
    0 & 1\\
    \end{pmatrix} 
    \otimes 
    \frac{1}{2}
    \begin{pmatrix}
    0 & 1\\
    1 & 0\\
    \end{pmatrix}
    \otimes
    \begin{pmatrix}
    1 & 0\\
    0 & 1\\
    \end{pmatrix}.
\end{equation*}

\heading{Calculation}
We begin by defining the following function:
\begin{lstlisting}
def Angular_Momentum_Operators(N):
    """
    Angular_Momentum_Operator produce the 4-D matrices containing all 
    the individual spin angular momentum operators.
    
    :Input N: "N" is the number of spin-1/2s of the molecule
    
    """
    
    #Generate a empty 4-D matrix with the dimention (N,3,2^N,2^N)
    S = np.empty((N, 3, 2 ** N, 2 ** N), dtype=np.complex128)
    
    #single spin-1/2 angular momentum operators
    s_x = np.array([[0, 1 / 2], [1 / 2, 0]])
    s_y = np.array([[0, -1j / 2], [1j / 2, 0]])
    s_z = np.array([[1 / 2, 0], [0, -1 / 2]])
    I_1 = np.array([[1, 0], [0, 1]])

    for i in range(N):
        Ix_current = 1
        Iy_current = 1
        Iz_current = 1
        for j in range(N):
            if j == i:
                Ix_current = np.kron(Ix_current, s_x)
                Iy_current = np.kron(Iy_current, s_y)
                Iz_current = np.kron(Iz_current, s_z)
            else:
                Ix_current = np.kron(Ix_current, I_1)
                Iy_current = np.kron(Iy_current, I_1)
                Iz_current = np.kron(Iz_current, I_1)
        S[i][0] = Ix_current
        S[i][1] = Iy_current
        S[i][2] = Iz_current
    return S
\end{lstlisting}

Here, the normalized Pauli matrices are specified as \lstinline{s_x}, \lstinline{s_y}, \lstinline{s_z}, and 
\lstinline{I_1}, respectively.
The inner loop iterates through the index \lstinline{j} and generates the matrices $\mathrm{I}_{j\alpha}$ ($\alpha=x,y,z$). These matrices are then stored in their respective positions within the 4-D matrix \lstinline{S}. Consequently, the generated 4D matrix S takes the form:
\begin{equation}
\mathrm{S}= \begin{bmatrix}
{\rm I}_{1x} & {\rm I}_{1y} & {\rm I}_{1z} \\
{\rm I}_{2x} & {\rm I}_{2y} & {\rm I}_{2z} \\
\vdots & \vdots & \vdots \\
{\rm I}_{Nx} & {\rm I}_{Ny} & {\rm I}_{Nz} \\
\end{bmatrix}.
\label{Eq:4DI}
\end{equation}
which contains all the matrix representations of individual spin angular momentum operators and has the dimension of $(N,3,2^N,2^N)$. Indeed, the 4D matrix S can understood as a $N$-by-$3$ matrix with the $2^N\times2^N$ square matrices $ {\rm I}_{i\alpha}$ as its elements. Specifically, $\mathrm{S}_{i\alpha mn} = (\mathrm{I}_{i\alpha})_{mn}$.

In practice, we call the defined function to generate the 4D matrix L:
\begin{lstlisting}
    N = NUM_SPINS  #Input: number of spins
    S = Angular_Momentum_Operator(N)
\end{lstlisting}

\subsection{Matrix representations of Hamiltonian operators}\label{Section: Hamiltonian}
\subsubsection{Zero-field Hamiltonian}
Many liquid-state NMR experiments focus on spin-1/2 systems, so our discussion will be limited to such systems. At zero field, the Hamiltonian consists solely of \textit{J}-coupling terms and can be represented as:
\begin{equation*}
    \hat{H}=\sum_{i<j} J_{ij} (\hat{\mathbf{I}}_i \cdot \hat{\mathbf{I}}_j),
\end{equation*}
where $J_{ij}$ is the scalar spin-spin \textit{J}-coupling between the $i^{\text{th}}$ and $j^{\text{th}}$ spins.

The Hamiltonian in Zeeman product basis can be represented as:
\begin{equation}
    \mathrm{H} = \sum_{i<j} J_{ij} \left( \sum_{\alpha=x,y,z} \mathrm{I}_{i\alpha}\cdot\mathrm{I}_{j\alpha}  \right).
\label{Eq: zero-field H}
\end{equation}
Here, the matrix $\mathrm{I}_{i\alpha}$ stands for the matrix representation of $\hat{I}_{i\alpha}$. These matrices have been calculated in Section~\ref{Section: Step1}.

\heading{Calculation}
For convenience, we store the \textit{J}-couplings, $J_{ij}$, in matrix form as follows:
\begin{equation*}
    \mathrm{J_0}=\begin{bmatrix}
    0 &  J_{12} &  \dots & J_{1N}\\
     0 &  0 &  \dots & J_{2N}\\
     \vdots & \vdots & \ddots & \vdots \\
     0 &  0 &  \dots & 0\\
    \end{bmatrix}.
\end{equation*}
The code snippet below generates the matrix representation of the zero-field Hamiltonian as defined in Eq.\,\eqref{Eq: zero-field H}. 
\begin{lstlisting}
    N,J0 = NUM_SPIN, MATRIX_J0      #Input: number of spins, matrix J0
    S = Angular_Momentum_Operator(N)  
    H = np.einsum('ij,iamk,jakn->mn',J0,S,S)
\end{lstlisting}
where the subscript string is derived from the elements of Eq.~\eqref{Eq: zero-field H}:
\begin{equation*}
    \mathrm{H}_{mn} = \sum_{i<j} J_{ij} \left( \sum_{\alpha=x,y,z} \sum_{k}( \mathrm{I}_{i\alpha })_{mk}(\mathrm{I}_{j\alpha })_{kn}  \right) = \sum_{i<j} (\mathrm{J}_0)_{ij} \left( \sum_{\alpha=x,y,z} \sum_{k} \mathrm{S}_{i\alpha  mk} \mathrm{S}_{j\alpha kn}  \right).
\end{equation*}

\subsubsection{Ultralow-field Hamiltonian}
When a magnetic field is applied, the Zeeman terms are added to the Hamiltonian:
\begin{equation*}
    \hat{H}= -\sum_{i} \nu_i \hat{I}_{iz} +\sum_{\substack{i<j}} J_{ij}(\hat{\mathbf{I}}_i \cdot \hat{\mathbf{I}}_j).
\end{equation*}
Here, $v_i$ represents the Larmor frequency of the $i^{\rm th}$ nuclear spin, which is given by $\gamma_i B_0 (1+\delta_i)$, where $\gamma_i$ is the gyromagnetic ratio of the spin, $\delta_i$ is its chemical shift, and $B_0$ is the applied magnetic field.

The matrix form of the Hamiltonian can be expressed as:
\begin{equation}
    \mathrm{H} = -\sum_{i}\nu_{i}\,\mathrm{I}_{iz}+\sum_{i<j} J_{ij} \left( \sum_{\alpha=x,y,z} \mathrm{I}_{i\alpha}\cdot\mathrm{I}_{j\alpha}  \right).
\label{Eq: Low-field H}
\end{equation}

\heading{Calculation}
The calculation of the scalar coupling term is the same as before. The focus here is on the calculation of the Zeeman interaction. We begin by calculating the Larmor frequency of each spin at field $B_0$:
\begin{lstlisting}
    B0,Gyro,CS = B0,MATRIX_GRYO,MATRIX_CS  #Input: B0, matrix Gyro and CS
    V = B0*Gyro*(1+CS)
\end{lstlisting}
where the matrix Gyro and matrix CS take the form:
\begin{equation*}
    \begin{split}
        \mathrm{Gyro} & = 
            \begin{bmatrix}
                \gamma_{1} & \gamma_{2} & \dots & \gamma_{N}\\
            \end{bmatrix},\\
        \mathrm{CS} & = 
            \begin{bmatrix}
                \delta_{1} & \delta_{2} & \dots & \delta_{N}\\
            \end{bmatrix}.
    \end{split}
\end{equation*}
The calculated Larmor frequency for each spin is stored in matrix V, which takes the form:
\begin{equation*}
    \mathrm{V}=\begin{bmatrix}
    \nu_{1} & \nu_{2} & \dots & \nu_{N}\\
    \end{bmatrix}.
\end{equation*}

Consequently, the Zeeman interaction term, stored as \lstinline{H_Z}, can be calculated as:
\begin{lstlisting}
    N,V = NUM_SPINS,MATRIX_V   #Input: number of spins, Matrix V
    S = Angular_Momentum_Operators(N)
    HZ = -np.einsum('i,imn->mn',V,S[:,2])
\end{lstlisting}
Here, the subscript string is based on the elements of the Zeeman term:
\begin{equation*}
     (\mathrm{H_{\mathrm{Z}}})_{mn} = -\sum_{i}\nu_{i}(\mathrm{I}_{iz})_{mn} = -\sum_{i}\nu_{i}\mathrm{S}_{i3mn},
\end{equation*}
where $ \mathrm{S}_{i3mn} = \mathrm{S}_{izmn}$. Furthermore, due to the 0-based indexing in Python, to access the elements $\mathrm{S}{i3mn}$ in coding, we would use \lstinline{S[:, 2]} instead.

Add the scalar \textit{J}-coupling term, the entire Hamiltonian is calculated as:
\begin{lstlisting}
    N,J0,V = NUM_SPIN,MATRIX_J0,MATRIX_V  #Input: number of spins, matrix J0 and V
    S = Angular_Momentum_Operator(N)  
    HZ = -np.einsum('i,imn->mn',V,S[:,2])
    HJ0 = np.einsum('ij,iamk,jakn->mn',J0,S,S)
    H = HZ+HJ0
\end{lstlisting}

\subsubsection{Rotating-frame Hamiltonian}
In the rotating frame, the precession frequency of the spin ($\nu_i = \gamma_i B_0 \delta_i$) is determined by subtracting the rotating frequency of the reference frame, which is $\gamma_i B_0$, from the actual Larmor frequency. In addition, in the defined rotating frame, the \textit{J}-couplings between heteronuclear spins only have the secular parts \cite{levitt2013spin}. The secular \textit{J}-coupling term is similar to that of the isotropic \textit{J}-coupling term but with $\hat{\mathbf{I}}_i\cdot \hat{\mathbf{I}}_j$ replaced by $\hat{I}_{iz}\hat{I}_{jz}$. The whole Hamiltonian in the rotating frame can be expressed as:
\begin{equation*}
    \hat{H}^{\rm (Rot )}= -\sum_{i} \nu_i \hat{I}_{iz} +\sum_{\substack{i<j \\ \text{(Homo)}}} J_{ij}(\hat{\mathbf{I}}_i \cdot \hat{\mathbf{I}}_j) + \sum_{\substack{i<j \\ \text{(Hetero)}}}J_{ij} (\hat{I}_{iz}\cdot\hat{I}_{jz}).
\end{equation*}
Here, $nu_i=\gamma_i B_0 \delta_i$.

The matrix representation of the rotating-frame Hamiltonian $\hat{H}^{\rm (Rot)}$ can thus be expressed as:
\begin{equation}
    \mathrm{H}^{\rm (Rot)} = -\sum_{k=1}\nu_{i}\,\mathrm{I}_{iz}+\sum_{\substack{ i<j \\ \text{(Homo) }}} J_{ij}\left( \sum_{\alpha=x,y,z} \mathrm{I}_{i\alpha}\cdot\mathrm{I}_{j\alpha}  \right) + \sum_{\substack{ r<s \\ \text{(Hetero) }}} J_{ij}\left( \sum_{\alpha=x,y,z} \mathrm{I}_{i\alpha}\cdot\mathrm{I}_{j\alpha}  \right) .
\label{Eq: high-field H}
\end{equation}

\heading{Calculation}
In this case, matrix V is calculated as below:
\begin{lstlisting}
    V = B0*Gyro*CS
\end{lstlisting}

Additionally, one has to distinguish the \textit{J}-couplings between homonuclear spins or heteronuclear spins. As a convention, the homonuclear and heteronuclear J coupling parameters are stored in matrices J$_0$ and J$_1$, respectively. Consequently, the Hamiltonian can be calculated with the code snippet below:
\begin{lstlisting}
    #Input: number of spins, matrix J0, J1 and V
    N,J0,J1,V = NUM_SPIN,MATRIX_J0,MATRIX_J1,MATRIX_V  
    S = Angular_Momentum_Operator(N)  
    HZ = -np.einsum('i,imn->mn',V,S[:,2])
    HJ0 = np.einsum('ij,iamk,jakn->mn',J0,S,S)
    HJ1 = np.einsum('ij,imk,jkn->mn',J1,S[:,2],S[:,2])
    H = HZ+HJ0+HJ1
\end{lstlisting}

\subsection{Matrix representations of density operators in Hilbert space}
In Hilbert space, the density operator is represented as a square matrix. The density operator, denoted as $\hat{\rho}$, can be expanded over the bases $\{|i\rangle\}$ as:
\begin{equation*}
    \hat{\rho}=\sum_{ij}^{d}\rho_{ij}|i\rangle\langle j|.
\end{equation*}
Here, $d$ represents the dimension of the density operator and $\rho_{ij}$ represents the elements of the density matrix in the chosen basis. The corresponding matrix representation of the density in the chosen basis in given by:
\begin{equation*}
        \begin{pmatrix}
        \rho_{11} &  \cdots & \rho_{1d}\\
        \vdots    &  \ddots & \vdots\\
        \rho_{d1} &  \cdots & \rho_{dd}
    \end{pmatrix},
\end{equation*}
where $\rho_{ij}$ is placed at the $i$-th row and $j$-th column position in the matrix.

The matrix representation of the concentration-normalized density operator is obtained by multiplying each element of the corresponding density matrix by the concentration.

\subsection{The action of superoperators in Hilbert space}
The action of a superoperator $\doublehat{O}$ on an operator $\hat{A}$ can be generally expressed as a linear combination of the transformations involving the operators $\hat{M}_{k}$ and $\hat{N}_{l}$ and the associated coefficients $o_{kl}$:
\begin{equation*}
    \doublehat{O}\hat{A} = \sum_{kl}o_{kl}\hat{M}_{k}\hat{A}\hat{N}_{l}.
\end{equation*}

Indeed, there are similarities in the definitions between operators and superoperators. Operators describe transformations between quantum states, while superoperators describe transformations between operators. Just as operators can be represented as matrices in Hilbert space, which is spanned by the quantum states, superoperators can also be represented by matrices in a higher-dimensional "superspace" known as Liouville space.  In Liouville space, the basis vectors are linear operators rather than quantum states. In the following, we would discuss the matrix representation of operators and superoperatros in Liouville space.

\subsection{Matrix representations of density operators in Liouville space}\label{Section: rho in L sapce}
The mapping of an density operator from Hilbert space into Liouville space involves the transformation:
% $\hat{\rho}$ in Hilbert space and in Liouville space \cite{Kuprov}:
\begin{equation*}
    \hat{\rho}=\sum_{i,j}\rho_{ij}|i\rangle\langle j| 
    % =\sum_{i,j}\rho_{ij}|i\rangle \otimes \langle j|
    \Longrightarrow |\rho \rrangle = \sum_{ij}\rho_{ij}|i j\rrangle,
\end{equation*}
where the basis operator $|i\rangle\langle j|$ in Hilbert space is transformed into the basis state $|i j\rrangle = |i\rangle \otimes |j\rangle$ in Liouville space. The mapped density operator is denoted by $|\rho\rrangle$.

As a result, the density operator in Liouville space is represented as a vector by ``flattening'' its matrix representation in Hilbert space. This is done in a row-wise manner, which means we concatenate the rows of the matrix to obtain a column vector:
\begin{equation*}
    \begin{pmatrix}
        \rho_{11} &  \cdots & \rho_{1d}\\
        \vdots    &  \ddots & \vdots\\
        \rho_{d1} &  \cdots & \rho_{dd}
    \end{pmatrix}
    \Longrightarrow
    \begin{pmatrix}
        \rho_{11} \\
        \rho_{12} \\
        \vdots  \\
        \rho_{1d} \\
        \vdots  \\
        \rho_{dd} \\
    \end{pmatrix}.
\end{equation*}
Here, the $d\times d $ square matrix is transformed into a $d^2\times 1$ column matrix,  where $d$ represents the dimensionality of the corresponding Hilbert space. Additionally, the density operator in Hilbert space is represented in the basis states $\{|i\rangle\}$. While, in Liouville space, the density operator is represented using the basis states $\{|ij\rrangle\}$. The basis states $\{|i\rangle\}$ span the Hilbert space of the spin system, while the basis states $\{|ij\rrangle\}$ span the Liouville space.

Note here, the transformation of the density matrix from Hilbert space to Liouville space remains unaffected by the specific normalization of the density operator. As a result, the mentioned procedure applies to concentration-normalized density operators as well.

\heading{Implementation}
The code below generates the matrix representation of a density operator in Liouville space given the input of the density matrix in Hilbert space (denoted as \lstinline{rho}).
\begin{lstlisting}
    rho = %*\textcolor{cyan}{rho}*).flatten().reshape((-1, 1))
\end{lstlisting}
where the function \lstinline{flatten()} enforces a row-wise flattening, and the code \lstinline{reshape((-1, 1))} transposes the flattened matrix from a 1D row-vector to a column-vector.

\subsection{Matrix representations of observables in Liouville space}\label{Section: O in L sapce}
The NMR signal measured at certain time, denoted as $\mathrm{S}$, can be calculated by taking the trace over the product of the observable $\hat{O}$ and the concentration-normalized density operator $\hat{\sigma}$:
\begin{equation*}
    \mathrm{S}  = \mathrm{Tr}(\hat{\sigma} \hat{O}) =\sum_{ij} \sigma_{ij}O_{ji} = \sum_{ij} \sigma_{ij}(O^{\dagger})_{ij}^{*}.
\end{equation*}
Since we have $\sigma_{ij}=\llangle ij|\sigma \rrangle$ and $(O^{\mathrm{\dagger}})_{ij}^{*} = \llangle O^{\mathrm{\dagger}} | ij \rrangle$, given $|\sigma \rrangle$ and $|O^{\mathrm{\dagger}} \rrangle$ mapped from $\hat{\sigma}$ and $\hat{O}^{\dagger}$. The measured signal can be calculated in Liouville space as
\begin{equation*}
    \mathrm{S}  = \sum_{ij}  \llangle O^{\mathrm{\dagger}} | ij \rrangle \llangle ij|\sigma \rrangle  = \llangle O^{\mathrm{\dagger}}| \left(\sum_{ij}  | ij \rrangle \llangle ij| \right) |\sigma \rrangle = \llangle O^{\dagger} |\sigma \rrangle.
\end{equation*}
Here, we have used the fact that $\sum_{ij}  | ij \rrangle \llangle ij|$ equals to an identity superoperator.

\heading{Calculation}
The measured signal, $\llangle O^{\dagger} |\sigma \rrangle$, is calculated by taking a matrix product between the column-matrix $|\sigma \rrangle$ and the row-matrix $\llangle O^{\dagger}|$. Specifically, the row-matrix can be expressed as
\begin{equation*}
    \llangle O^{\dagger}| = |O^{\dagger}\rrangle^{\dagger}= |O^{\mathrm{T}}\rrangle^{\mathrm{T}}.
\end{equation*}
Consequently, the row matrix corresponding to the observable is calculated as:
\begin{lstlisting}
    O = MATRIX_O #Input: the Observable matrix in Hilbert space
    O = O.flatten('F')
\end{lstlisting}
Here, the function \lstinline{flatten('F')} enforces the column-wise compression (which is equivalent to taking a transpose of the matrix and performing a row-wise compression) and produces the row-matrix representing $\llangle O^{\dagger}|$.

Finally, the measured NMR signal is calculated as,
\begin{lstlisting}
    rho,O = COLUMN_RHO,COLUMN_O #Input: the density column and the observable column
    S = np.matmul(O,rho)[0][0].re
\end{lstlisting}
Here the code \lstinline{[0][0]} converts the calculated results from a 2D array into a number and the codes \lstinline{re} takes its real part.

\subsection{Matrix representations of linear transformations in Liouville space}\label{Section: Operator mapping}
Let us check the mapping of the linear transformation involving a left multiplication by $\hat{M}$ and a right multiplication by $\hat{N}$.
\begin{equation*}
\begin{aligned}
    \quad\hat{M}\cdot\hat{\rho}\cdot\hat{N}
    &=\sum_{ij}\rho_{ij}\hat{M}|i\rangle\langle j|\hat{N}\\
    &=\sum_{ij}\rho_{ij}\left (\sum_{k}|k\rangle\langle k|\right)\hat{M}|i\rangle\langle j|\hat{N}\left (\sum_{l}|l\rangle\langle l|\right)\\
    &=\sum_{ijkl}\rho_{ij}|k\rangle\langle k|\hat{M}|i\rangle\langle j|\hat{N}|l\rangle\langle l|\\
    &=\sum_{ijkl}\rho_{ij}M_{ki}N_{jl}|k\rangle\langle l|.
\end{aligned}
\end{equation*}
The transformed operator can be mapped into Liouville space with the substitution $|k\rangle\langle l|=|kl\rangle$. We denote the mapped superoperator as $|M\rho N\rrangle$.
\begin{equation*}
    |M\rho N\rrangle = \sum_{ijkl}\rho_{ij}M_{ki}N_{jl}|kl\rrangle.
\end{equation*}
Take $\rho_{ij}=\llangle ij | \rho\rrangle$,
\begin{equation*}
\begin{aligned}
    |M\rho N\rrangle & = \sum_{ijkl}\llangle ij | \rho\rrangle M_{ki}N_{jl}|kl\rrangle \\
                     & = \sum_{ijkl} M_{ki}N_{jl} |kl\rrangle\llangle ij|     \rho\rrangle \\
                     & = \left(\sum_{ijkl} M_{ki}N_{jl} |k\rangle\langle i| \otimes |l\rangle\langle j|  \right)  |\rho\rrangle
\end{aligned}       
\end{equation*}
With $N_{jl}=N^{\mathrm{T}}_{lj}$,
\begin{equation*}
    \begin{aligned}
        |M\rho N\rrangle &= \left(\sum_{ijkl} M_{ki}N^{\mathrm{T}}_{lj} |k\rangle\langle i| \otimes |l\rangle\langle j|  \right)  |\rho\rrangle \\
        &= \left(\sum_{ik} M_{ki}|k\rangle\langle i|\right)\otimes  \left( \sum_{lj}N^{\mathrm{T}}_{lj} |l\rangle\langle j| \right) |\rho\rrangle\\
    \end{aligned}
\end{equation*}
Since $\hat{M}=\sum_{ik} M_{ki}|k\rangle\langle i|$ and $\hat{N}^{\mathrm{T}}=\sum_{lj}N^{\mathrm{T}}_{lj} |l\rangle\langle j|$,
\begin{equation*}
    |M\rho N\rrangle = \hat{M}\otimes\hat{N}^{\mathrm{T}}|\rho\rrangle.
\end{equation*}

To summarize,
\begin{equation}
    \hat{M}\cdot\hat{\rho}\cdot\hat{N} \Longrightarrow  \hat{M}\otimes\hat{N}^{\mathrm{T}}|\rho\rrangle.
\label{Eq: Linear mapping}
\end{equation}

In some cases, either the left operator $\hat{M}$ or the right operator $\hat{N}$ alone acts on the density operator $\hat{\rho}$ while another operator is the identity operator $\hat{\textbf{\textit{1}}\,}$ (with the same dimension as $\hat{M}$ or $\hat{N}$). In such cases, the mapping law can be expressed by including the identity operator $\hat{\textbf{\textit{1}}\,}$ to complete the multiplication:
\begin{equation*}
\begin{split}
\hat{M}\cdot\hat{\rho} &= \hat{M}\cdot\hat{\rho}\cdot \hat{\textbf{\textit{1}}\,} \Longrightarrow \left(\hat{M}\otimes\hat{\textbf{\textit{1}}\,}\right)|\rho\rrangle,\\
\hat{\rho} \cdot \hat{N} &= \hat{\textbf{\textit{1}}\,}\cdot\hat{\rho}\cdot \hat{N} \Longrightarrow \left(\hat{\textbf{\textit{1}}\,}\otimes\hat{N}^{\mathrm{T}}\right)|\rho\rrangle.
\end{split}
\end{equation*}

The matrix representation of the superoperator can be derived by taking the Kronecker product between the matrix $\hat{M}$ and $\hat{N}^{\mathrm{T}}$. This is one of the most important algebraic properties used for constructing matrix representations of superoperators necessary for calculating chemical exchange in NMR.

\heading{Implementation}
 The code below generates a matrix representation of the mapped superoperator given the input matrix representations of $\hat{M}$ and $\hat{N}$.
\begin{lstlisting}
    np.kron(M,N.transpose()) #Input: Matrix M and N
\end{lstlisting}
 where the command \lstinline{N.transpose())} returns the matrix of $\hat{N}^{\mathrm{T}}$.

\subsection{Matrix representations Hamiltonian superoperators}\label{Section: Hamiltonian Superoperator}
In the Hilbert space, the Hamiltonian $\hat{H}$ acts on the density operator $\hat{\rho}$ through the commutator
$$
[\hat{H},\hat{\rho}] = \hat{H}\hat{\rho}-\hat{\rho}\hat{H}.
$$
According to the law of mapping (Eq.~\eqref{Eq: Linear mapping}), we write the Hamiltonian superoperator $\doublehat{H}$ as:
\begin{equation*}
    \doublehat{H}=\hat{H}\otimes\hat{\textbf{\textit{1}}\,}-\hat{\textbf{\textit{1}}\,}\otimes\hat{H}^{\mathrm T},
\label{Eq: SuperH}
\end{equation*}
where $\hat{\textbf{\textit{1}}\,}$ is the identity operator acting on the Hilbert space of the molecule.

\heading{Implementation}
The code below calculates the matrix representation of the Hamiltonian superoperator. 

\begin{lstlisting}
    N,H = NUM_SPIN,MATRIX_H #Input: The number of spins, and the matrix H 
    Id = np.eye(2**N)
    HS = np.kron(H,Id)-np.kron(Id,H.transpose())
\end{lstlisting}

\subsection{Matrix representations of relaxation superoperators}\label{Section: Relaxation Superoperator}
The evolution of spins is governed by the LvN equation,
\begin{equation*}
    \frac{\partial}{\partial t}\hat{\rho}(t)=-i[\hat{H}+\hat{H}_1(t),\hat{\rho}(t)],
\end{equation*}
where the time-independent component $\hat{H}$ governs the coherent dynamics and the time-dependent part $\hat{H}_1(t)$ produces relaxation effects.

To deal with the random fluctuating part of the Hamiltonian $\hat{H}_1(t)$, the crucial assumption (known as \textit{motional narrowing} \cite{kowalewski2017nuclear}) is made: the correlation time scale of the fluctuation, $\tau_c$, is sufficiently short that its product with the root-mean-square of the fluctuation of $\hat{H}_{1}(t)$ is much less than a unity. The resulting master equation of the density operator can be simplified as:
\begin{equation}
    \frac{\partial}{\partial t}\hat{\rho}(t)= -i[\hat{H},\hat{\rho}(t)] -\frac{1}{2}\int_{-\infty}^{\infty} d\tau \overline{[\hat{H}_1(t),[e^{-i\hat{H}_0\tau}\hat{H}_1(t-\tau)e^{i\hat{H}_0\tau},\hat{\rho}(t)]]},
    \label{Eq: VN}
\end{equation}
where the overline stands for the average over the fluctuations of the noise fields.

The time-dependent part of the Hamiltonian stems from the coupling between the molecule and its surroundings and the stochastic interaction of the molecule with itself. Despite the complexity, one can consider a simple case where each spin only relaxes due to the random fluctuating local fields. The resulting time-dependent Hamiltonian takes the form:
\begin{equation*}
    \hat{H}_1(t)=-\sum_{i} \gamma_{i}\left( B_{ix}(t)\hat{I}_{ix}+B_{iy}(t)\hat{I}_{iy}+B_{iz}(t)\hat{I}_{iz}\right),
\end{equation*}
where $B_{ix}$, $B_{iy}$, and $B_{iz}$ denote the random fluctuating fields acting on the $i$-th spin along different directions. Additionally, the fluctuations of $B_{ix}$, $B_{iy}$, and $B_{iz}$ will be treated as independent. For simplicity, we make an additional assumption that the time scale of the fluctuation $\tau_c$ is considerably shorter than the time scale of coherent dynamics, known as ``\textit{extreme narrowing}'' \cite{ivanov2008high}. As a result, we can assume that the noise fields' fluctuations are delta-correlated in time,
\begin{equation*}
    \gamma_i^2  \overline{B_{i\alpha}(t) B_{j\beta}(t-\tau)} = \frac{1}{T_{1i}}\delta(\tau)\delta_{ij}\delta_{\alpha\beta}.
\end{equation*}
 Here $T_{1i}$ is the longitudinal relaxation time of the $i$-th spin (assumed to be known, for example by a measurement performed at high field). We note the correlations of the fluctuating fields are independent of time variable $t$ assuming the noise field is stationary. 

Insert these assumptions into Equation~\eqref{Eq: VN}, the LvN equation can be expressed as:
\begin{equation*}
    \frac{\partial}{\partial t}\hat{\rho}(t)= -i[\hat{H},\hat{\rho}(t)] +\doublehat{R}\hat{\rho}(t),
\end{equation*}
where the action of the relaxation superoperator takes the form:
\begin{equation*}
\doublehat{R}\hat{\rho}(t)=\sum_{i}\frac{1}{2T_{1i}}\sum_{\alpha=x,y,z}\left[2\hat{I}_{i\alpha} \hat{\rho}(t) \hat{I}_{i\alpha}-\hat{\rho}(t)\hat{I}_{i\alpha}^2-\hat{I}_{i\alpha}^2 \hat{\rho}(t)\right].
\label{Eq: Hilbert form Relaxation}
\end{equation*}

According to Eq.~\eqref{Eq: Linear mapping}, in Liouville space, the relaxation superoperator $\doublehat{R}$ is expressed as follows:
\begin{equation*}
    \doublehat{R}=\sum_{i}\frac{1}{2T_{1i}}\sum_{\alpha=x,y,z}\left[2\hat{I}_{i\alpha}\otimes \hat{I}_{i\alpha}^{\mathrm T}-\hat{\textbf{\textit{1}}\,}\otimes\hat{I}_{i\alpha}^2-\hat{I}_{i\alpha}^2\otimes\hat{\textbf{\textit{1}}\,}\right].
\end{equation*}
Moreover, assume the $i$-th spin is with an angular momentum quantum number $L_i$, i.e., $\sum_{\alpha}\hat{I}_{i\alpha}^2 = L_i(L_i+1)\hat{\textbf{\textit{1}}\,}$, the form of the relaxation superoperator can be further simplified as follows:
\begin{equation*}
    \doublehat{R}=\sum_{i}\frac{1}{T_{1i}} \left[ \left(\sum_{\alpha=x,y,z} \hat{I}_{i\alpha}\otimes \hat{I}_{i\alpha}^{\mathrm T} \right) - L_i(L_i+1) \hat{\textbf{\textit{1}}\,}\!\otimes\hat{\textbf{\textit{1}}\,} \right].
\end{equation*}

Specifically, for a system consisted of only spin-1/2 particles, the relaxation superoperator $\doublehat{R}$ takes the form:
\begin{equation}
    \doublehat{R}=\sum_{i}\frac{1}{T_{1i}} \left(\sum_{\alpha=x,y,z} \hat{I}_{i\alpha}\otimes \hat{I}_{i\alpha}^{\mathrm T} \right) - \frac{3}{4}\sum_{i}\frac{1}{T_{1i}}  \hat{\textbf{\textit{1}}\,}\!\otimes\hat{\textbf{\textit{1}}\,}.
\label{Eq: R}
\end{equation}

\heading{Calculation}\label{A: R code}
The matrix form of the relaxation superoperator is expressed as follows:
\begin{equation}
    \mathrm{R} = \sum_{i} \frac{1}{T_{1i}} \left(\sum_{\alpha=x,y,z} \mathrm{I}_{i\alpha}  \otimes \mathrm{I}_{i\alpha }^{\mathrm T} \right)- \frac{3}{4}\sum_{i}\frac{1}{T_{1i}}   \textbf{1}_{2N},
\label{Eq: Matrix_R}
\end{equation}
where by definition $\textbf{1}_{2N}$ represents the identity matrix with the size $2^{2N}=4^{N}$.

In the context of the calculation, we store the individual spin relaxation time in matrix T$_1$,
\begin{equation*}
    \mathrm{T}_1=\begin{bmatrix}
        T_{11} & T_{12} &  \dots & T_{1N}
    \end{bmatrix}.
\end{equation*}

Consequently, the code snippet below calculates Eq.~\eqref{Eq: Matrix_R}.
\begin{lstlisting}
    N,T1 = NUM_SPIN,MATRIX_T1 #Input: number of spins, matrix T1
    S = Angular_Momentum_Operator(N)  
    R = np.einsum("i,iamn,iaqp>mpnq",1/T1,S,S).reshape(4**N,4**N)
        -3/4*np.eye(4**N)*np.sum(1/T1)
\end{lstlisting}
Here, the subscript string in the first summation is based on the elements of Eq.~\eqref{Eq: Matrix_R} as shown below:
\begin{equation*}
\begin{aligned}
    \left(\mathrm{R}_1\right)_{mp,nq} &=  \sum_{i=1}^{N} \sum_{\alpha=1}^{3}\left(\frac{1}{\mathrm{T}_1}\right)_i               (\mathrm{I}_{i\alpha})_{mn} (\mathrm{I}_{i\alpha}^{\mathrm{T}})_{pq} \\
     &= \sum_{i=1}^{N} \sum_{\alpha=1}^{3}\left(\frac{1}{\mathrm{T}_1}\right)_i\mathrm{S}_{i\alpha mn}\mathrm{S}_{i\alpha qp} ,
\end{aligned}
\end{equation*}
where we have used $(\mathrm{I}_{i\alpha}^{\mathrm{T}})_{pq} = \mathrm{S}_{i\alpha qp}$.

\subsection{Matrix representation of partial trace superoperators}\label{Section: Partial trace superoperator}
Let's consider an example where a molecule XY, described by the density operator $\hat{\rho}_{\mathrm{XY}}$, dissociates into molecules X and Y:
\begin{equation*}
    \rm XY \longrightarrow \underline{X}+Y 
\end{equation*}

Assume the density operator, denoted as $\hat{\rho}_{\mathrm{X}}$, takes the form:
\begin{equation*}
    \hat{\rho}_{\mathrm{X}} = \sum_{xx^{'}} \rho_{xx^{'}} |x\rangle \langle x^{'}|,
\end{equation*}
where the polarization moments $\rho_{xx^{'}}$ can be obtained by assuming a pseudo-measurement on  $\hat{\rho}_{\mathrm{XY}}$ using the observable $|x^{'}\rangle\langle x|\otimes \hat{\textbf{\textit{1}}\,}_{\!\mathrm{Y}}$,
\begin{equation*}
    \rho_{xx^{'}} = \mathrm{Tr}\left[ (|x^{'}\rangle\langle x|\otimes \hat{\textbf{\textit{1}}\,}_{\!\mathrm{Y}})\hat{\rho}_{\mathrm{XY}}  \right].
\end{equation*}
Since the identity operator can be expanded as $\hat{\textbf{\textit{1}}\,}_{\!\mathrm{Y}}  = \sum_{y} |y\rangle \langle y|$ where $|y\rangle$ represents a basis state for molecule Y, the observable can be re-expressed as,
\begin{equation*}
    |x^{'}\rangle\langle x|\otimes \hat{\textbf{\textit{1}}\,}_{\!\mathrm{Y}} = \sum_{y} (|x^{'}\rangle \otimes |y\rangle)   (\langle x|\otimes \langle y|) = \sum_{y} |x^{'}y\rangle\langle xy|
\end{equation*}
where we have used the mixed-product property of the Kronecker product. Based on the expansion, the trace operation can be reformulated as: 
\begin{equation*}
    \mathrm{Tr}\left[ \left(|x^{'}\rangle\langle x|\otimes \hat{\textbf{\textit{1}}\,}_{\!\mathrm{Y}}\right)\hat{\rho}_{\mathrm{XY}}  \right] = \sum_{y} \langle xy| \hat{\rho}_{\mathrm{XY}} |x^{'}y\rangle .
\end{equation*}

Consequently, the density operator of the dissociated molecule X is given by:
\begin{equation*}
\begin{aligned}
     \hat{\rho}_{\mathrm{X}} &= \sum_{xx^{'}}  |x\rangle \langle x^{'}| \sum_{y} \langle xy| \hat{\rho}_{\mathrm{XY}} |x^{'}y\rangle \\
     &= \sum_{y} \sum_{xx^{'}}  |x\rangle  \langle xy| \hat{\rho}_{\mathrm{XY}} |x^{'}y\rangle  \langle x^{'}|.
\end{aligned}
\end{equation*}
Insert,
\begin{equation*}
\begin{aligned}
    |x\rangle  \langle xy| &= (|x\rangle\otimes 1)(\langle x|\otimes \langle y|)= (|x\rangle\langle x|)\otimes \langle y|, \\
    |x^{'}y\rangle  \langle x^{'}| &= (|x^{'}\rangle \otimes |y\rangle) (\langle x^{'}| \otimes 1) =(|x^{'}\rangle \langle x^{'}|) \otimes |y\rangle,
\end{aligned}
\end{equation*}
the form of $\hat{\rho}_{\mathrm{X}}$ can be further simplified as,
\begin{equation*}
\begin{aligned}
    \hat{\rho}_{\mathrm{X}} &= \sum_{y} \left[\left(\sum_{x} |x\rangle\langle x|\right) \otimes \langle y|\right] \hat{\rho}_{\mathrm{XY}} | \left[\left(\sum_{x^{'}} |x^{'}\rangle\langle x^{'}|\right) \otimes |y\rangle \right] \\
    &= \sum_{y} \left(\hat{\textbf{\textit{1}}\,}_{\!\mathrm{X}}\otimes \langle  y|\right)\hat{\rho}_{\mathrm{XY}}\left(\hat{\textbf{\textit{1}}\,}_{\!\mathrm{X}}\otimes |y\rangle \right).
\end{aligned}
\end{equation*}

One can quickly check the validity of the equation by testing it on a a separable density operator, $\hat{\rho}_{\rm X} \otimes \hat{\rho}_{\rm Y}$,
\begin{equation*}
\begin{aligned}
    &\quad  \sum_{y} \left(\hat{\textbf{\textit{1}}\,}_{\!\mathrm{X}}\otimes \langle y|\right)\hat{\rho}_{\rm X} \otimes \hat{\rho}_{\rm Y}\left(\hat{\textbf{\textit{1}}\,}_{\!\mathrm{X}}\otimes |y\rangle \right) \\
    &= \left(\hat{\textbf{\textit{1}}\,}_{\!\mathrm{X}}\, \hat{\rho}_{\rm X}\, \hat{\textbf{\textit{1}}\,}_{\!\mathrm{X}} \right) \otimes \left( \sum_{y}  \langle y| \hat{\rho}_{\rm Y} | y\rangle \right) \\
    & =  \hat{\rho}_{\rm X}  \mathrm{Tr}( \hat{\rho}_{\rm Y} ) = \hat{\rho}_{\rm X}.
\end{aligned}
\end{equation*}
The transformation does produce the expected the result. Indeed, $\hat{\rho}_{\mathrm{X}}$ is obtained by taking the partial trace over the degree of freedom associated with molecule Y, and therefore we denote the transformation by the partial trace superoperator,
\begin{equation}
    \doublehat{T}_{\mathrm{Y}}^{\mathrm{(XY)}} \hat{\rho}_{\mathrm{XY}}= \sum_{y}\left(\hat{\textbf{\textit{1}}\,}_{\!\mathrm{X}}\otimes \langle y|\right)\hat{\rho}_{\mathrm{XY}}\left(\hat{\textbf{\textit{1}}\,}_{\!\mathrm{X}}\otimes |y\rangle \right).
\label{Eq: Tr1_H}
\end{equation}

In Liouville space, the form of $\doublehat{T}_{\mathrm{Y}}^{\mathrm{(XY)}}$ is obtained based to the mapping law given in Section~\ref{Section: Operator mapping},
\begin{equation}
    \doublehat{T}^{\mathrm{(XY)}}_{\mathrm{Y}}= \sum_{y}\left(\hat{\textbf{\textit{1}}\,}_{\!\mathrm X}\otimes \langle y|\right) \otimes \left(\hat{\textbf{\textit{1}}\,}_{\!\mathrm{X}}\otimes \langle y|^{*}\right).
\label{Eq: Tr1_L}
\end{equation}
Here, we have employed $|y\rangle^{\mathrm{T}}=\langle y|^{*}$.

Similarly, the discussion can be extended to the partial trace superoperator
$\doublehat{T}_{\mathrm{X}}^{\mathrm{(XY)}}$, which is used to calculate the density operator of the dissociated molecule Y from the molecule XY. In this case, the partial trace is performed over the degree of freedom associated with molecule X.

Furthermore, in the context of a molecule XYZ dissociating into a molecule XZ and Y, the density operator of the dissociated molecule XZ can be determined using the partial trace superoperator $\doublehat{T}_{\mathrm{Y}}^{\mathrm{(XYZ)}}$. Here, the partial trace is taken over the degree of freedom associated with molecule Y.

Below, we summarize the form of the partial trace superoperators for the three different cases:\newline

\noindent\textbf{Case 1: \textbf{XY} $\longrightarrow$ \underline{\textbf{X}}+\textbf{Y}} \newline
In Case 1, we consider the reaction: 
\begin{equation*}
    \rm XY \longrightarrow \underline{X}+Y 
\end{equation*}

The density operator of the dissociated molecule X, can be obtained by applying the partial trace superoperator, $\doublehat{T}_{\mathrm{Y}}^{\mathrm{(XY)}}$, to the density operator of molecule XY.

The forms of $\doublehat{T}_{\mathrm{Y}}^{\mathrm{(XY)}}$ in Hilbert space and Liouville space are given by Eq.~\eqref{Eq: Tr1_H} and Eq.~\eqref{Eq: Tr1_L}, respectively. \newline

\noindent\textbf{Case 2: \textbf{XY} $\longrightarrow$ \textbf{X} + \underline{\textbf{Y}}}
In Case 2, we consider the reaction:
\begin{equation*}
    \rm XY \longrightarrow X + \underline{Y} 
\end{equation*}
Similarly, the density operator of the dissociated molecule Y, denoted as $\hat{\rho}_{\mathrm{Y}}$, is obtained by applying the partial superoperator $\doublehat{T}_{\mathrm{X}}^{\mathrm{(XY)}}$ to $\hat{\rho}_{\mathrm{XY}}$,
\begin{equation}
    \hat{\rho}_{\mathrm{Y}} = \doublehat{T}_{\mathrm{X}}^{\mathrm{(XY)}}\hat{\rho}_{\mathrm{XY}} =  \sum_{x} \left( \langle x|\otimes \hat{\textbf{\textit{1}}\,}_{\!\mathrm Y} \right)\hat{\rho}_{\mathrm{XY}}\left(|x\rangle \otimes \hat{\textbf{\textit{1}}\,}_{\!\mathrm Y} \right) .
\label{Eq: Tr2_H}
\end{equation}
Here, $\{ |x\rangle \}$ forms a basis for molecule X and $\hat{\textbf{\textit{1}}\,}_{\mathrm Y}$ is the identity operator acting on the Hilbert space associated with molecule Y.

The action of $\doublehat{T}_{\mathrm{X}}^{\mathrm{(XY)}}$ can be interpreted as performing a partial trace over the degree of freedom associated with molecule X. This can be easily seen by apply it to a separable density operator,
\begin{equation*}
    \doublehat{T}_{\mathrm{X}}^{\mathrm{(XY)}} \hat{\rho}_{\rm X} \otimes \hat{\rho}_{\rm Y} =  \mathrm{Tr}(\hat{\rho}_{\rm X}) \hat{\rho}_{\rm Y} = \hat{\rho}_{\rm Y}.
\end{equation*}

In Liouville space, $\doublehat{T}_{\mathrm{X}}^{\mathrm{(XY)}}$ takes the form:
\begin{equation}
    \doublehat{T}_{\mathrm X}^{\mathrm{(XY)}}= \sum_{x}\left(\langle x|\otimes \hat{\textbf{\textit{1}}\,}_{\!\mathrm Y}\right) \otimes \left(\langle x|^{*}\otimes \hat{\textbf{\textit{1}}\,}_{\!\mathrm Y}\right).
\label{Eq: Tr2_L}
\end{equation}

\noindent\textbf{Case 3: \textbf{XYZ} $\longrightarrow$ \underline{\textbf{XZ}}+\textbf{Y}} \newline
In Case 3, we consider the reaction:
\begin{equation*}
    \rm XYZ \longrightarrow  \underline{XZ} +Y
\end{equation*}
The density operator of the dissociated molecule XZ, denoted as $\hat{\rho}_{\mathrm{XZ}}$, is calculated by applying the partial trace superoperator $\doublehat{T}_{\mathrm{Y}}^{\mathrm{(XYZ)}}$ to the density operator $\hat{\rho}_{\mathrm{XYZ}}$,
\begin{equation}
    \hat{\rho}_{\mathrm{XZ}} = \doublehat{T}_{\mathrm{Y}}^{\mathrm{(XYZ)}}\hat{\rho}_{\mathrm{XYZ}}= \sum_{y}\left(\hat{\textbf{\textit{1}}\,}_{\!\mathrm{X}}\otimes \langle y| \otimes \hat{\textbf{\textit{1}}\,}_{\!\mathrm Z} \right)\hat{\rho}_{\mathrm{XYZ}}\left(\hat{\textbf{\textit{1}}\,}_{\!\mathrm{X}}\otimes |y\rangle \otimes \hat{\textbf{\textit{1}}\,}_{\!\mathrm Z} \right),
\label{Eq: Tr3_H}
\end{equation}
Here, $\{ |y \rangle \}$ spans a basis for molecule fragments Y. The operators $\hat{\textbf{\textit{1}}\,}_{\!\mathrm{X}}$ and $\hat{\textbf{\textit{1}}\,}_{\mathrm Z}$ are the identity operators acting on the Hilbert space associated with molecule fragments X and Z, respectively.

The action of $\doublehat{T}_{\mathrm{Y}}^{\mathrm{(XYZ)}}$ can be interpreted as performing a partial trace over the degree of freedom associated with molecule fragment Y. Its effects on a separable density operator is given by, 
\begin{equation*}
    \doublehat{T}_{\mathrm{Y}}^{\mathrm{(XYZ)}} \hat{\rho}_{\rm X} \otimes \hat{\rho}_{\rm Y} \otimes \hat{\rho}_{\rm Z}=  \hat{\rho}_{\rm X} \otimes \hat{\rho}_{\rm Z} \mathrm{Tr}(\hat{\rho}_{\rm Y})  = \hat{\rho}_{\rm X} \otimes \hat{\rho}_{\rm Z}. 
\end{equation*}

In Liouville space, $\doublehat{T}_{\mathrm{Y}}^{\mathrm{(XYZ)}}$ takes the form:
\begin{equation}
    \doublehat{T}_{\mathrm Y}^{\mathrm{(XYZ)}}= \sum_{y}\left(\hat{\textbf{\textit{1}}\,}_{\!\mathrm{X}}\otimes \langle y|\otimes \hat{\textbf{\textit{1}}\,}_{\!\mathrm Z} \right) \otimes \left(\hat{\textbf{\textit{1}}\,}_{\!\mathrm{X}}\otimes \langle y|^{*} \otimes \hat{\textbf{\textit{1}}\,}_{\!\mathrm Z} \right).
\label{Eq: Tr3_L}
\end{equation}

\heading{Calculation}
The Zeeman product basis is adopted for representing the partial trace superoperators. In addition, we assume that the relative order of the Hilbert spaces associated with the undissociated spins remains unaltered before and after dissociation. This particular assumption ensures the consistency of the Hilbert spaces used to describe the same spins both before and after dissociation, eliminating the requirements for an additional basis transformation.

In Liouville space and with the defined basis, the matrix representations of the discussed partial trace superoperators can be derived based on to Eq.~\eqref{Eq: Tr1_L}, Eq.~\eqref{Eq: Tr2_L}, and Eq.~\eqref{Eq: Tr3_L}. In general, the identity operators are replaced with their corresponding identity matrices of the appropriate size. Furthermore, each Dirac's bra is substituted with the appropriate row vector from the identity matrix. For instance, the basis bra denoted as $\langle y|$ is represented by the row vector corresponding to the $y$-th row of the identity matrix with the appropriate size. The Kronecker products are then calculated using \lstinline{np.einsum} as introduced in Section~\ref{Section: einsum}. The corresponding code snippets for the calculations are given below.\newline

\noindent\textbf{Case 1: \textbf{XY} $\longrightarrow$ \underline{\textbf{X}}+\textbf{Y}} \newline
The elements of Eq.~\eqref{Eq: Tr1_L} represented in Zeeman product basis is given by:
\begin{equation*}
\begin{aligned}
    \left(\mathrm{T}^{\mathrm{(XY)}}_{\rm Y}\right)_{il,jkmn} &=
    \sum_{y}(\textbf{1}_{\mathrm{X}})_{ij}\,(\langle y|)_{k}\,(\textbf{1}_{\mathrm{X}})_{lm}\,(\langle y|)_{n} \\
    &=\sum_{y}(\textbf{1}_{\mathrm{X}})_{ij}\,(\textbf{1}_{\mathrm{Y}})_{yk}\,(\textbf{1}_{\mathrm{X}})_{lm}\,(\textbf{1}_{\mathrm{Y}})_{yn}
\end{aligned}
\end{equation*}
Here, $(\textbf{1}_{\mathrm{X}})$ and $(\textbf{1}_{\mathrm{Y}})$ refer to identity matrices.  Moreover, since $\langle y|$ is represented by the $y$-th row of $(\textbf{1}_{\mathrm{Y}})$, their elements are expressed as $(\langle y|)_{k} = (\textbf{1}_{\mathrm{Y}})_{yk}$.

% Their dimensions equal to $d_{\mathrm{X}}$ and $d_{\mathrm{Y}}$, respectively.

The summation is calculated with the code snippet below:
\begin{lstlisting}
    dX,dY = d_X,d_Y #Input: the dimensions of molecule X and Y
    IX,IY = np.eye(dX),np.eye(dY)
    Tr = np.einsum("ij,yk,lm,yn->iljkmn",IX,IY,IX,IY).reshape(dX**2,(dX*dY)**2)
\end{lstlisting}

\noindent\textbf{Case 2: \textbf{XY} $\longrightarrow$ \textbf{X} + \underline{\textbf{Y}}}
The elements of Eq.~\eqref{Eq: Tr2_L} represented in Zeeman product basis is given by:
\begin{equation*}
\begin{aligned}
        \left(\mathrm{T}_{\mathrm X}^{\mathrm{(XY)}}\right)_{jm,ikln} &= \sum_{x} (\langle x|)_{i}\,(\textbf{1}_{\mathrm{Y}})_{jk}\,(\langle x|)_{l}\,(\textbf{1}_{\mathrm{Y}})_{mn} \\
        &= \sum_{x} (\textbf{1}_{\mathrm{X}})_{xi}\,(\textbf{1}_{\mathrm{Y}})_{jk}\,(\textbf{1}_{\mathrm{X}})_{xl}\,(\textbf{1}_{\mathrm{Y}})_{mn}
\end{aligned}
\end{equation*}

Consequently, it is calculated as:
\begin{lstlisting}
    dX,dY = d_X,d_Y #Input: the dimensions of molecule X and Y
    IX,IY = np.eye(dX),np.eye(dY)
    Tr = np.einsum("xi,jk,xl,mn->jmikln",IX,IY,IX,IY).reshape(dY**2,(dX*dY)**2)
\end{lstlisting}

\noindent\textbf{Case 3: \textbf{XYZ} $\longrightarrow$ \underline{\textbf{XZ}}+\textbf{Y}} \newline
The elements of Eq.~\eqref{Eq: Tr3_L} represented in the defined Zeeman product basis is calculated as:
\begin{equation*}
\begin{aligned}
    \left(\mathrm{T}_{\mathrm{X}}^{\mathrm{(XYZ)}}\right)_{ilnr,jkmpqs} &= \sum_{y} (\textbf{1}_{\mathrm{X}})_{ij
    }\,(\langle y|)_{
    k}\,(\textbf{1}_{\mathrm{Z}})_{lm
    }\,(\textbf{1}_{\mathrm{X}})_{np
    }\,(\langle y|)_{
    q}\,(\textbf{1}_{\mathrm{Z}})_{rs
    } \\ 
    &= \sum_{y} (\textbf{1}_{\mathrm{X}})_{ij
    }\,(\textbf{1}_{\mathrm{Y}})_{
    yk}\,(\textbf{1}_{\mathrm{Z}})_{lm
    }\,(\textbf{1}_{\mathrm{X}})_{np
    }\,(\textbf{1}_{\mathrm{Y}})_{
    yq}\,(\textbf{1}_{\mathrm{Z}})_{rs
    } 
\end{aligned}
\end{equation*}

Consequently, it is calculated as:
\begin{lstlisting}
    dX,dY,dZ = d_X,d_Y,d_Z #Input: the dimensions of molecule X, Y, and Z
    IX,IY,IZ = np.eye(dX),np.eye(dY),np.eye(dZ)
    Tr = np.einsum('ij,yk,lm,np,yq,rs->ilnrjkmpqs',
         IX,IY,IZ,IX,IY,IZ).reshape((dX*dZ)**2,(dX*dY*dZ)**2)
\end{lstlisting}

To illustrate the construction of the subscript string, let's consider Case 3 as an example. In this case, the input labels are obtained by extracting the indices of the matrices involved. These input labels are then rearranged to form the output labels. The letter ``y" is used to represent the summation index and thus disappears in the output labels. The indices ``ilnr" correspond to the row indices and are arranged first in the output labels. Subsequently, the indices ``jkmpqs" representing the column indices are appended after the row indices in the output labels.

In spite of the intricate nature of the subscript strings, it is worth mentioning that users only need to pass the dimensions of each molecules as inputs to the code. No additional modifications to the code snippets are required when unitizing them.

\heading{Example}
We would like to take the dissociation of $^{15}$N-ammonium as an example to illustrate the procedure of calculating the corresponding partial trace superoperators. NMR spectra for this spin system are presented in the section~\ref{Section: ZF_NH4} and section~\ref{Section: HF_NH4}.

To begin with, we denote the ammonium molecule as $[^{15}\mathrm{NH_{A}H_{B}H_{C}H_{D}}]^{+}$. Since ammonium has four equivalent protons, each with an equal probability of dissociation. We need to analyze the form of the partial trace superoperator for different cases. Let us first discuss the dissociation of proton H$_{\mathrm{A}}$. The corresponding chemical reaction is represented as:
\begin{equation*}
    [^{15}\mathrm{NH_{A}H_{B}H_{C}H_{D}}]^{+} \longrightarrow \mathrm{^{15}NH_{B}H_{C}H_{D}} +\mathrm{H_{A}}^{+}
\end{equation*}

The given reaction is of the same form as discussed in Case 3, where the molecules X, Y, and Z are identified as $^{15}\mathrm{N}$, $\mathrm{H_{A}}$, $\mathrm{H_{B}H_{C}H_{D}}$, respectively. The form of the corresponding partial trace superoperator is given by Eq.~\eqref{Eq: Tr3_H}.

It is important to emphasize that the obtained density operator, derived from Eq.~\eqref{Eq: Tr3_H}, is described within a reduced Hilbert space, where the relative ordering of the Hilbert spaces for the remaining individual spins is preserved. However, it is crucial to note that the consistency between this reduced Hilbert space and the Hilbert space based on the molecule notation of the dissociated molecule is not always guaranteed. In certain reactions, where the molecular configuration changes and the ordering of spins is altered, the reduced Hilbert space after the partial trace may not align with the Hilbert space based on the molecule notation. Consequently, an additional mapping of states between the two differently defined Hilbert spaces becomes necessary. 

For example, in the reaction:
\begin{equation*}
[^{15}\mathrm{NH_{A}H_{B}H_{C}H_{D}}]^{+} \longrightarrow \mathrm{^{15}NH_{C}H_{D}H_{B}} +\mathrm{H_{A}}^{+}
\end{equation*}
the reduced Hilbert space after the partial trace does not correspond to the Hilbert space based on the molecule notation $\mathrm{^{15}NH_{C}H_{D}H_{B}}$. Therefore, an additional mapping of states is required.

Go back to our earlier discussed reaction, the two Hilbert spaces are indeed consistent and the dissociation process is correctly described by Eq.~\eqref{Eq: Tr3_H}. In the context of the matrix representation of Eq.~\eqref{Eq: Tr3_H}, we choose the Zeeman product basis as the set of basis states which are separable with respect to molecules X, Y, and Z, allowing for the matrix representations of the operators for the entire molecule to be obtained by taking Kronecker products between the matrix representations of the involved individual molecule operators. The identity operators $\hat{\textbf{\textit{1}}\,}_{\!\mathrm{X}}$ and $\hat{\textbf{\textit{1}}\,}_{\mathrm Z}$, are consistently represented as identity matrices. In the case of molecule fragment X, which encompasses a single spin-1/2 nuclear spin, $\hat{\textbf{\textit{1}}\,}_{\!\mathrm{X}}$ is represented as a $2\times 2$ identity matrix (denoted as $\mathbf{1}_1$):
\begin{equation*}
    \hat{\textbf{\textit{1}}\,}_{\!\mathrm{X}} = \mathbf{1}_1  =\begin{pmatrix}
        1 & 0 \\
        0 & 1 \\
    \end{pmatrix}.
\end{equation*}
Similarly, molecule fragment Z is composed of three spin-1/2 spins, therefore $\hat{\textbf{\textit{1}}\,}_{\mathrm Z}$ is represented as a $8\times 8$ identity matrix (denoted as $\mathbf{1}_3 = \mathbf{1}_1 \otimes \mathbf{1}_1 \otimes \mathbf{1}_1$):
\begin{equation*}
    \hat{\textbf{\textit{1}}\,}_{\mathrm Z} =  \mathbf{1}_3 = \begin{pmatrix}
1 & 0 & 0 & 0 & 0 & 0 & 0 & 0 \\
0 & 1 & 0 & 0 & 0 & 0 & 0 & 0 \\
0 & 0 & 1 & 0 & 0 & 0 & 0 & 0 \\
0 & 0 & 0 & 1 & 0 & 0 & 0 & 0 \\
0 & 0 & 0 & 0 & 1 & 0 & 0 & 0 \\
0 & 0 & 0 & 0 & 0 & 1 & 0 & 0 \\
0 & 0 & 0 & 0 & 0 & 0 & 1 & 0 \\
0 & 0 & 0 & 0 & 0 & 0 & 0 & 1 \\
\end{pmatrix},
\end{equation*}

The choice of $|y\rangle$ for molecule Y is not unique as long as they form a complete set of orthogonal and normalized basis states. For simplicity, we set $|y\rangle$ to be either $|\alpha\rangle$ or $|\beta\rangle$. The matrix form of these basis states are expressed as:
\begin{equation*}
    | \alpha \rangle = \begin{pmatrix}
        1 \\ 0
    \end{pmatrix}
    \quad
    | \beta \rangle = \begin{pmatrix}
        0 \\ 1
    \end{pmatrix}
    \quad
    \langle \alpha | = \begin{pmatrix}
        1 & 0
    \end{pmatrix}
    \quad
    \langle \beta | = \begin{pmatrix}
        0 & 1
    \end{pmatrix}
\end{equation*}

Finally, the summation in Eq.~\eqref{Eq: Tr3_H} is performed over all possible $|y\rangle$. In this case, the summation consists two terms, where $|y\rangle=|\alpha\rangle$ or $|y\rangle=|\beta\rangle$. Combing the results, the density matrix for the dissociated ammonia molecule is given by,
\begin{equation*}
\begin{aligned}
    \rho_{\mathrm{NH_3}}
    &= \left(\mathbf{1}_1 \otimes \begin{pmatrix}
        1 & 0
    \end{pmatrix} \otimes \mathbf{1}_3 \right) \rho_{\mathrm{NH_4}} \left(\mathbf{1}_1 \otimes \begin{pmatrix}
        1 \\ 0
    \end{pmatrix} \otimes \mathbf{1}_3 \right) \\
    & +\left(\mathbf{1}_1 \otimes \begin{pmatrix}
        0 & 1
    \end{pmatrix} \otimes \mathbf{1}_3 \right) \rho_{\mathrm{NH_4}} \left(\mathbf{1}_1 \otimes \begin{pmatrix}
        0 \\ 1
    \end{pmatrix} \otimes \mathbf{1}_3 \right),
\end{aligned}
\end{equation*}
where the density matrices for the dissociated ammonia and the dissociation-consumed ammonium are denoted as $\rho_{\mathrm{NH_3}}$ and $\rho_{\mathrm{NH_4}}$, respectively. By examining the dimensions of the matrices involved, one can readily confirm the consistency of the equations. The matrices multiplied on the left from the density matrix have dimensions of $16 \times 32$, while those on the right from the density matrix have dimensions of $32 \times 16$. Additionally, the density matrix $\rho_{\mathrm{NH_4}}$ has a dimension of $32 \times 32$. As a result, the resulting matrix will be of size $16 \times 16$, which matches the dimension of $\rho_{\mathrm{NH_3}}$.

In Liouville space, the transformation is described by a partial trace superoperator, denoted by $\mathrm{T}_1$. The form of $\mathrm{T}_1$ is based on Eq.~\eqref{Eq: Tr3_L} and is expressed as follows:
\begin{equation*}
\begin{aligned}
    \mathrm{T}_1 &= \mathbf{1}_1 \otimes \begin{pmatrix}
        1 & 0
    \end{pmatrix} \otimes \mathbf{1}_3 \otimes \mathbf{1}_1 \otimes \begin{pmatrix}
        1 & 0
    \end{pmatrix} \otimes \mathbf{1}_3 \\
    &+ \mathbf{1}_1 \otimes \begin{pmatrix}
        0 & 1
    \end{pmatrix} \otimes \mathbf{1}_3 \otimes \mathbf{1}_1 \otimes \begin{pmatrix}
        0 & 1
    \end{pmatrix} \otimes \mathbf{1}_3.
\end{aligned}
\end{equation*}
The matrix $\mathrm{T}_1$ has a dimension of $256 \times 1024$ which transforms the density vector for ammonium ($1024\times 1$) to a density vector for ammonia ($256\times 1$).

The matrix $\mathrm{T}_1$ can be calculated by performing the Kronecker products using the \lstinline{np.kron}. Alternatively, one can employ the code snippet provided for calculating the partial trace superoperator in Case 3. The dimensions of molecules X, Y, and Z are $d_{\mathrm{X}}= 2$, $d_{\mathrm{Y}}=2$, and $d_{\mathrm{Z}}=8$, respectively.
\begin{lstlisting}
    dX,dY,dZ = 2,2,8
    IX,IY,IZ = np.eye(dX),np.eye(dY),np.eye(dZ)
    T1 = np.einsum("ij,yk,lm,np,yq,rs->ilnrjkmpqs", 
                   IX,IY,IZ,IX,IY,IZ).reshape((dX*dZ)**2, (dX*dY*dZ)**2)
\end{lstlisting}

In the case of the dissociation of proton $\mathrm{H_{B}}$, the corresponding chemical reaction is given by:
\begin{equation*}
[^{15}\mathrm{NH_{A}H_{B}H_{C}H_{D}}]^{+} \longrightarrow \mathrm{^{15}NH_{A}H_{C}H_{D}} +\mathrm{H_{B}}^{+}
\end{equation*}
In this reaction, the molecules X, Y, and Z are assigned as $^{15}\mathrm{N}\mathrm{H_{A}}$, $\mathrm{H_{B}}$, $\mathrm{H_{C}H_{D}}$, respectively. The dimensions of their respective Hilbert spaces are 4, 2, and 4, respectively. 

The mathematical form of the partial trace superoperator in Liouville space, denoted as T$_2$, is given by,
\begin{equation*}
\begin{aligned}
    \mathrm{T}_2 &= \mathbf{1}_2 \otimes \begin{pmatrix}
        1 & 0
    \end{pmatrix} \otimes \mathbf{1}_2 \otimes \mathbf{1}_2 \otimes \begin{pmatrix}
        1 & 0
    \end{pmatrix} \otimes \mathbf{1}_2 \\
    &+ \mathbf{1}_2 \otimes \begin{pmatrix}
        0 & 1
    \end{pmatrix} \otimes \mathbf{1}_2 \otimes \mathbf{1}_2 \otimes \begin{pmatrix}
        0 & 1
    \end{pmatrix} \otimes \mathbf{1}_2,
\end{aligned}
\end{equation*}
Where $\mathbf{1}_2 = \mathbf{1}_1 \otimes \mathbf{1}_1$ represents the $4\times 4$ identity matrix. The matrix T$_2$ can be calculated using the following code snippet:
\begin{lstlisting}
    dX,dY,dZ = 4,2,4
    IX,IY,IZ = np.eye(dX),np.eye(dY),np.eye(dZ)
    T2 = np.einsum("ij,yk,lm,np,yq,rs->ilnrjkmpqs", 
                   IX,IY,IZ,IX,IY,IZ).reshape((dX*dZ)**2, (dX*dY*dZ)**2)
\end{lstlisting}

In the case of the dissociation of proton $\mathrm{H_{C}}$, the corresponding chemical reaction takes the form:
\begin{equation*}
[^{15}\mathrm{NH_{A}H_{B}H_{C}H_{D}}]^{+} \longrightarrow \mathrm{^{15}NH_{A}H_{B}H_{D}} +\mathrm{H_{C}}^{+}
\end{equation*}
In this reaction, the molecules X, Y, and Z are assigned as $^{15}\mathrm{N}\mathrm{H_{A}H_{B}}$, $\mathrm{H_{C}}$, $\mathrm{H_{D}}$, respectively. The dimensions of their respective Hilbert spaces are 8, 2, and 2, respectively.

The mathematical form of the partial trace superoperator, denoted as T$_3$, is given by:
\begin{equation*}
\begin{aligned}
    \mathrm{T}_3 &= \mathbf{1}_3 \otimes \begin{pmatrix}
        1 & 0
    \end{pmatrix} \otimes \mathbf{1}_1 \otimes \mathbf{1}_3 \otimes \begin{pmatrix}
        1 & 0
    \end{pmatrix} \otimes \mathbf{1}_1 \\
    &+ \mathbf{1}_3 \otimes \begin{pmatrix}
        0 & 1
    \end{pmatrix} \otimes \mathbf{1}_1 \otimes \mathbf{1}_3 \otimes \begin{pmatrix}
        0 & 1
    \end{pmatrix} \otimes \mathbf{1}_1.
\end{aligned}
\end{equation*}
The matrix T$_3$ can be calculated using the code snippet:
\begin{lstlisting}
    dX,dY,dZ = 8,2,2
    IX,IY,IZ = np.eye(dX),np.eye(dY),np.eye(dZ)
    T3 = np.einsum("ij,yk,lm,np,yq,rs->ilnrjkmpqs", 
                   IX,IY,IZ,IX,IY,IZ).reshape((dX*dZ)**2, (dX*dY*dZ)**2)
\end{lstlisting}

Lastly, in the case of the dissociation of proton $\mathrm{H_{D}}$, the following reaction is assumed:
\begin{equation*}
[^{15}\mathrm{NH_{A}H_{B}H_{C}H_{D}}]^{+} \longrightarrow \mathrm{^{15}NH_{A}H_{B}H_{C}} +\mathrm{H_{D}}^{+}
\end{equation*}
Conversely, this reaction corresponds to Case 1, where the molecules X, and Y are assigned as $^{15}\mathrm{N}\mathrm{H_{A}H_{B}H_{C}}$ and $\mathrm{H_{D}}$, respectively. The dimensions of their respective Hilbert spaces are 16 and 2, respectively. 

The mathematical form of the partial trace superoperator, denoted as T$_4$, is given by (according to Eq.~\eqref{Eq: Tr1_L}):
\begin{equation*}
\begin{aligned}
    \mathrm{T}_4 &= \mathbf{1}_4 \otimes \begin{pmatrix}
        1 & 0
    \end{pmatrix} \otimes  \mathbf{1}_4 \otimes \begin{pmatrix}
        1 & 0
    \end{pmatrix} \\
    &+ \mathbf{1}_4 \otimes \begin{pmatrix}
        0 & 1
    \end{pmatrix} \otimes \mathbf{1}_4 \otimes \begin{pmatrix}
        0 & 1
    \end{pmatrix}.
\end{aligned}
\end{equation*}
Where $\mathbf{1}_4 = \mathbf{1}_1 \otimes \mathbf{1}_1 \otimes \mathbf{1}_1 \otimes \mathbf{1}_1$ represents the $16\times 16$ identity matrix. The matrix T$_4$ can be calculated using the code snippet:
\begin{lstlisting}
    dX,dY = 16,2
    IX,IY = np.eye(dX),np.eye(dY)
    T4 = np.einsum("ij,yk,lm,yn->iljkmn",IX,IY,IX,IY).reshape(dX**2, (dX*dY)**2)
\end{lstlisting}

\subsection{Matrix representations of Kronecker product superoperators}\label{Section: Kronecker product Superoperator}
Consider a reaction where molecules X and Y associate, resulting in the formation of a molecule XY. In this context, the density operator of the associated molecule XY, denoted as $\hat{\rho}_{\mathrm{XY}}$, is expressed as:
\begin{equation*}
    \hat{\rho}_{\mathrm{XY}} = \hat{\rho}_{\mathrm{X}} \otimes \hat{\rho}_{\mathrm{Y}}.
\end{equation*}
Here, $\hat{\rho}_{\mathrm{X}}$ and $\hat{\rho}_{\mathrm{Y}}$ represent the density operators for the association-consumed molecules X and Y, respectively.  Their ordering in the Kronecker product follows the arrangement of X and Y as indicated by the molecule notation ``XY".

In fact, the Kronecker product $\hat{\rho}_{\mathrm{X}} \otimes \hat{\rho}_{\mathrm{Y}}$ can be expressed as,
\begin{equation*}
\begin{aligned}
    \hat{\rho}_{\mathrm{X}} \otimes \hat{\rho}_{\mathrm{Y}} = \sum_{yy^{'}} \rho_{yy^{'}} \left( \hat{\textbf{\textit{1}}\,}_{\!\mathrm{X}} \otimes |y \rangle \right)\hat{\rho}_{\mathrm{X}} \left( \hat{\textbf{\textit{1}}\,}_{\!\mathrm{X}} \otimes \langle y^{'} | \right).
\end{aligned}
\end{equation*}
Here, $|y\rangle$ forms a basis for molecule Y. The operator $\hat{\textbf{\textit{1}}\,}_{\!\mathrm{X}}$ represents the identity operator acting on the Hilbert space associated with molecule X.

To prove the equivalency, we expand $\hat{\rho}_{\mathrm{X}}=\sum_{x,x^{'}} \rho_{x,x^{'}}|x\rangle \langle x^{'}|$,
\begin{equation*}
    \begin{aligned}
        &\quad\sum_{yy^{'}} \rho_{yy^{'}} \left( \hat{\textbf{\textit{1}}\,}_{\!\mathrm{X}} \otimes |y \rangle \right)\hat{\rho}_{\mathrm{X}} \left( \hat{\textbf{\textit{1}}\,}_{\!\mathrm{X}} \otimes \langle y^{'} | \right)\\
        &=\sum_{xx^{'}yy^{'}} \rho_{xx^{'}} \rho_{yy^{'}} \left( \hat{\textbf{\textit{1}}\,}_{\!\mathrm{X}} \otimes |y \rangle \right) |x\rangle \langle x^{'}| 
        \left( \hat{\textbf{\textit{1}}\,}_{\!\mathrm{X}}  \otimes \langle y^{'} | \right).
    \end{aligned}
\end{equation*}
Since $|x\rangle\langle x^{'}|=(|x\rangle\langle x^{'}|)\otimes 1$, we use the mixed-product property:
\begin{equation*}
    \left( \hat{\textbf{\textit{1}}\,}_{\!\mathrm{X}} \otimes |y \rangle \right) |x\rangle\langle x^{'}| \left( \hat{\textbf{\textit{1}}\,}_{\!\mathrm{X}} \otimes \langle y^{'} | \right) = \left(\hat{\textbf{\textit{1}}\,}_{\!\mathrm{X}} |x\rangle\langle x^{'}| \hat{\textbf{\textit{1}}\,}_{\!\mathrm{X}}   \right) \otimes \left( |y\rangle\langle y^{'}| \right) = |x\rangle\langle x^{'}| \otimes  |y\rangle\langle y^{'}|.
\end{equation*}
Insert it into the summation,
\begin{equation*}
    \begin{aligned}
        &\quad\sum_{xx^{'}yy^{'}} \rho_{xx^{'}} \rho_{yy^{'}} \left( \hat{\textbf{\textit{1}}\,}_{\!\mathrm{X}} \otimes |y \rangle \right) |x\rangle\langle x^{'}| \left( \hat{\textbf{\textit{1}}\,}_{\!\mathrm{X}} \otimes \langle y | \right) \\
        &= \sum_{xx^{'}yy^{'}} \rho_{xx^{'}} \rho_{yy^{'}} |x\rangle\langle x^{'}| \otimes  |y\rangle\langle y^{'}| \\
        &= \left(\sum_{xx^{'}}  \rho_{xx^{'}} |x\rangle\langle x^{'}|\right) \otimes \left( \sum_{yy^{'}}  \rho_{yy^{'}} |y\rangle\langle y^{'}| \right)\\
        &= \hat{\rho}_{\mathrm{X}} \otimes \hat{\rho}_{\mathrm{Y}},
    \end{aligned}
\end{equation*}
which verifies the validity of the expansion. Indeed, the Kronecker product can be understood as a result of the linear transformation acting on $\hat{\rho}_{\mathrm{X}}$. We denote this linear transformation by the Kronecker product superoperator, $\doublehat{D}_{\mathrm{Y}}^{\mathrm{(XY)}}$, whose effect is given by:
\begin{equation}
    \doublehat{D}_{\mathrm{Y}}^{\mathrm{(XY)}} \hat{\rho}_{\mathrm{X}} = \sum_{yy^{'}} \rho_{yy^{'}} \left( \hat{\textbf{\textit{1}}\,}_{\!\mathrm{X}} \otimes |y \rangle \right)\hat{\rho}_{\mathrm{X}} \left( \hat{\textbf{\textit{1}}\,}_{\!\mathrm{X}} \otimes \langle y^{'} | \right).
\label{Eq: Kron1_H}
\end{equation}

The form of $\doublehat{D}_{\mathrm{Y}}^{\mathrm{(XY)}}$ in Liouville space can be obtained using the mapping law introduced in Section~\ref{Section: Operator mapping}:
\begin{equation}
    \doublehat{D}_{\mathrm Y}^{\mathrm {(XY)}}=\sum_{yy^{'}} \rho_{yy^{'}} \left(\hat{\textbf{\textit{1}}\,}_{\!\mathrm{X}}\otimes |y\rangle\right)\otimes \left(\hat{\textbf{\textit{1}}\,}_{\!\mathrm{X}}\otimes |y^{'}\rangle^{*}\right).
\label{Eq: Kron1_L}
\end{equation}
Here, $|y^{'}\rangle^{*}  = \langle y^{'}|^{\mathrm{T}}$.

In association reactions, it is essential to take into account the relative arrangement of the Hilbert spaces associated with individual spins when constructing the Hilbert space for the resulting associated molecule. The specific ordering of the Hilbert spaces associated with different molecules affects the forms of the Kronecker product superoperators. Here, we provide a summary of the different forms of the Kronecker product superoperators for a few common cases.\newline

\noindent\textbf{Case 1: \underline{\textbf{X}} + \textbf{Y} $\longrightarrow$ \textbf{XY}}\newline
Case 1, we consider the reaction:
\begin{equation*}
    \rm \underline{X} + Y \longrightarrow XY 
\end{equation*}
where molecule X transforms into molecule XY by associating with molecule Y.

The density operator of the associated molecule $\rm XY$ can be obtained by applying the Kronecker product superoperator, $ \doublehat{D}_{\mathrm{Y}}^{\mathrm{(XY)}}$, to the density operator of molecule X.

In this case, the forms of $ \doublehat{D}_{\mathrm{Y}}^{\mathrm{(XY)}}$ in Hilbert space and in Liouville space are given by Eq.~\eqref{Eq: Kron1_H} and Eq.~\eqref{Eq: Kron1_L}, respectively. \newline

\noindent\textbf{Case 2: \textbf{X} + \underline{\textbf{Y}}   $\longrightarrow$  \textbf{XY}}\newline
In Case 2, we consider the reaction:
\begin{equation*}
    \rm X  + \underline{Y} \longrightarrow XY 
\end{equation*}
where molecule Y transforms into molecule XY by associating with molecule X.
 
The density operator of the the associated molecule $\rm XY$, denoted as $\hat{\rho}_{\mathrm{XY}}$ is calculated by
applying the Kronecker product superoperator, $ \doublehat{D}_{\mathrm{X}}^{\mathrm{(XY)}}$, to the density operator of molecule Y:
\begin{equation*}
    \hat{\rho}_{\mathrm{XY}} = \hat{\rho}_{\mathrm{X}} \otimes \hat{\rho}_{\mathrm{Y}} = \doublehat{D}_{\mathrm{X}}^{\mathrm{(XY)}} \hat{\rho}_{\mathrm{Y}}.
\end{equation*}

The action of $ \doublehat{D}_{\mathrm{X}}^{\mathrm{(XY)}}$ on $ \hat{\rho}_{\mathrm{Y}}$ can be expanded as,
\begin{equation}
    \doublehat{D}_{\mathrm{X}}^{\mathrm{(XY)}} \hat{\rho}_{\mathrm{Y}} = \sum_{xx^{'}}\rho_{xx^{'}} \left(|x\rangle\otimes \hat{\textbf{\textit{1}}\,}_{\!\mathrm Y}\right) \hat{\rho}_{\mathrm{Y}}  \left(\langle x^{'}|\otimes \hat{\textbf{\textit{1}}\,}_{\!\mathrm Y}\right).
\label{Eq: Kron2_H}
\end{equation}
Here, $|x\rangle$ forms a basis for molecule X. The operator $\hat{\textbf{\textit{1}}\,}_{\mathrm Y}$ represents the identity operator acting on the Hilbert space associated with molecule Y.

In Liouville space, $ \doublehat{D}_{\mathrm{X}}^{\mathrm{(XY)}}$ takes the form:
\begin{equation}
    \doublehat{D}_{\mathrm X}^{\mathrm {(XY)}}=\sum_{xx^{'}}\rho_{xx^{'}} \left(|x\rangle\otimes \hat{\textbf{\textit{1}}\,}_{\!\mathrm Y}\right) \otimes \left(|x^{'}\rangle^{*}\otimes \hat{\textbf{\textit{1}}\,}_{\!\mathrm Y}\right).
\label{Eq: Kron2_L}
\end{equation}

\noindent\textbf{Case 3: \underline{\textbf{XZ}}+\textbf{Y} $\longrightarrow$ \textbf{XYZ}}\newline
In Case 3, we consider the reaction:
\begin{equation*}
    \rm  \underline{XZ} + Y \longrightarrow XYZ 
\end{equation*}
where molecule XZ transforms into molecule XYZ by associating with molecule Y.

The density operator of the the associated molecule $\rm XYZ$, denoted as $\hat{\rho}_{\mathrm{XYZ}}$, is derived by
applying the Kronecker product superoperator, $ \doublehat{D}_{\mathrm Y}^{\mathrm {(XYZ)}}$, to the density operator of molecule XZ:
\begin{equation*}
    \hat{\rho}_{\mathrm{XYZ}}  = \doublehat{D}_{\mathrm{Y}}^{\mathrm{(XYZ)}} \hat{\rho}_{\mathrm{XZ}}.
\end{equation*}

The action of $ \doublehat{D}_{\mathrm Y}^{\mathrm {(XYZ)}}$ on a density operator for molecule XZ can be expanded as,
\begin{equation}
    \doublehat{D}_{\mathrm Y}^{\mathrm {(XYZ)}}=\sum_{yy^{'}} \rho_{yy^{'}} \left(\hat{\textbf{\textit{1}}\,}_{\!\mathrm{X}}\otimes |y\rangle \otimes \hat{\textbf{\textit{1}}\,}_{\!\mathrm Z} \right) \hat{\rho}_{\mathrm{XZ}} \left(\hat{\textbf{\textit{1}}\,}_{\!\mathrm{X}}\otimes \langle y^{'}| \otimes \hat{\textbf{\textit{1}}\,}_{\!\mathrm Z} \right).
\label{Eq: Kron3_H}
\end{equation}
Here, $|y\rangle$ forms a basis for molecule Y. The operators $\hat{\textbf{\textit{1}}\,}_{\!\mathrm{X}}$ and  $\hat{\textbf{\textit{1}}\,}_{\!\mathrm Z}$ represent the identity operators acting on the Hilbert spaces associated with molecule X and Z, respectively.

For instance, the action of $ \doublehat{D}_{\mathrm Y}^{\mathrm {(XYZ)}}$ on a separable density operator for molecule XZ, $ \hat{\rho}_{\mathrm{X}}\otimes\hat{\rho}_{\mathrm{Z}}$, is straightforward and is given by,
\begin{equation*}
    \doublehat{D}_{\mathrm Y}^{\mathrm {(XYZ)}} (\hat{\rho}_{\mathrm{X}}\otimes\hat{\rho}_{\mathrm{Z}}) = \hat{\rho}_{\mathrm{X}} \otimes \hat{\rho}_{\mathrm{Y}} \otimes \hat{\rho}_{\mathrm{Z}}.
\end{equation*}

In Liouville space, $ \doublehat{D}_{\mathrm Y}^{\mathrm {(XYZ)}}$ takes the form:
\begin{equation}
    \doublehat{D}_{\mathrm Y}^{\mathrm {(XYZ)}}=\sum_{yy^{'}} \rho_{yy^{'}} \left(\hat{\textbf{\textit{1}}\,}_{\!\mathrm{X}}\otimes |y\rangle \otimes \hat{\textbf{\textit{1}}\,}_{\!\mathrm Z} \right)\otimes \left(\hat{\textbf{\textit{1}}\,}_{\!\mathrm{X}}\otimes |y^{'}\rangle^{*} \otimes \hat{\textbf{\textit{1}}\,}_{\!\mathrm Z} \right).
\label{Eq: Kron3_L}
\end{equation}

\noindent\textbf{Case 4: \underline{\textbf{XYZ}}+\textbf{UV} $\longrightarrow$ \textbf{XUYVZ}}\newline
In Case 4, we consider the reaction:
\begin{equation*}
    \rm  \underline{XYZ} + UV \longrightarrow XUYVZ 
\end{equation*}
where molecule XYZ transforms into molecule XUYVZ by associating with molecule UV. The hydrogenation reaction in parahydrogen-induced polarization (PHIP) experiments is indeed of the same kind as the reaction described in Case 4.

The density operator of the the associated molecule $\rm XUYVZ$, denoted as $\hat{\rho}_{\mathrm{XUYVZ}}$ is calculated by applying the Kronecker product superoperator, $ \doublehat{D}_{\mathrm{UV}}^{\mathrm{(XUYVZ)}}$, to the density operators for molecule XYZ:
\begin{equation*}
    \hat{\rho}_{\mathrm{XUYVZ}}  = \doublehat{D}_{\mathrm{UV}}^{\mathrm{(XUYVZ)}} \hat{\rho}_{\mathrm{XYZ}}.
\end{equation*}
% Similarly, we represent unconventional Kronecker product as the action of the Kronecker product superoperator, denoted as , on the density operator $\hat{\rho}_{\mathrm{XYZ}}$.

Generally, the action of $ \doublehat{D}_{\mathrm{UV}}^{\mathrm {(XUYVZ)}}$ on $ \hat{\rho}_{\mathrm{XYZ}}$ is expanded as,
\begin{equation}
    \hat{\rho}_{\mathrm{XUYVZ}} = \sum_{uu^{'}vv^{'}}\rho_{uu^{'}vv^{'}} \left(\hat{\textbf{\textit{1}}\,}_{\!\mathrm{X}} \otimes |u\rangle \otimes \hat{\textbf{\textit{1}}\,}_{\!\mathrm Y} \otimes |v\rangle \otimes \hat{\textbf{\textit{1}}\,}_{\!\mathrm Z} \right) \hat{\rho}_{\mathrm{XYZ}}  \left(\hat{\textbf{\textit{1}}\,}_{\!\mathrm{X}} \otimes \langle u^{'}| \otimes \hat{\textbf{\textit{1}}\,}_{\!\mathrm Y} \otimes \langle v^{'}| \otimes \hat{\textbf{\textit{1}}\,}_{\!\mathrm Z} \right)
\label{Eq: Kron4_H}
\end{equation}
Given the density operator for the molecule UV expressed as $\hat{\rho}_{\mathrm{UV}}=\sum_{uu^{'}vv^{'}}\rho_{uu^{'}vv^{'}} |u\rangle\langle u^{'}|\otimes |v\rangle \langle v^{'}|$. The state sets $\{|u\rangle \}$ and  $\{|v\rangle \}$ form a basis for molecule fragments U and V, respectively. The operators $\hat{\textbf{\textit{1}}\,}_{\!\mathrm{X}}$,  $\hat{\textbf{\textit{1}}\,}_{\!\mathrm Y}$ and $\hat{\textbf{\textit{1}}\,}_{\!\mathrm Z}$ represent the identity operators acting on the Hilbert spaces associated with molecule X, Y, and Z, respectively.

For instance, the action of $ \doublehat{D}_{\mathrm UV}^{\mathrm {(XUYVZ)}}$ on a separable density operator for molecule XYZ, $ \hat{\rho}_{\mathrm{X}}\otimes\hat{\rho}_{\mathrm{Y}}\otimes\hat{\rho}_{\mathrm{Z}}$, is straightforwardly expressed as:
\begin{equation*}
    \doublehat{D}_{\mathrm{UV}}^{\mathrm {(XUYVZ)}} (\hat{\rho}_{\mathrm{X}}\otimes\hat{\rho}_{\mathrm{Y}}\otimes\hat{\rho}_{\mathrm{Z}}) = \hat{\rho}_{\mathrm{X}} \otimes\hat{\rho}_{\mathrm{U}} \otimes  \hat{\rho}_{\mathrm{Y}} \otimes \hat{\rho}_{\mathrm{V}}  \otimes \hat{\rho}_{\mathrm{Z}}.
\end{equation*}

In Liouville space, $ \doublehat{D}_{\mathrm UV}^{\mathrm {(XUYVZ)}}$ takes the form:
\begin{equation}
    \doublehat{D}_{\mathrm{UV}}^{\mathrm {(XUYVZ)}}= \sum_{uu^{'}vv^{'}}\rho_{uu^{'}vv^{'}} \left(\hat{\textbf{\textit{1}}\,}_{\!\mathrm{X}} \otimes |u\rangle \otimes \hat{\textbf{\textit{1}}\,}_{\!\mathrm Y} \otimes |v\rangle \otimes \hat{\textbf{\textit{1}}\,}_{\!\mathrm Z} \right) \otimes \left(\hat{\textbf{\textit{1}}\,}_{\!\mathrm{X}} \otimes |u^{'}\rangle^{*} \otimes \hat{\textbf{\textit{1}}\,}_{\!\mathrm Y} \otimes |v^{'}\rangle^{*} \otimes \hat{\textbf{\textit{1}}\,}_{\!\mathrm Z} \right).
\label{Eq: Kron4_L}
\end{equation}

\heading{Calculation}
The Zeeman product basis would be adopted for the matrix representations. Furthermore, we assume the relative ordering of the Hilbert spaces associated with spins from the same molecules remain unaltered within the Hilbert space for the associated molecule. In other words, if we have multiple spins from the same molecule, we maintain their original ordering within the composite Hilbert space for the associated molecule.

In the defined Liouville spaces, the matrix representations of the Kronecker product superoperators are calculate based on Eq.~\eqref{Eq: Kron1_L}, Eq.~\eqref{Eq: Kron2_L}, Eq.~\eqref{Eq: Kron3_L}, and Eq.~\eqref{Eq: Kron4_L}. The identity operators are represented with identity matrices. Moreover, each ket is represented with the appropriate column vector from the identity matrix. For instance, the basis ket $|y\rangle$ is represented by the column vector corresponding to the $y$-th column of the identity matrix with the appropriate size.\newline

\noindent\textbf{Case 1: \underline{\textbf{X}} + \textbf{Y} $\longrightarrow$ \textbf{XY}}
The elements of Eq.~\eqref{Eq: Kron1_L} represented in Zeeman product basis is given by,
\begin{equation*}
\begin{aligned}
    \left(\mathrm{D}_{\mathrm Y}^{\mathrm {(XY)}}\right)_{ikln,jm} &=\sum_{yy^{'}}  (\textbf{1}_{\mathrm{X}})_{ij}\, (|y\rangle)_{k} \, (\textbf{1}_{\mathrm{X}})_{lm}\, (|y^{'}\rangle)_{n}\, \rho_{yy^{'}}\\
    &=\sum_{yy^{'}} (\textbf{1}_{\mathrm{X}})_{ij}\, (\textbf{1}_{\mathrm{Y}})_{ky} \, (\textbf{1}_{\mathrm{X}})_{lm}\, (\textbf{1}_{\mathrm{Y}})_{ny^{'}} \,\rho_{yy^{'}} .
\end{aligned}
\end{equation*}

Consequently, it is calculated as,
\begin{lstlisting}
    #Input: The dimensions of molecule X and Y; 
    #       The density matrix of molecule Y
    dX,dY,rho = d_X,d_Y,rho_Y  
    IX,IY = np.eye(dX),np.eye(dY)
    Kron = np.einsum("ij,ky,lm,nY,yY->iklnjm",
                     IX,IY,IX,IY,rho).reshape((dX*dY)**2,dX**2)
\end{lstlisting}
Here, we have changed the summation index from $y^{'}$ to $Y$.\newline

\noindent\textbf{Case 2: \textbf{X} + \underline{\textbf{Y}}   $\longrightarrow$  \textbf{XY}}
The elements of Eq.~\eqref{Eq: Kron2_L} represented in Zeeman product basis is given by:
\begin{equation*}
\begin{aligned}
    \left(\mathrm{D}_{\mathrm X}^{\mathrm {(XY)}}\right)_{ijlm,kn} & =\sum_{x,x^{'}} (|x\rangle)_{i} \, (\textbf{1}_{\mathrm{Y}})_{jk}\, (|x^{'}\rangle)_{l} \, (\textbf{1}_{\mathrm{Y}})_{mn}  \,\rho_{xx^{'}}\\
    & =\sum_{x,x^{'}} (\textbf{1}_{\mathrm{X}})_{ix} \, (\textbf{1}_{\mathrm{Y}})_{jk}\, (\textbf{1}_{\mathrm{X}})_{lx^{'}} \, (\textbf{1}_{\mathrm{Y}})_{mn}\, \rho_{xx^{'}} 
\end{aligned}
\end{equation*}

Consequently, it is calculated as,
\begin{lstlisting}
    #Input: The dimensions of molecule X and Y;
    #       The density matrix of molecule X.
    dY,dY,rho = d_X,d_Y,rho_X
    IX,IY = np.eye(%*\textcolor{cyan}{dX}*)),np.eye(%*\textcolor{cyan}{dY}*))
    Kron = np.einsum("ix,jk,lX,mn,xX->ijlmkn",
                     IX,IY,IX,IY,rho).reshape((dX*dY)**2,dY**2)
\end{lstlisting}
Here, we change the summation index from $x^{'}$ to $X$.\newline

\noindent\textbf{Case 3: \underline{\textbf{XZ}}+\textbf{Y} $\longrightarrow$ \textbf{XYZ}}
The elements of Eq.~\eqref{Eq: Kron3_L} represented in Zeeman product basis is given by:
\begin{equation*}
\begin{aligned}
    \left(\mathrm{D}_{\mathrm Y}^{\mathrm {(XYZ)}}\right)_{iklnqr,jmps} &= \sum_{y,y^{'}}  (\textbf{1}_{\mathrm{X}})_{ij}\, (| y \rangle)_{k}\, (\textbf{1}_{\mathrm{Z}})_{lm}\, (\textbf{1}_{\mathrm{X}})_{np}\, (| y^{'} \rangle)_{q}\, (\textbf{1}_{\mathrm{Z}})_{rs} \,\rho_{yy^{'}} \\
    & = \sum_{y,y^{'}}  (\textbf{1}_{\mathrm{X}})_{ij}\, (\textbf{1}_{\mathrm{Y}})_{ky}\, (\textbf{1}_{\mathrm{Z}})_{lm}\, (\textbf{1}_{\mathrm{X}})_{np}\, (\textbf{1}_{\mathrm{Y}})_{qy^{'}}\, (\textbf{1}_{\mathrm{Z}})_{rs} \,\rho_{yy^{'}}
\end{aligned}
\end{equation*}

Consequently, it is calculated with the code snippet below:
\begin{lstlisting}
    #Input: The dimensions of molecule X, Y and Z; 
    #       The density matrix of molecule Y.
    dY,dY,dZ,rho = d_X,d_Y,d_Z,rho_Y
    IX,IY,IZ =  np.eye(dX),np.eye(dY),np.eye(dZ)
    Kron = np.einsum("ij,ky,lm,np,qY,rs,yY->iklnqrjmps",
                     IX,IY,IZ,IX,IY,IZ,rho).reshape((dX*dY*dZ)**2,(dX*dZ)**2)
\end{lstlisting}
Here, we change the summation index from $y^{'}$ to $Y$.\newline

\noindent\textbf{Case 4: \underline{\textbf{XYZ}}+\textbf{UV} $\longrightarrow$ \textbf{XUYVZ}}
The elements of Eq.~\eqref{Eq: Kron4_L} represented in Zeeman product basis is given by:
\begin{equation*}
\begin{aligned}
    &\quad\left(\mathrm{D}_{\mathrm{UV}}^{\mathrm {(XUYVZ)}} \right)_{acdfgiklnp,behjmq} \\
    &= \sum_{uu^{'}vv^{'}} (\textbf{1}_{\mathrm{X}})_{ab}\, (| u \rangle)_{c}\, (\textbf{1}_{\mathrm{Y}})_{de}\,(| v \rangle)_{f}\, (\textbf{1}_{\mathrm{Z}})_{gh} \, (\textbf{1}_{\mathrm{X}})_{ij}\, (| u^{'} \rangle)_{k}\, (\textbf{1}_{\mathrm{Y}})_{lm}\,(| v^{'} \rangle)_{n}\, (\textbf{1}_{\mathrm{Z}})_{pq}  \,\rho_{uu^{'}vv^{'}} \\
    & = \sum_{uu^{'}vv^{'}}  (\textbf{1}_{\mathrm{X}})_{ab}\, (\textbf{1}_{\mathrm{U}})_{cu}\, (\textbf{1}_{\mathrm{Y}})_{de}\,(\textbf{1}_{\mathrm{V}})_{fv}\, (\textbf{1}_{\mathrm{Z}})_{gh} \, (\textbf{1}_{\mathrm{X}})_{ij}\, (\textbf{1}_{\mathrm{U}})_{ku^{'}}\, (\textbf{1}_{\mathrm{Y}})_{lm}\,(\textbf{1}_{\mathrm{V}})_{nv^{'}}\, (\textbf{1}_{\mathrm{Z}})_{pq} \,\rho_{uu^{'}vv^{'}}
\end{aligned}
\end{equation*}

Consequently, it is calculated as (Substitute $u^{'},v^{'}$ with $U,V$, respectively):
\begin{lstlisting}
    #Input: The dimensions of molecule X, Y, Z, U, and V;
    #       The density matrix of molecule UV.
    dX,dY,dZ,dU,dV,rho = d_X,d_Y,d_Z,d_U,d_V,rho_UV
    rho = rho.reshape((dU,dV,dU,dV))).transpose((0,2,1,3))
    IX,IY,IZ,IU,IV =np.eye(dX),np.eye(dY),np.eye(dZ),np.eye(dU),np.eye(dV)
    Kron = np.einsum("ab,cu,de,fv,gh,ij,kU,lm,nV,pq,uUvV->acdfgiklnpbehjmq",
                     IX,IU,IY,IV,IZ,IX,IU,IY,IV,IZ,
                     rho_uUvV).reshape((dX*dY*dZ*dU*dV)**2,(dX*dY*dZ)**2)
\end{lstlisting}
In this scenario, the input density matrix is a two-dimensional matrix that stores the elements $\rho_{(uv)(u^{'}v^{'})}$. The row and column indices are represented by 'uv' and '$u^{'}v^{'}$, respectively. To obtain the desired 4-D matrix $\rho_{uu^{'}vv^{'}}$, we implement two steps. First, we reshape the input matrix using the function \lstinline{reshape((dU,dV,dU,dV)))} to obtain $\rho_{uvu^{'}v^{'}}$. Second, we swap the second and third dimensions using the function \lstinline{transpose((0,2,1,3))} to derive the required matrix $\rho_{uu^{'}vv^{'}}$.\newline

\heading{Example}
Let us discuss association of protons to ammonia as an example. Similarly, we need to consider four different cases where the newly added proton can enter ammonium as proton $\mathrm{H_{A}}$, $\mathrm{H_{B}}$, $\mathrm{H_{C}}$, and $\mathrm{H_{D}}$, respectively.

In the context where the newly added proton enters as proton $\mathrm{H_{A}}$, the reaction takes the form:
\begin{equation*}
    \mathrm{^{15}NH_{B}H_{C}H_{D}} +\mathrm{H_{A}}^{+}  \longrightarrow [^{15}\mathrm{NH_{A}H_{B}H_{C}H_{D}}]^{+}
\end{equation*}

The reaction is of the same form as introduced in Case 3, where the molecule fragments or entities X, Y, and Z are assigned as $\mathrm{^{15}}$, $\mathrm{H_{D}}$, and $\mathrm{H_{A}H_{B}H_{C}}$, respectively. The operator form of the of the Kronecker product superoperator is given by Eq.~\eqref{Eq: Kron3_L}. According to the equation, the matrix representation of the Kronecker product superoperator, denoted as $\mathrm{Kron_1}$, within the given basis is given by,
\begin{equation*}
\begin{aligned}
    \mathrm{Kron}_1 &= \mathbf{1}_1 \otimes \begin{pmatrix}
        1 \\ 0
    \end{pmatrix} \otimes \mathbf{1}_3 \otimes \mathbf{1}_1 \otimes \begin{pmatrix}
        1 \\ 0
    \end{pmatrix} \otimes \mathbf{1}_3 \, \rho_{\alpha\alpha}\\
    &+ \mathbf{1}_1 \otimes \begin{pmatrix}
        1 \\ 0
    \end{pmatrix} \otimes \mathbf{1}_3 \otimes \mathbf{1}_1 \otimes \begin{pmatrix}
        0 \\ 1
    \end{pmatrix} \otimes \mathbf{1}_3  \, \rho_{\alpha\beta}\\
    &+ \mathbf{1}_1 \otimes \begin{pmatrix}
        0 \\ 1
    \end{pmatrix} \otimes \mathbf{1}_3 \otimes \mathbf{1}_1 \otimes \begin{pmatrix}
        1 \\ 0
    \end{pmatrix} \otimes \mathbf{1}_3  \, \rho_{\beta\alpha}\\
    &+ \mathbf{1}_1 \otimes \begin{pmatrix}
        0 \\ 1
    \end{pmatrix} \otimes \mathbf{1}_3 \otimes \mathbf{1}_1 \otimes \begin{pmatrix}
        0 \\ 1
    \end{pmatrix} \otimes \mathbf{1}_3 \, \rho_{\beta\beta}.
\end{aligned}
\end{equation*}
Here, we have selected $|y\rangle$ to be either $|\alpha\rangle$ or $|\beta\rangle$. The density matrix for molecule Y in this bases is given by:
\begin{equation*}
    \rho_{\mathrm{Y}}= \begin{pmatrix}
        \rho_{\alpha\alpha} & \rho_{\alpha\beta} \\
        \rho_{\beta\alpha}  & \rho_{\beta\beta}
    \end{pmatrix}.
\end{equation*}

The matrix Kron$_1$ can be calculated as,
\begin{lstlisting}
    dY,dY,dZ,rho = 2,2,8,rho_Y
    IX,IY,IZ =  np.eye(dX),np.eye(dY),np.eye(dZ)
    Kron1 = np.einsum("ij,ky,lm,np,qY,rs,yY->iklnqrjmps",
                     IX,IY,IZ,IX,IY,IZ,rho).reshape((dX*dY*dZ)**2,(dX*dZ)**2)
\end{lstlisting}
Here, the variable \lstinline{rho_Y} represents the density matrix for molecule Y as shown above.\newline

Similarly, if the newly added proton enters as proton $\mathrm{H_{B}}$, the reaction takes the form:
\begin{equation*}
    \mathrm{^{15}NH_{A}H_{C}H_{D}} +\mathrm{H_{B}}^{+}  \longrightarrow [^{15}\mathrm{NH_{A}H_{B}H_{C}H_{D}}]^{+}
\end{equation*}

The reaction is also of the same form as introduced in Case 3, where the molecule fragments or entities X, Y, and Z are assigned as  $\mathrm{^{15}NH_{A}}$, $\mathrm{H_{D}}$, and $\mathrm{H_{B}H_{C}}$, respectively. The matrix form of the Kronecker product superoperator, denoted as $\mathrm{Kron_2}$, within the given basis is expressed as:
\begin{equation*}
\begin{aligned}
    \mathrm{Kron}_2 &= \mathbf{1}_2 \otimes \begin{pmatrix}
        1 \\ 0
    \end{pmatrix} \otimes \mathbf{1}_2 \otimes \mathbf{1}_2 \otimes \begin{pmatrix}
        1 \\ 0
    \end{pmatrix} \otimes \mathbf{1}_2 \, \rho_{\alpha\alpha}\\
    &+ \mathbf{1}_2 \otimes \begin{pmatrix}
        1 \\ 0
    \end{pmatrix} \otimes \mathbf{1}_2 \otimes \mathbf{1}_2 \otimes \begin{pmatrix}
        0 \\ 1
    \end{pmatrix} \otimes \mathbf{1}_2  \, \rho_{\alpha\beta}\\
    &+ \mathbf{1}_2 \otimes \begin{pmatrix}
        0 \\ 1
    \end{pmatrix} \otimes \mathbf{1}_2 \otimes \mathbf{1}_2 \otimes \begin{pmatrix}
        1 \\ 0
    \end{pmatrix} \otimes \mathbf{1}_2  \, \rho_{\beta\alpha}\\
    &+ \mathbf{1}_2 \otimes \begin{pmatrix}
        0 \\ 1
    \end{pmatrix} \otimes \mathbf{1}_2 \otimes \mathbf{1}_2 \otimes \begin{pmatrix}
        0 \\ 1
    \end{pmatrix} \otimes \mathbf{1}_2 \, \rho_{\beta\beta}\\
\end{aligned}
\end{equation*}

The matrix Kron$_2$ can be calculated using the code snippet:
\begin{lstlisting}
    dY,dY,dZ,rho = 4,2,4,rho_Y
    IX,IY,IZ =  np.eye(dX),np.eye(dY),np.eye(dZ)
    Kron2 = np.einsum("ij,ky,lm,np,qY,rs,yY->iklnqrjmps",
                     IX,IY,IZ,IX,IY,IZ,rho).reshape((dX*dY*dZ)**2,(dX*dZ)**2)
\end{lstlisting}

Moreover, if the newly added proton enters as proton $\mathrm{H_{C}}$, the reaction takes the form:
\begin{equation*}
    \mathrm{^{15}NH_{A}H_{B}H_{D}} +\mathrm{H_{C}}^{+}  \longrightarrow [^{15}\mathrm{NH_{A}H_{B}H_{C}H_{D}}]^{+}
\end{equation*}

The reaction is also of the same form as introduced in Case 3, where the molecule fragments or entities X, Y, and Z are assigned as  $\mathrm{^{15}NH_{A}H_{B}}$, $\mathrm{H_{D}}$, and $\mathrm{H_{C}}$, respectively. The matrix form of the Kronecker product superoperator, denoted as $\mathrm{Kron_3}$, within the given basis is expressed as:
\begin{equation*}
\begin{aligned}
    \mathrm{Kron}_3 &= \mathbf{1}_3 \otimes \begin{pmatrix}
        1 \\ 0
    \end{pmatrix} \otimes \mathbf{1}_1 \otimes \mathbf{1}_3 \otimes \begin{pmatrix}
        1 \\ 0
    \end{pmatrix} \otimes \mathbf{1}_1 \, \rho_{\alpha\alpha}\\
    &+ \mathbf{1}_3 \otimes \begin{pmatrix}
        1 \\ 0
    \end{pmatrix} \otimes \mathbf{1}_1 \otimes \mathbf{1}_3 \otimes \begin{pmatrix}
        0 \\ 1
    \end{pmatrix} \otimes \mathbf{1}_1  \, \rho_{\alpha\beta}\\
    &+ \mathbf{1}_3 \otimes \begin{pmatrix}
        0 \\ 1
    \end{pmatrix} \otimes \mathbf{1}_1 \otimes \mathbf{1}_3 \otimes \begin{pmatrix}
        1 \\ 0
    \end{pmatrix} \otimes \mathbf{1}_1  \, \rho_{\beta\alpha}\\
    &+ \mathbf{1}_3 \otimes \begin{pmatrix}
        0 \\ 1
    \end{pmatrix} \otimes \mathbf{1}_1 \otimes \mathbf{1}_3 \otimes \begin{pmatrix}
        0 \\ 1
    \end{pmatrix} \otimes \mathbf{1}_1 \, \rho_{\beta\beta}\\
\end{aligned}
\end{equation*}

The matrix Kron$_3$ can be calculated using the code snippet:
\begin{lstlisting}
    dY,dY,dZ,rho = 8,2,2,rho_Y
    IX,IY,IZ =  np.eye(dX),np.eye(dY),np.eye(dZ)
    Kron3 = np.einsum("ij,ky,lm,np,qY,rs,yY->iklnqrjmps",
                     IX,IY,IZ,IX,IY,IZ,rho).reshape((dX*dY*dZ)**2,(dX*dZ)**2)
\end{lstlisting}

Lastly, if the newly added proton enters as proton $\mathrm{H_{D}}$, the reaction takes the form:
\begin{equation*}
    \mathrm{^{15}NH_{A}H_{B}H_{C}} +\mathrm{H_{D}}^{+}  \longrightarrow [^{15}\mathrm{NH_{A}H_{B}H_{C}H_{D}}]^{+}
\end{equation*}

The reaction is of the same form as introduced in Case 1, where the molecule fragments or entities X, and Y are assigned as  $\mathrm{^{15}NH_{A}H_{B}H_{C}}$, and $\mathrm{H_{D}}$, respectively. The matrix form of the Kronecker product superoperator, denoted as $\mathrm{Kron_4}$, within the given basis is expressed as:
\begin{equation*}
\begin{aligned}
    \mathrm{Kron}_4 &= \mathbf{1}_4 \otimes \begin{pmatrix}
        1 \\ 0
    \end{pmatrix} \otimes \mathbf{1}_4 \otimes \begin{pmatrix}
        1 \\ 0
    \end{pmatrix}   \, \rho_{\alpha\alpha}\\
    &+ \mathbf{1}_4 \otimes \begin{pmatrix}
        1 \\ 0
    \end{pmatrix} \otimes \mathbf{1}_4 \otimes \begin{pmatrix}
        0 \\ 1
    \end{pmatrix}  \, \rho_{\alpha\beta}\\
    &+ \mathbf{1}_4 \otimes \begin{pmatrix}
        0 \\ 1
    \end{pmatrix} \otimes \mathbf{1}_4 \otimes \begin{pmatrix}
        1 \\ 0
    \end{pmatrix}  \, \rho_{\beta\alpha}\\
    &+ \mathbf{1}_4 \otimes \begin{pmatrix}
        0 \\ 1
    \end{pmatrix} \otimes \mathbf{1}_4 \otimes \begin{pmatrix}
        0 \\ 1
    \end{pmatrix} \, \rho_{\beta\beta}\\
\end{aligned}
\end{equation*}

The matrix Kron$_4$ can be calculated using the code snippet:
\begin{lstlisting}
    dX,dY,rho = 16,2,rho_Y  
    IX,IY = np.eye(dX),np.eye(dY)
    Kron = np.einsum("ij,ky,lm,nY,yY->iklnjm",
                     IX,IY,IX,IY,rho).reshape((dX*dY)**2,dX**2)
\end{lstlisting}

\subsection{Solving the LvN equation for chemical exchanging systems}\label{Section: Solution}
In this section, we discuss the procedure in solving Eq.~\eqref{Eq: LvN with chemical exchange}.  

\subsubsection{Calculate the coefficient matrix}\label{Section: generate coefficient matrix}
First, we calculate the coefficient matrix, which governs the dynamics of the system and takes the following form:
\begin{equation*}
    \begin{pmatrix}
    \mathrm{L}_{\rm A}-\tilde{k}_a \mathbf{1}_{2\mathrm A} & +k_d \mathrm{T}_{\mathrm B}^{\mathrm {(AB)}}\\
    +\tilde{k}_a\mathrm{D}_{\mathrm B}^{\mathrm {(AB)}} &  \mathrm{L}_{\mathrm{AB}}-k_d \mathbf{1}_{2\mathrm{AB}}\\
    \end{pmatrix}.
\end{equation*}
Here, the matrices $\mathbf{1}_{2\mathrm A}$ and $\mathbf{1}_{2\mathrm AB}$ are the matrix representations of $\doublehat{\textit{\textbf{1}}\,}_{\mathrm{A}}$ and $\doublehat{\textit{\textbf{1}}\,}_{\mathrm{AB}}$, respectively. The matrices $\mathrm{L}_{\rm A}$ and $\mathrm{L}_{\mathrm{AB}}$ denote the matrix representations of the Liouvillians for molecule A and AB, respectively. They are computed as:
\begin{lstlisting}
    L_A = -1j*HS_A + R_A
    L_AB = -1j*HS_AB + R_AB
\end{lstlisting}
Here, \lstinline{HS_A} and \lstinline{HS_AB} store the matrix representations of the Hamiltonian superoperators for molecule A and AB, respectively. Additionally, $\mathrm{R}_{\mathrm{A}}$ and $\mathrm{R}_{\mathrm{AB}}$ represent the matrix representations of the relaxation superoperators for molecule A and AB, respectively. Note here the dimensions of $\mathrm{L}_{\rm A}$ and $\mathrm{L}_{\mathrm{AB}}$ are $d^2_{\mathrm{A}}\times d^2_{\mathrm{A}}$ and $d^2_{\mathrm{AB}}\times d^2_{\mathrm{AB}}$, respectively.

To perform the calculation, one needs to determine the type of the chemical reaction and identify the dimensions of the involved molecule fragments or entities. These dimensions are then used as inputs in the corresponding code snippets. For the partial superoperator, the resulting matrix $\mathrm{T}_{\mathrm{B}}^{\mathrm {(AB)}}$ has dimensions of $d_{\mathrm{A}}^2 \times d_{\mathrm{AB}}^2$, and we store it as \lstinline{Tr}. On the other hand, for the Kronecker product superoperator, the matrix $\mathrm{D}_{\mathrm B}^{\mathrm {(AB)}}$ has dimensions of $d_{\mathrm{AB}}^2 \times d_{\mathrm{A}}^2$ and is stored as \lstinline{Kron}.

Given the calculated matrices, we generate the coefficient matrix of Eq.~\eqref{Eq: LvN with chemical exchange} (Stored as \lstinline{M}) with the code snippet below:
\begin{lstlisting}
    M = np.block([[L_A-ka*np.eye(d_A**2), kd*Tr],
                  [ka*Kron, L_AB-kd*np.eye(d_AB**2)]])              
\end{lstlisting}
Here, $\tilde{k}_a$ and $k_d$ are stored as \lstinline{k_a} and \lstinline{k_d}, respectively.

\subsubsection{Determine the initial condition of the equation}\label{Section: Intial rho}
Secondly, the initial condition for the equation are determined by the initial concentration-normalized density operators for both molecules. These concentration-normalized density operators are calculated as follows:
\begin{lstlisting}
    sigma_A = A*rho_A   #A: the concentration of molecule A
    sigma_AB = AB*rho_AB  #AB: the concentration of molecule AB
\end{lstlisting}
Here, \lstinline{rho_A} and \lstinline{rho_AB} store the initial density matrix for molecule A and AB, respectively. Next, they are mapped into the corresponding density columns, as described in Section~\ref{Section: rho in L sapce}.
\begin{lstlisting}
    sigma_A = sigma_A.flatten().reshape((-1,1))   
    sigma_AB = sigma_AB.flatten().reshape((-1,1))   
\end{lstlisting}
Specifically, the density column for molecule A (stored as \lstinline{sigma_A}) has the dimensions of $d_{\mathrm{A}}^2 \times 1$, and the density column for molecule AB (stored as \lstinline{sigma_AB}) has the dimensions of $d_{\mathrm{AB}}^2 \times 1$.  

Finally, the initial density column for the entire system (Store as \lstinline{sigma_0}) is obtained by concatenating the calculated density columns for each molecule:
\begin{lstlisting}
    sigma_0 = np.block([[sigma_A0],
                        [sigma_AB0]])   
\end{lstlisting}

\subsubsection{Solve the evolution of the system} \label{Section: evolve rho}
As a result, we solve the matrix differential equation as illustrated in Fig.~\ref{fig:LvN Eq} with the initial condition specified by the column matrix \lstinline{sigma_0}.

To solve the equation from $t=0$ to $t=t_{\mathrm{end}}$, we discretize the time domain by creating a sequence of time points from $t=0$ to $t=t_{\mathrm{end}}$ with step size $\Delta t$. Next, we use a time-stepping loop to iterate through each time step, updating the state vector of the system using: 
\begin{equation*}
    \begin{pmatrix}
        |\sigma_{\mathrm{A}}(t+\Delta t)\rrangle \\
        |\sigma_{\mathrm{AB}}(t+\Delta t)\rrangle \\
    \end{pmatrix} 
    = \exp{(\mathrm{M} \Delta t)}     
    \begin{pmatrix}
        |\sigma_{\mathrm{A}}(t)\rrangle \\
        |\sigma_{\mathrm{AB}}(t)\rrangle \\
    \end{pmatrix}.
\end{equation*}
Numerically, this iteration of the state vector is done with following code snippet:
\begin{lstlisting}
    U_evol = expm(M*dt)  
    sigma = np.matmul(U_evol,sigma) 
\end{lstlisting}
Here, \lstinline{expm} is a function from the NumPy library that computes the matrix exponential.

\subsubsection{Generate the observable for the composite system}\label{Section: solve O}
With the solved state vector of the entire system, the measured time-domain NMR signal is calculated as,
\begin{lstlisting}
    S=np.matmul(O,sigma)[0][0].real
\end{lstlisting}
Here, the row-matrix \lstinline{O} stores the observable for the composite system. To calculate this matrix, we first convert the matrix representations of the observables (e.g., $\hat{O}$) for both molecules into their corresponding row-forms (e.g., $\llangle O^{\dagger}|)$ according to Section~\ref{Section: O in L sapce},
\begin{lstlisting}
    O_A = O_A.flatten('F')
    O_AB = O_AB.flatten('F')
\end{lstlisting}
The row-matrix \lstinline{O} is derived by connecting the two row-matrices.
\begin{lstlisting}
    O = np.block([[O_A, O_AB]])
\end{lstlisting}

\subsubsection{Simulation the NMR signals} \label{Section: simulation of NMR}
The following code snippet simulates the time-domain NMR signal by sampling the measured NMR signal at each time point and storing the them in the array denoted by \lstinline{S}. 
\begin{lstlisting}
    U_evol = expm(M*dt) 
    for n in range(N_steps):
        S[n]=np.matmul(O,sigma)[0][0].real  # calcuate the measured signal
        sigma = np.matmul(U_evol,sigma)    # Evolve the system for dt
\end{lstlisting}
where \lstinline{N_steps} is the total number of time steps for the simulation.

The NMR spectral is simulated by taking a Fourier transform of the calculated time-domain data:
\begin{lstlisting}
    ZF = 32768    # Define the zero filling factor
    xf = np.arange(-Bandwidth/2,Bandwidth/2,Bandwidth/ZF)  # Calculate the frequency axis

    # Implement the zero filling of the FID data
    S_ZF = np.zeros(ZF)
    S_ZF[:N_steps] = S

    # Implement the Fourier transformation
    F[n] = np.real(fft(S_ZF)) # Perform FFT
    F[n] = scipy.fft.fftshift(S[n])  #Apply frequency shift
\end{lstlisting}

\subsubsection{Solve the steady state of the system}\label{Section: steady rho}
There are cases where only the steady state solution is required. Instead of solving the evolution of the system, its steady state corresponds to the null-space of the coefficient matrix M, i.e., the eigen-vector that have a zero eigenvalue.
\begin{lstlisting}
    sigma = null_space(M)
    rho_A = np.reshape(sigma[:d_A**2],(d_A,d_A),order='F'])
    rho_A = rho_A/np.trace(rho_A)
    rho_AB = np.reshape(sigma[d_A**2:d_AB**2],(d_AB,d_AB),order='F')
    rho_AB = rho_AB/np.trace(rho_AB)
\end{lstlisting}
Here, \lstinline{null_space} is a function from the NumPy library that computes the null space of a matrix.

\section{Results}
In this section, we provide three examples of applying the code for problems involving chemical exchange between NH$_3$ and NH$_4^{+}$ at zero and high magnetic field, as well as for the polarization transfer from parahydrogen in SABRE (signal amplification by reversible exchange) at low magnetic fields (0-20\,mT).

\subsection{Zero-field NMR of ammonia/ammonium}\label{Section: ZF_NH4}
In the example of chemical exchange between NH$_3$ and NH$_4^{+}$ at zero field \cite{barskiy2019zero}, molecules A, B, and C refer to NH$_3$, free H$^{+}$ in the solvent, and NH$_{4}^{+}$, respectively. The essential parameters for the simulation are listed in Table~\ref{Table: NH4_param_matrix_ZF}.

\begin{figure}[!ht]
    \centering
    \includegraphics{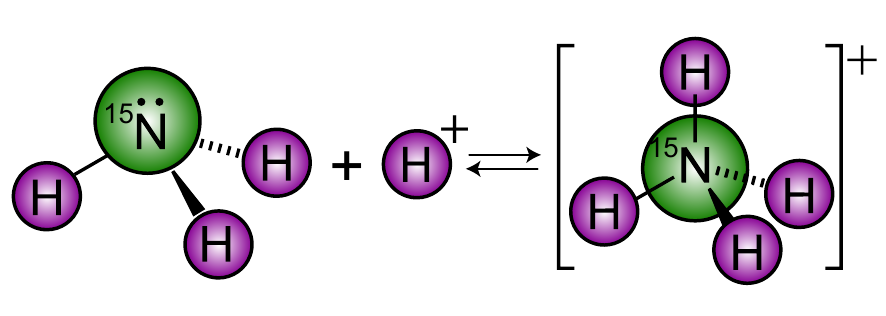}
    \caption{The chemical diagram for the chemical exchange between ammonia and ammonium.}
    \label{fig: Chemical exhcange ammonia}
\end{figure}

The chemical diagram of the chemical exchange is shown in Fig.~\ref{fig: Chemical exhcange ammonia}. 

Below, we outline the steps for the simulation and the full code is provided as the appendix.

\begin{table}
\begin{center}
\footnotesize
\begin{tabular}{|c |c |c |} 
 \hline
      & NH$_3$  &NH$_{4}^{+}$  \\ 
 \hline
 Number of spins & 4  & 5 \\ 
 \hline
  J$_0$ (Hz) & $\begin{pmatrix}
    0 &  J_{\rm NH} &  
    J_{\rm NH} & J_{\rm NH}\\
     0 &  0 &  J_{\rm HH} & J_{\rm HH}\\
     0 & 0 & 0 & J_{\rm HH} \\
     0 &  0 &  0 & 0\\
    \end{pmatrix}$ &  
    $\begin{pmatrix}
    0 &  J_{\rm NH} &  J_{\rm NH} & J_{\rm NH} & J_{\rm NH}\\
     0 &  0 &  J_{\rm HH} & J_{\rm HH} & J_{\rm HH}\\
     0 & 0 & 0 & J_{\rm HH} & J_{\rm HH} \\
     0 & 0 & 0 & 0 & J_{\rm HH} \\
     0 &  0 &  0 & 0 & 0\\
    \end{pmatrix}$ \\ 
 \hline
T$_1$ (s) & $\begin{pmatrix}
    5 &  1 &  1 & 1\\
    \end{pmatrix}$ 
    & 
    $\begin{pmatrix}
    5 &  1 &  1 & 1 & 1 \\
    \end{pmatrix}$ \\
\hline
Initial density matrix & $\gamma_{\rm 15N}$I$_{1z}$+$\gamma_{\rm 1H}$ (I$_{2z}$+I$_{3z}$+I$_{4z}$+I$_{5z}$)  &  $\gamma_{\rm 15N}$I$_{1z}$+$\gamma_{\rm 1H}$ (I$_{2z}$+I$_{3z}$+I$_{4z}$) \\
\hline
Observable &  $\gamma_{\rm 15N}\mathrm{I}_{1z}+\gamma_{\rm 1H} (\mathrm{I}_{2z}+\mathrm{I}_{3z}+\mathrm{I}_{4z}+\mathrm{I}_{5z})  $  &    $\gamma_{\rm 15N}\mathrm{I}_{1z}+\gamma_{\rm 1H} (\mathrm{I}_{2z}+\mathrm{I}_{3z}+\mathrm{I}_{4z}) $  \\
\hline
\end{tabular} 
\end{center}
\caption{Parameters used for calculating the zero-field NMR spectra of NH$_3$/NH$_4^{+}$ molecules undergoing chemical exchange. The table includes information on the number of spins, the \textit{J}-coupling constants, $T_1$ relaxation times, initial density matrices, and observables. The \textit{J}-couplings applied are $J_{\mathrm{HH}}=-16.9~\mathrm{Hz}$ and $J_{\mathrm{NH}}=-73.4~\mathrm{Hz}$ \cite{barskiy2019zero}.}
\label{Table: NH4_param_matrix_ZF}
\end{table}

\noindent\textbf{Step 1: Specify the spin systems} \newline
In this step, we define the Hilbert spaces for each molecule and choose the bases spanning the corresponding Hilbert spaces. 

The Hilbert spaces for ammonium and ammonia are defined based on the molecule notations [$^{15}$NHHHH]$^{+}$ and $^{15}$NHHH, respectively. The Zeeman product basis within the defined Hilbert spaces are adopted to span the corresponding Hilbert spaces.\newline

\noindent\textbf{Step 2: Construct the Hamiltonian superoperator} \newline
In this step, we first calculate the Hamiltonian operators for both molecules using the code snippet for the zero-field Hamiltonian. The required inputs are the matrix J$_0$ for each molecule and they are provided in the third row of the table.

The calculated Hamiltonians are then map them into the related Hamiltonian superoperators using the code snippet provided in Section~\ref{Section: Hamiltonian Superoperator}.\newline

\noindent\textbf{Step 3: Construct the relaxation superoperator}  \newline
In this step, the relaxation superoperator is generated using the code snippet given in Section~\ref{Section: Relaxation Superoperator}. The required matrices T$_1$ for both molecules can be found in the fourth row of the table.\newline

\noindent\textbf{Step 4: Construct the chemical exchange superoperator}  \newline
The calculation of the partial trace superoperators and the Kronecker product superoperators accounting for the chemical exchange between ammonia and ammonium has been discussed as the example in Section~\ref{Section: Partial trace superoperator} and Section~\ref{Section: Kronecker product Superoperator}.\newline

\noindent\textbf{Step 5: Construct the coefficient matrix for the composite system}\newline
In this step, we calculate the coefficient matrix of Eq.~\eqref{Eq: LvN with chemical exchange} according to Section~\ref{Section: generate coefficient matrix}. Note here we express the effective association rate as $\tilde{k}_a = k_d W$ with $W=\mathrm{[NH_4^{+}]/[NH_3]}$ in the code.\newline

\noindent\textbf{Step 6: Construct the initial state vector for the composite system} \newline
In this step, we calculate the initial condition of Eq.~\eqref{Eq: LvN with chemical exchange} according to Section~\ref{Section: Intial rho}. The forms of the initial density operators for both molecules can be found in the fifth row of the table. For simplicity, we assume the concentrations of ammonium and ammonium are $1$ and $1/W$, respectively, when constructing the initial concentration-normalized density operator. \newline

\noindent\textbf{Step 7: Construct the observable for the composite system} \newline
In this step, we generate the row-matrix representing the observable $\llangle O|$ for the composite system according to Section~\ref{Section: solve O}. \newline

\begin{figure}[ht]
    \centering
    \includegraphics[scale=.75]{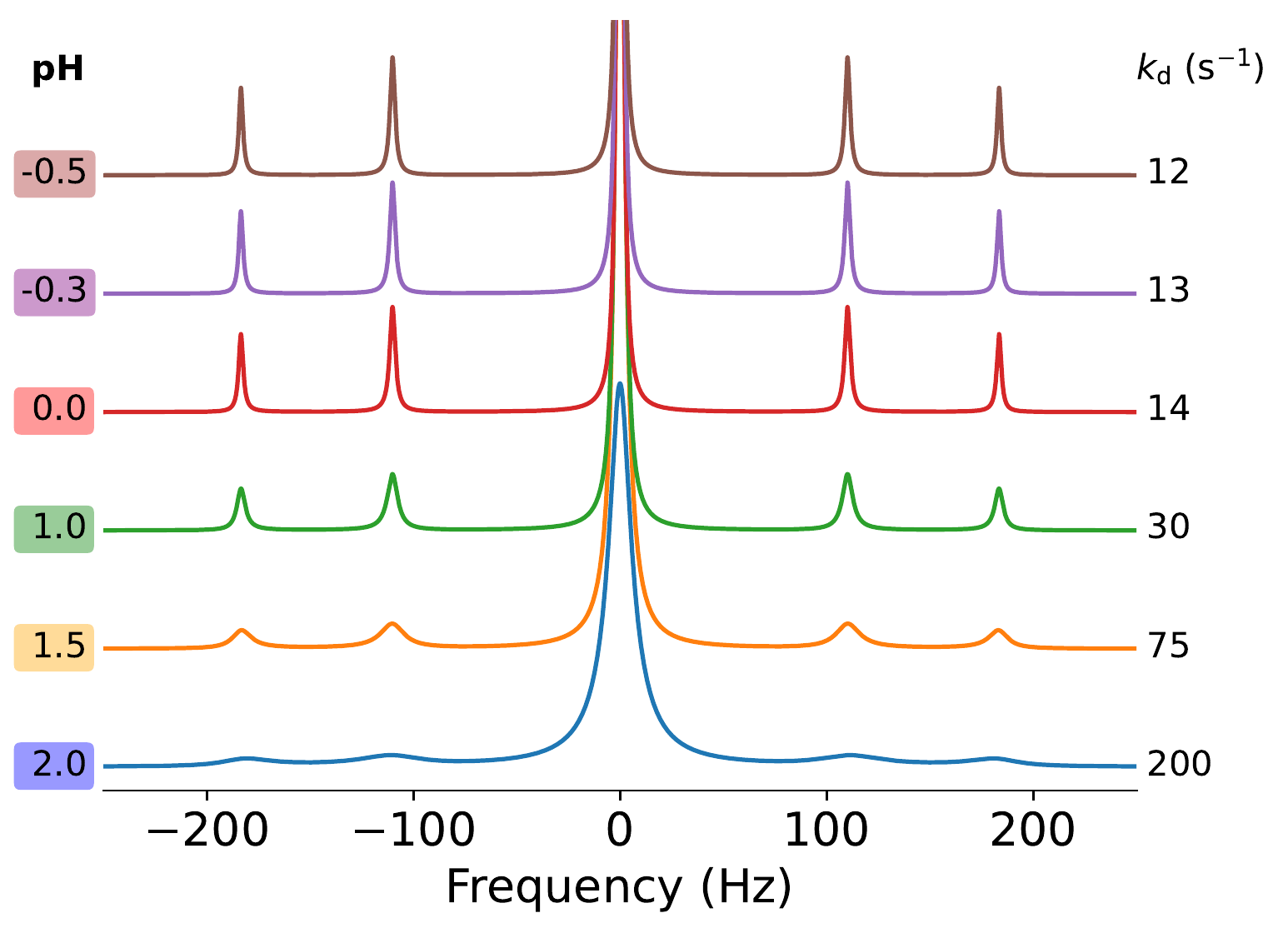}
    \caption{Zero-field NMR spectra of NH$_3$/NH$_4^{+}$ molecules in an exchanging system as a function of pH.}
    \label{fig:NH4_ZF}
\end{figure}

\noindent\textbf{Step 9: Simulation of the spectra} \newline
Finally, the measured NMR spectra are simulated according to Section~\ref{Section: simulation of NMR} and the results are presented in Fig.~\ref{fig:NH4_ZF}.

\subsection{High-field NMR of ammonia/ammonium}\label{Section: HF_NH4}
\begin{table}
\begin{center}
\footnotesize
\begin{tabular}{|c |c |c |} 
 \hline
      & NH$_3$  &NH$_{4}^{+}$  \\ 
 \hline
 Num of spins & 4  & 5 \\ 
 \hline
 Gyro & $\begin{pmatrix}
    \gamma_{15\mathrm{N}} &  \gamma_{1\mathrm{H}} &  \gamma_{1\mathrm{H}} & \gamma_{1\mathrm{H}}\\
    \end{pmatrix}$  
    & $\begin{pmatrix}
    \gamma_{15\mathrm{N}} &  \gamma_{1\mathrm{H}} &  \gamma_{1\mathrm{H}} & \gamma_{1\mathrm{H}} & \gamma_{1\mathrm{H}} \\
    \end{pmatrix}$ \\
 \hline
CS (ppm) & $\begin{pmatrix}
    0 &  1 &  1 & 1\\
    \end{pmatrix}$ 
    & $\begin{pmatrix}
    0 &  0 &  0 & 0 & 0 \\
    \end{pmatrix}$ \\
  \hline
  J$_0$ (Hz) & $\begin{pmatrix}
    0 &  0 &  0 & 0\\
     0 &  0 &  J_{\rm HH} & J_{\rm HH}\\
     0 & 0 & 0 & J_{\rm HH} \\
     0 &  0 &  0 & 0\\
    \end{pmatrix}$ &  
    $\begin{pmatrix}
    0 &  0 &  0 & 0 & 0\\
     0 &  0 &  J_{\rm HH} & J_{\rm HH} & J_{\rm HH}\\
     0 & 0 & 0 & J_{\rm HH} & J_{\rm HH} \\
     0 & 0 & 0 & 0 & J_{\rm HH} \\
     0 &  0 &  0 & 0 & 0\\
    \end{pmatrix}$ \\ 
 \hline
   J$_1$ (Hz)& $\begin{pmatrix}
    0 &  J_{\rm NH} &  
    J_{\rm NH} & J_{\rm NH}\\
     0 &  0 &  0 & 0\\
     0 &  0 &  0 & 0 \\
     0 &  0 &  0 & 0\\
    \end{pmatrix}$ &  
    $\begin{pmatrix}
    0 &  J_{\rm NH} &  J_{\rm NH} & J_{\rm NH} & J_{\rm NH}\\
     0 &  0 &  0 & 0 & 0\\
     0 &  0 &  0 & 0 & 0 \\
     0 &  0 &  0 & 0 & 0\\
     0 &  0 &  0 & 0 & 0\\
    \end{pmatrix}$ \\ 
 \hline
T$_1$ (s) & $\begin{pmatrix}
    5 &  1 &  1 & 1\\
    \end{pmatrix}$ 
    & 
    $\begin{pmatrix}
    5 &  1 &  1 & 1 & 1 \\
    \end{pmatrix}$ \\
\hline
Initial density matrix & $\gamma_{\rm 15N}$I$_{1x}$+$\gamma_{\rm 1H}$ (I$_{2z}$+I$_{3z}$+I$_{4z}$+I$_{5z}$)  &  $\gamma_{\rm 15N}$I$_{1x}$+$\gamma_{\rm 1H}$ (I$_{2z}$+I$_{3z}$+I$_{4z}$) \\
\hline
Observable &  $\mathrm{I}_{1x}-i\mathrm{I}_{1y}$  &    $\mathrm{I}_{1x}-i\mathrm{I}_{1y}$  \\
\hline
\end{tabular}
\end{center}
\caption{Parameters used for simulating the high-field (18.8~T) $^{15}$N NMR spectra of NH$_3$/NH$_4^{+}$ molecules in an exchanging system. The table includes information on the number of spins, the matrix Gyro and CS, the \textit{J}-coupling constants, $T_1$ relaxation times, initial density matrices, and observables. The \textit{J}-couplings applied are $J_{\mathrm{HH}}=-16.9~\mathrm{Hz}$ and $J_{\mathrm{NH}}=-73.4~\mathrm{Hz}$ \cite{barskiy2019zero}.}
\label{Table: NH4_param_matrix_HF}
\end{table}

The steps for simulating the high-field $^{15}$N NMR spectra of ammonia/ammonium \cite{barskiy2019zero} are the same as that in the low-field case except for calculation of the Hamiltonian. In this example, the Hamiltonian should be calculate using the code snippet discussed for rotating-frame Hamiltonian.  All related parameters are given in Table~\ref{Table: NH4_param_matrix_HF} and the full code is provided in appendix. The simulated results are shown in Fig.~\ref{fig:NH4_HF}.

\begin{figure}[ht]
    \centering
    \includegraphics[scale=.75]{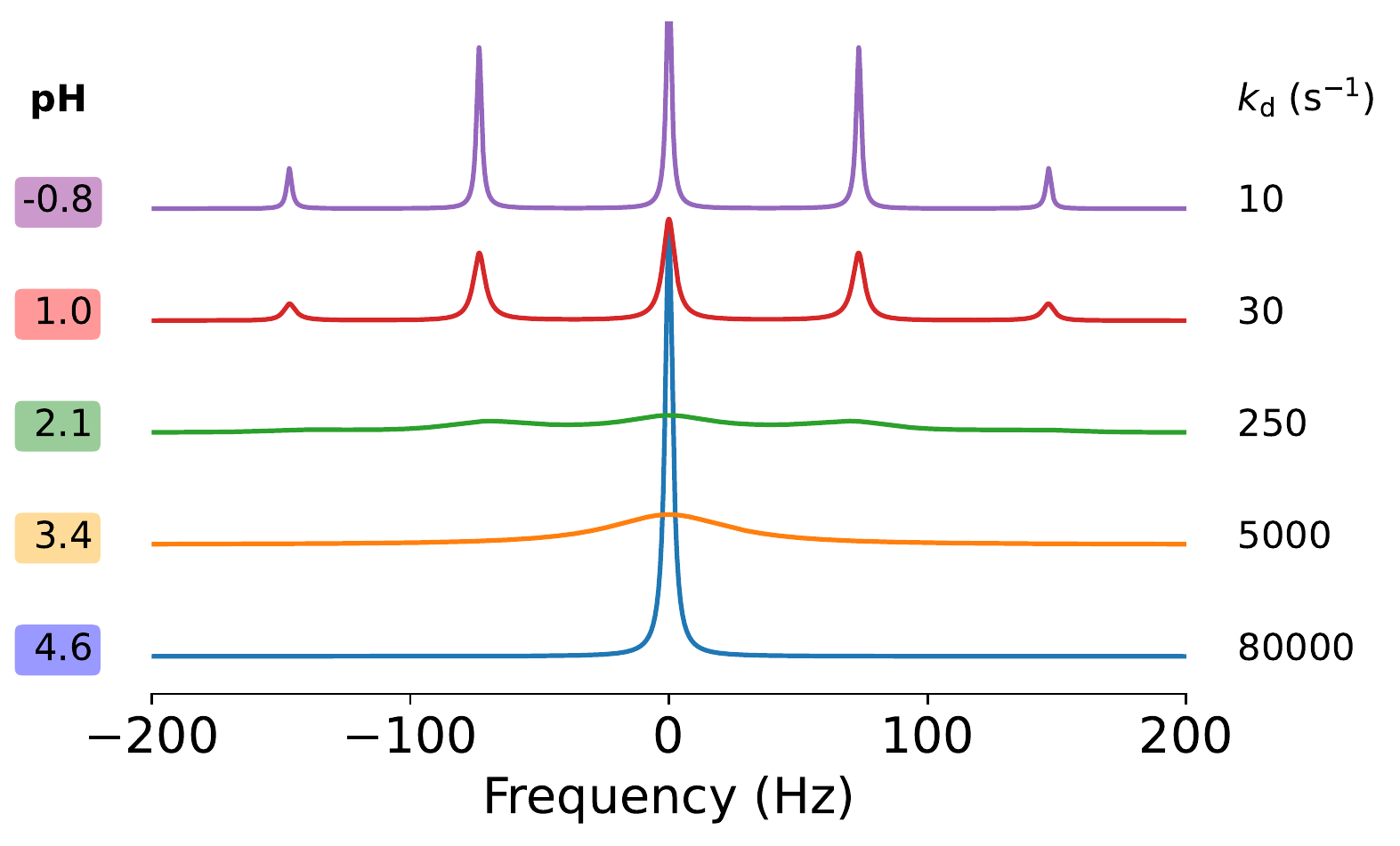}
    \caption{High-field (18.8 T) $^{15}$N NMR spectra of NH$_3$/NH$_4^{+}$ molecules in an exchanging system as a function of pH.}
    \label{fig:NH4_HF}
\end{figure}

\subsection{Proton-SABRE field-dependence profiles}

In this section, we aim to explore the magnetic field dependence of the polarization transfer from parahydrogen to substrate protons in SABRE experiments \cite{pravdivtsev2015spin}. 

\begin{figure}[ht]
    \centering
    \includegraphics[scale=.75]{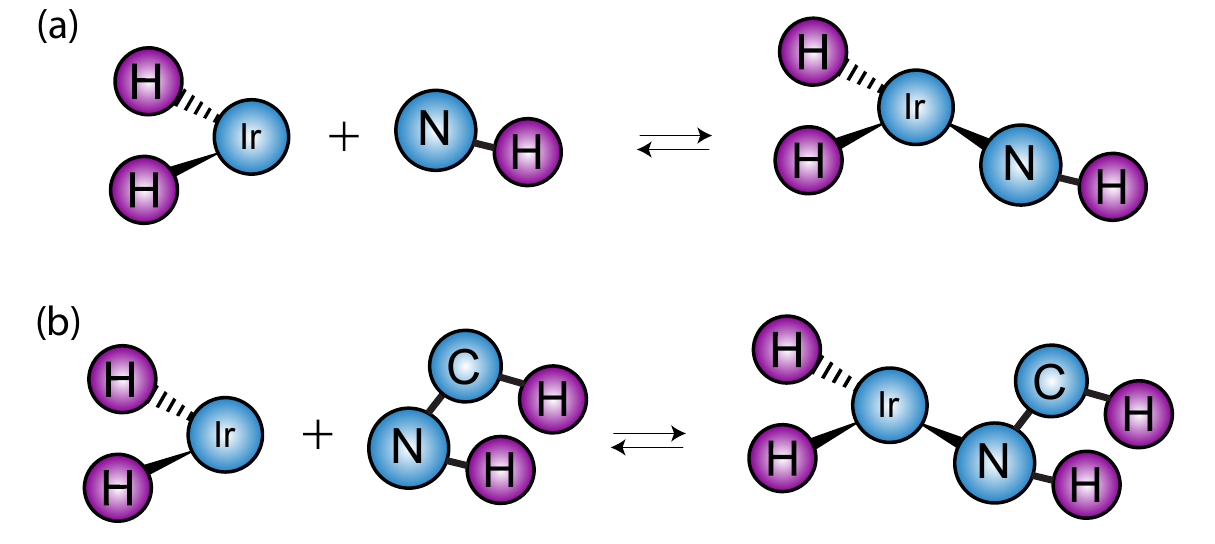}
    \caption{Chemical exchange diagram illustrating the SABRE experiment with substrates containing (a) a single proton and (b) two protons coupled through $J$-coupling. Atoms that do not participate in spin dynamical evolution are represented by blue balls. }
    \label{fig:SABRE diagram}
\end{figure}

We consider two distinct situations where the substrate contains (a) a single proton, and (b) two protons that are coupled through \textit{J}-coupling (Fig.~\ref{fig:SABRE diagram}). In the first case, we expect a single maximum at around 6~mT given proton-proton coupling of -7\,Hz between the hydrides and a difference between chemical shift of about 24 ppm between the hydride and a bound-substrate's proton \cite{barskiy2019sabre}. In the second case, the two maxima are observed: in addition to the first peak at a similar field of 7\,mT, a second peak is present at around 13~mT. The shift of the main peak and the emergence of the second peak are attributed to the presence of internal \textit{J}-couplings within the substrate. 
The chemical diagram of the involved reaction is illustrated in Fig.~\ref{fig:SABRE diagram}.

To facilitate the discussion, we label the hydrides as H$_1$ and H$_2$, respectively. For the substrate in Fig.~\ref{fig:SABRE diagram}(a), we label its proton as H$_{\mathrm{S1}}$. For the substrate in Fig.~\ref{fig:SABRE diagram}(b), we label its protons as H$_{\mathrm{S1}}$ and H$_{\mathrm{S2}}$.

Below we outline the key steps for the simulation while the complete codes are included in the appendix. In this particular example, molecules A and AB refer to the free substrate, and the SABRE polarization-transfer catalyst (PTC) complex, respectively. Table~\ref{Table: SABRE substrate 1} and Table~\ref{Table: SABRE substrate 2} provide all the relevant NMR parameters.

The main assumption of this calculation is that the state of the hydride spins remain constant during the SABRE process (i.e., singlet state). Moreover, we assume only one substrate molecule per complex. More complicated situations accounting for the evolution of the spin state of the hydrides and the presence of more than one substrate per complex have been demonstrated \cite{pravdivtsev2019simulating}. 

\noindent\textbf{Step 1: Specify the spin systems}  \newline
In the case of the chemical exchange depicted in Fig.~\ref{fig:SABRE diagram}(a), we define the Hilbert spaces for the substrate and PTC complex according to their molecule notations H$_{\mathrm{S1}}$ and H$_1$H$_2$H$_{\mathrm{S1}}$, respectively. 

While in the case of the chemical exchange depicted in Fig.~\ref{fig:SABRE diagram}(b), the Hilbert spaces for the substrate and PTC complex are defined according to their molecule notations H$_{\mathrm{S1}}$H$_{\mathrm{S2}}$ and H$_1$H$_2$H$_{\mathrm{S1}}$H$_{\mathrm{S1}}$, respectively. \newline

\noindent\textbf{Step 2: Construct the Hamiltonian superoperator} \newline
We use the code snippet for the rotating-frame Hamiltonian. Since there are only protons in the system, there is no hetero-nuclear \textit{J}-coupling, and, thus, the J$_1$ matrix is zero. \newline

\noindent\textbf{Step 3: Construct the relaxation superoperator}\newline
We calculate the relaxation superoperators given the input matrix T$_1$. \newline

\noindent\textbf{Step 4: Construct the chemical exchange superoperators}  \newline
Let us discuss the calculation of the partial trace superoperators first. For both cases, the reactions are of the form as discussed in Case 2 under section~\ref{Section: Partial trace superoperator}. 

In the case of the single-proton substrate, the molecules X, Y are assigned as H$_1$H$_2$, H$_{\mathrm{S1}}$, respectively. Consequently, the dimensions of their respective Hilbert space are $d_\mathrm{X}=4$ and $d_\mathrm{Y}=2$.

Similarly, in the case of the two-proton substrate, the molecules X and Y are assigned as H$_1$H$_2$, H$_{\mathrm{S1}}$H$_{\mathrm{S2}}$, respectively, leading to dimensions $d_\mathrm{X}=4$ and $d_\mathrm{Y}=4$ for their corresponding Hilbert spaces.

These dimensions enter as the inputs for calculating the partial trace superoperators using the code snippet discussed in Case 2 under section~\ref{Section: Partial trace superoperator}.

The main assumption is that the state of the hydride spins remain constant during the SABRE process (i.e., singlet state). Other computation approaches allow the states of the hydride spins to evolve as well \cite{pravdivtsev2019simulating}. 

The calculation of the Kronecker product superoperators are based on the discussion in Case 2 under the section~\ref{Section: Kronecker product Superoperator}.

The assignments and dimensions for molecules X and Y remain consistent with those used for calculating the partial trace superoperator. An additional input required is the density matrix of molecule X, which is calculated as: 
\begin{lstlisting}
    f_pH2 = 1  #pH2 ratio
    #density operator of pure pH2
    rho_H2 = np.array([[0,0,0,0],[0,0.5,-0.5,0],[0,-0.5,0.5,0],[0,0,0,0]]) 
    rhoX = (4*f_pH2-1)/3*rho_H2+(1-f_pH2)/3*np.eye(4)
\end{lstlisting}

We enter the dimensions as well the density matrix of molecule X to calculate the Kronecker product superoperators using the code snippet discussed in Case 2 under section~\ref{Section: Kronecker product Superoperator}.

\noindent\textbf{Step 5: Calculate the coefficient matrix for the composite system}\newline
The coefficient matrix is calculate according to Section~\ref{Section: generate coefficient matrix}.\newline

\noindent\textbf{Step 6: Calculate the steady-state of the composite system}\newline
In this step, we calculate the steady state of the system according to Section~\ref{Section: steady rho}.

\noindent\textbf{Step 7: Calculate the steady-state polarization level}\newline
The polarization level is calculated as:
\begin{equation*}
    P = 2 \cdot \mathrm{Tr}(\hat{\rho} \hat{I}_z),
\end{equation*}
where $\hat{\rho}$ is the steady-state density operator of the substrate and $\hat{I}_z$ is the angular momentum operator of the substrate proton of interest.\newline

In the code snippet below, the steady-state density operator is stored as \lstinline{rho}. For single-proton substrate, the calculated polarization level is saved in the array \lstinline{P}.
\begin{lstlisting}
    P = 2*np.trace(np.matmul(rho,I_S[2][0])).real
\end{lstlisting}

For the two-proton substrate, the calculated polarization levels are saved in the arrays \lstinline{P1} and \lstinline{P2} for the amino proton and the methylene proton, respectively. 
\begin{lstlisting}
    P1 = 2*np.trace(np.matmul(rho,I_S[2][0])).real
    P2 = 2*np.trace(np.matmul(rho,I_S[2][1])).real
\end{lstlisting}

\begin{table}[ht]
\begin{center}
\footnotesize
\begin{tabular}{|c |c |c |} 
 \hline
      & Substrate  & Complex  \\ 
 \hline
 Num of spins & 1  & 3 \\ 
 \hline
%  Specification &  H$_1$   & H$_a$, H$_b$, H$_1$  \\
%  \hline
CS (ppm) & $\begin{pmatrix}
    8.3\\
    \end{pmatrix}$ 
    & $\begin{pmatrix}
    -22 &  -22 &  8.3  \\
    \end{pmatrix}$ \\
  \hline
  J$_0$ (Hz)& $\begin{pmatrix}
    0 \\
    \end{pmatrix}$ &  
    $\begin{pmatrix}
     0 &  -7 &  1 \\
     0 &  0  &  0 \\
     0 &  0  &  0  \\
    \end{pmatrix}$ \\ 
 \hline
T$_1$ (s) & $\begin{pmatrix}
    10\\
    \end{pmatrix}$ 
    & 
    $\begin{pmatrix}
    1 &  1 & 3 \\
    \end{pmatrix}$ \\
% \hline
% Partial trace &    & XY$\rightarrow$Y: d$_{\rm X}$=4, d$_{\rm Y}$=2   \\
% \hline
% Kronecker product & Y$\rightarrow$XY:  d$_{\rm X}$=4, d$_{\rm Y}$=2, $\hat{\rho}_{\rm X}=\hat{\rho}_{\rm H_2}$ &  \\
\hline
\end{tabular}
\end{center}
\caption{Parameters used for simulating the field dependence of polarization transfer in SABRE from parahydrogen to the substrate containing one proton. The table includes information on the number of spins, \textit{J}-coupling constants, and matrix T$_1$.}
\label{Table: SABRE substrate 1}
\end{table}

\begin{table}[ht]
\begin{center}
\footnotesize
\begin{tabular}{|c |c |c |} 
 \hline
      & Substrate  & Complex  \\ 
 \hline
 Num of spins & 2  & 4 \\ 
 \hline
%  Specification &  H$_1$, H$_2$   & H$_a$, H$_b$, H$_1$, H$_2$ \\
%  \hline
CS (ppm) & $\begin{pmatrix}
    1.74 & 3.9 \\
    \end{pmatrix}$ 
    & $\begin{pmatrix}
    -22 &  -22 &  1.74 & 3.9  \\
    \end{pmatrix}$ \\
  \hline
  J$_0$ (Hz) & $\begin{pmatrix}
    0 & 6.6\\
    0 &  0 \\
    \end{pmatrix}$ &  
    $\begin{pmatrix}
     0 &  -7 &  1  &  0 \\
     0 &  0  &  0  &  0 \\
     0 &  0  &  0  &  6.6\\
     0 &  0  &  0  &  0 \\
    \end{pmatrix}$ \\ 
 \hline
T$_1$ (s)& $\begin{pmatrix}
    10 & 10\\
    \end{pmatrix}$ 
    & 
    $\begin{pmatrix}
    1 &  1 & 3 & 3\\
    \end{pmatrix}$ \\
\hline
% Partial trace &    & XY$\rightarrow$Y: d$_{\rm X}$=4, d$_{\rm Y}$=4   \\
% \hline
% Kronecker product & Y$\rightarrow$XY:  d$_{\rm X}$=4, d$_{\rm Y}$=4, $\hat{\rho}_{\rm X}=\hat{\rho}_{\rm H_2}$ &  \\
% \hline
\end{tabular}
\end{center}
\caption{Parameters used for simulating the field dependence of polarization transfer in SABRE from parahydrogen to the substrate containing two protons. The table includes information on the number of spins, the \textit{J}-coupling constants, the matrix T$_1$.}
\label{Table: SABRE substrate 2}
\end{table}

\begin{figure}[ht]
    \centering
    \includegraphics[scale=.75]{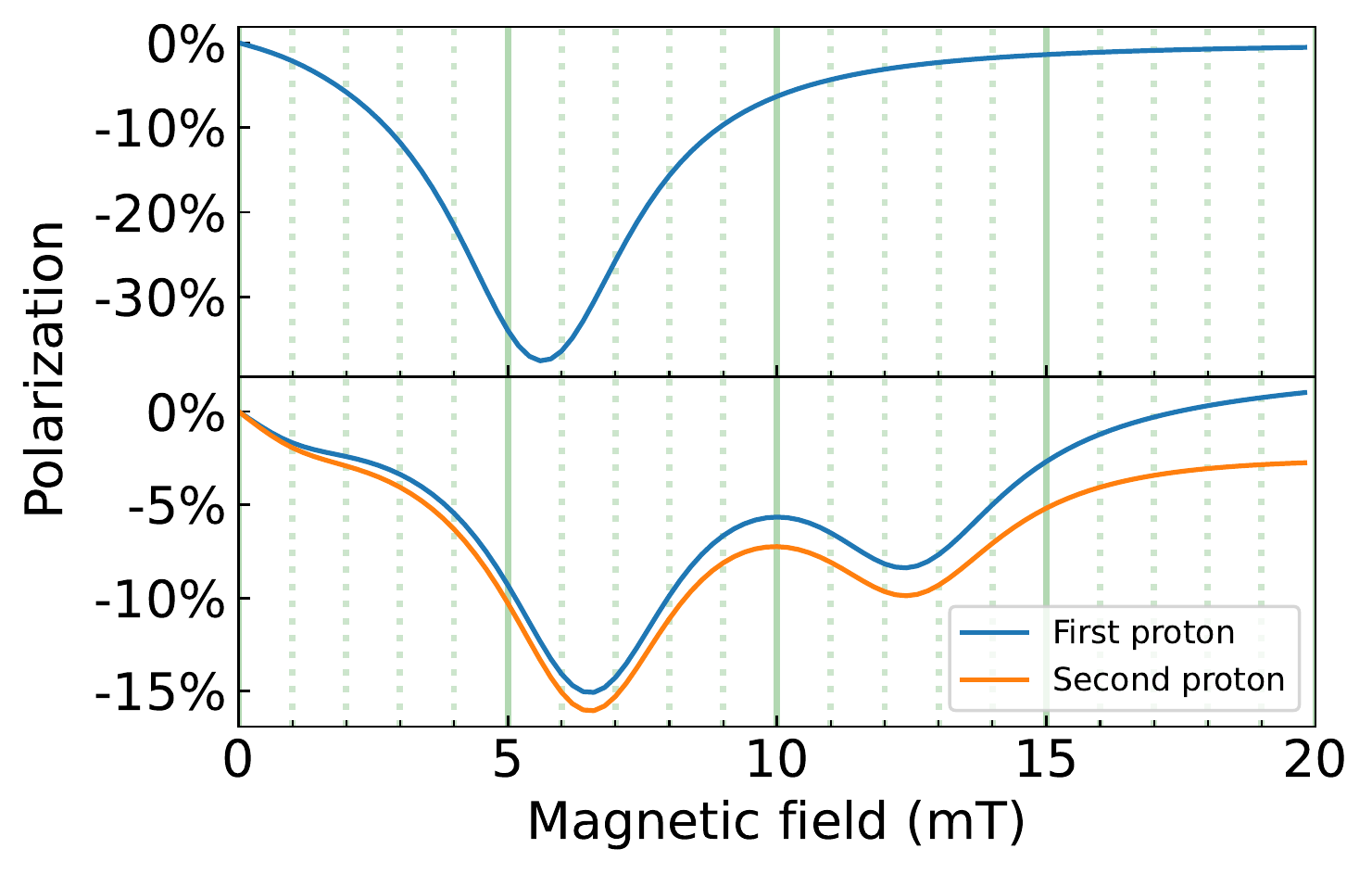}
    \caption{Magnetic-field dependence of the polarization transfer in SABRE (signal amplification by reversible exchange) from parahydrogen to the substrate containing one proton (top) and the substrate containing two protons (bottom).}
    \label{fig:SABRE}
\end{figure}

\section{Conclusions}

In this work, we demonstrate the essential tools of linear algebra (matrix multiplication, Kronecker product, mapping between the matrices in the Hilbert space and the Liouville space, etc.) for calculating full nuclear spin dynamics in the presence of chemical exchange. As examples, we show how to compute the NMR spectra at high and zero field for the chemically exchanging ammonia/ammonium system as well as polarization transfer in SABRE. The presented approach is general, it allows to describe spin dynamics fully and can be extended further by incorporating the effects of RF pulses. However, since dimensions of the matrices in the Liouville space grow exponentially ($16^{N}$, where $N$ is the number of spins-1/2), utilizing the presented methods for large spin systems is challenging. Further studies and/or different approaches are necessary when dealing with large spin systems and equations containing non-linear terms.

\printbibliography

\end{document}